\documentclass[manuscript]{geophysics}
\usepackage[T1]{fontenc}
\usepackage[utf8]{inputenc}
\usepackage{amsmath,amssymb,amsfonts}
\usepackage{booktabs,multirow,makecell,tabularx,adjustbox}
\usepackage{float,placeins}
\usepackage{url}
\usepackage[unicode,hidelinks,hypertexnames=false]{hyperref}

\renewcommand{\today}{November 8, 2026}
\begin{document}
\title{Geometry-Aware Transformer for Five-Dimensional Seismic Noise Attenuation and First-Arrival Picking}
\address{
\footnotemark[1]School of Mathematics, Harbin Institute of Technology, Harbin, 150001, China\\
\footnotemark[2]School of Earth and Space Sciences, Peking University, Beijing, 100871, China\\
\textsuperscript{*}Corresponding author. E-mail: jma@hit.edu.cn.
}
\author{Tianxiang Gao\footnotemark[1], Jianwei Ma\footnotemark[1]\textsuperscript{,}\footnotemark[2]}
\righthead{GeoFormer}
\maketitle
\thispagestyle{empty}
\begin{abstract}
We propose GeoFormer, a Geometry-Aware Transformer architecture specifically designed for five-dimensional (5D) prestack seismic data. Unlike Vision Transformer, whose tokens are extracted from 2D patches and primarily encode visual patterns, GeoFormer is designed for prestack seismic data by explicitly incorporating acquisition geometry. Each seismic trace is represented by its waveform and associated acquisition-geometry information. The latter includes four source--receiver coordinates and two geometric attributes, offset and relative elevation, where relative elevation is the receiver elevation minus the source elevation. GeoFormer therefore performs trace-level tokenization, where each token integrates the seismic waveform with its associated geometric attributes. To exploit these geometric attributes, GeoFormer employs positional encoding of source and receiver coordinates to capture spatial relationships among traces and introduces three geometry injection mechanisms operating at different levels of the Transformer pipeline. At the token level, GeomMLP replaces the classification token with a per-trace geometric representation derived from offset and elevation. At the normalization level, GeomAdaLN replaces uniform layer normalization with geometry-conditioned feature modulation. At the attention level, GeomAttnBias injects a parameter-free physical prior that geometrically proximate traces should attend more to each other. We validate GeoFormer on two representative seismic processing tasks: seismic noise attenuation and first-arrival picking. Together, these tasks evaluate the capability of GeoFormer for both regression and segmentation in seismic data processing.
Extensive experiments demonstrate that GeoFormer consistently outperforms the baseline methods on these two tasks. Ablation studies demonstrate that GeomMLP, GeomAdaLN, and GeomAttnBias each contribute to the performance improvements of GeoFormer.
\end{abstract}

\section{Introduction}

Transformer architecture was originally introduced for sequence transduction and has since become a widely adopted framework in deep learning \citep{vaswani2017attention}. Its self-attention mechanism models pairwise interactions between tokens, enabling the capture of long-range dependencies. This design has subsequently been extended beyond natural language processing. Vision Transformer (ViT) demonstrated that a standard Transformer could be applied directly to sequences of image patches for visual recognition \citep{dosovitskiy2020image}. 

Despite its success, the standard Transformer faces challenges in computational efficiency and structural inductive bias. FlashAttention improves attention efficiency through IO-aware computation \citep{dao2022flashattention}, while Mamba explores an alternative sequence modeling architecture based on selective state spaces \citep{gu2024mamba}. In computer vision, Swin Transformer employs shifted-window attention and hierarchical feature representations to improve efficiency and capture multi-scale structures \citep{liu2021swin}. Restormer redesigns attention and feed-forward modules to enable efficient long-range interactions for high-resolution image restoration \citep{zamir2022restormer}. In medical image analysis, TransUNet combines convolutional features with Transformer-based global context modeling to capture local details and long-range dependencies for medical image segmentation \citep{chen2024transunet}, while UNETR formulates 3D medical image segmentation as a sequence-to-sequence prediction task using a Transformer encoder and a U-shaped decoder with multiscale skip connections \citep{hatamizadeh2022unetr}. In remote sensing, SatMAE adapts Transformer pretraining to satellite imagery through specialized positional encodings for temporal and multispectral information and tailored masking strategies \citep{cong2022satmae}. Similarly, SpectralFormer introduces group-wise spectral embeddings to capture local dependencies across neighboring spectral bands in hyperspectral image classification \citep{hong2022spectralformer}.
In embodied intelligence, BAKU develops an efficient Transformer architecture for multi-task robot policy learning by combining multimodal observations, action chunking, and specialized action heads \citep{haldar2024baku}

Transformer architectures have also been explored for seismic data processing. Transformer variants have been applied to seismic denoising and random-noise attenuation \citep{li2024swin,zhao2026seismic,sun2025removing}, while multiscale and Swin-based architectures, as well as latent diffusion Transformers, have been investigated for seismic interpolation and reconstruction \citep{guo2023seismic,liu2023consecutively,zhang2024stugan,wang2026seisrdt}. Transformer-based feature extraction has also been explored for two- and three-dimensional first-arrival picking \citep{jiang2023seismic,jiang20243}. These studies demonstrate the potential of self-attention to capture nonlocal seismic structures and long-range dependencies. However, the effectiveness of Transformer-based methods depends on the task and architectural design, and they have not consistently demonstrated clear advantages over other methods. We attribute this limitation to the image-centric representations adopted by existing Transformers, which overlook the inherent geometric properties of seismic data.

Many existing Transformer-based approaches adopt an image-centric representation, dividing seismic data into two-dimensional patches and treating them as visual tokens. Following the Vision Transformer paradigm \citep{dosovitskiy2020image}, this strategy represents seismic data as patches on a regular two-dimensional grid. However, prestack seismic data are organized on a fundamentally different domain. They are defined over a five-dimensional (5D) acquisition domain \((s_x,s_y,r_x,r_y,t)\in\mathbb{R}^5\), where \((s_x,s_y)\) and \((r_x,r_y)\) are the source and receiver coordinates and \(t\) is time. Each trace fixes the four acquisition coordinates and samples the waveform along \(t\), so a shot gather is a three-dimensional slice of this domain at a fixed source position. Because this domain is irregularly sampled, reshaping it into a regular 2D patch grid implicitly fixes or averages the source and receiver coordinates and retains only a spatial-temporal slice. Each trace is associated with a waveform and its acquisition geometry, including source and receiver coordinates and derived attributes such as offset and relative elevation. The waveform describes the recorded seismic signal, whereas the geometric attributes characterize the trace's acquisition position and spatial relationships with other traces. Consequently, representing prestack seismic data solely as two-dimensional image patches may fail to fully exploit their underlying geometry.

In our experiments, Vision Transformers underperform on two representative seismic processing tasks, seismic noise attenuation and first-arrival picking, compared with the proposed geometry-aware approach. These results suggest that the limitation lies not necessarily in the Transformer architecture itself, but in the mismatch between conventional image-based representations and the structure of prestack seismic data. This observation motivates a trace-level representation that explicitly incorporates acquisition geometry.

We therefore propose GeoFormer, a geometry-aware Transformer for seismic data processing. Rather than treating a seismic gather as a collection of two-dimensional image patches, GeoFormer represents each trace as a token that combines its signal with per-trace geometric attributes. This design enables acquisition geometry to be incorporated throughout the Transformer through three complementary mechanisms. GeomMLP provides a geometric representation for each trace token, GeomAdaLN modulates feature representations based on trace geometry, and GeomAttnBias introduces a geometry-dependent attention bias that favors interactions between geometrically proximate traces. Together, these mechanisms enable self-attention to model inter-trace relationships using both seismic signal content and acquisition geometry.

The main contributions of this work are:

\begin{enumerate}
\item \textbf{A geometry-aware representation for prestack seismic data.} We propose GeoFormer, a Transformer architecture that represents each seismic trace as a token by jointly encoding its waveform and acquisition geometry. This trace-level representation explicitly accounts for the irregular geometric structure of prestack seismic data.

\item \textbf{A multi-level geometry injection framework.} We introduce three complementary mechanisms, GeomMLP, GeomAdaLN, and GeomAttnBias, to incorporate geometric information at the token, normalization, and attention levels, respectively. Together, they enable the Transformer to exploit acquisition geometry throughout the feature extraction process.

\item \textbf{A unified evaluation across representative seismic processing tasks.} We evaluate GeoFormer on seismic noise attenuation and first-arrival picking, representing regression and segmentation tasks, respectively. Experiments show consistent improvements over baselines on both tasks, while ablation studies verify the contribution of each geometry module.

\end{enumerate}

\section{Method}

\subsection{Architecture of the GeoFormer}

\begin{figure}[!htbp]
\centering
\begin{adjustbox}{max totalsize={\textwidth}{0.76\textheight},center}
\begin{minipage}{\textwidth}
\centering
\includegraphics[width=\textwidth]{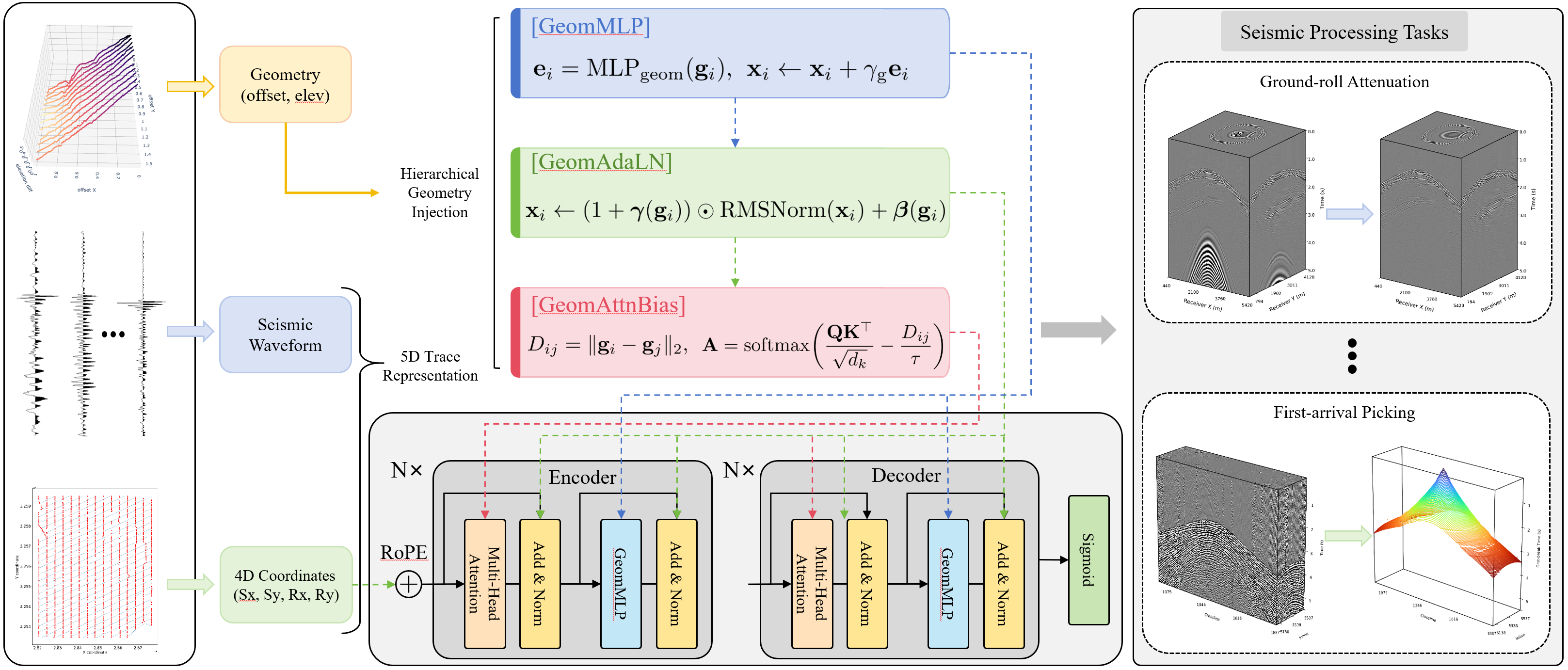}
\end{minipage}
\end{adjustbox}
\caption{Architecture of the proposed GeoFormer. The model adopts a multi-stage Transformer encoder-decoder backbone with three complementary geometry injection mechanisms: GeomMLP, GeomAdaLN, and GeomAttnBias.}
\label{fig:architecture}
\end{figure}

GeoFormer takes as input seismic trace waveforms and their associated acquisition-geometry information, as illustrated in Figure~\ref{fig:architecture}. For a shot gather with $L$ traces, each trace is represented by three components, including its full waveform, four source-receiver coordinates $(s_x, s_y, r_x, r_y)$ that locate the trace in the survey, and two geometric attributes, namely offset and relative elevation. Together, the four acquisition coordinates and the time dimension define the five-dimensional (5D) acquisition domain.

These three types of input are processed through dedicated pathways. The waveform is projected to a per-trace token embedding. The source-receiver coordinates are processed by a geometry-aware positional encoding that makes self-attention aware of each trace's spatial location in the survey. The geometric attributes are injected into the Transformer through three complementary geometry injection mechanisms operating at the token, normalization, and attention levels. Together, these pathways produce a geometry-aware per-trace representation in which each token carries both waveform and physical acquisition context. Self-attention then models global inter-trace relationships informed by both waveform similarity and geometric proximity, and a final prediction head produces per-sample outputs.

The following subsections first describe how GeoFormer organizes each seismic trace waveform and its associated geometric information, then introduce the geometry-aware positional encoding for the four source--receiver coordinates, and finally present the three complementary geometry injection mechanisms for incorporating geometric attributes.

\subsection{5D Trace Representation}

Let a shot gather consist of $L$ traces, each containing $T$ time samples. In a standard Vision Transformer, the shot gather is divided into 2D patches, each of which is linearly projected into a token. This operation implicitly fixes or averages the source and receiver coordinates, thereby collapsing the 5D domain into a 2D grid. In GeoFormer, by contrast, each token corresponds to exactly one seismic trace, anchored at its own source-receiver coordinates $(s_x, s_y, r_x, r_y)$ and sampled along the time dimension $t$. The token representation combines the waveform with its source-receiver coordinates:
\begin{equation}
\mathbf{t}_i = \left( \mathbf{w}_i,\; \mathbf{c}_i \right), \quad i = 1, \ldots, L,
\label{eq:trace_5d}
\end{equation}
where $\mathbf{c}_i = (s_x^i, s_y^i, r_x^i, r_y^i)$ denotes the 4D source-receiver coordinates that uniquely locate each trace. Together, the waveform and the four coordinate dimensions constitute the trace representation. The offset and relative elevation of each trace are treated as its per-trace geometric attributes:
\begin{equation}
\mathbf{g}_i = \left( \delta x_i,\; \delta z_i \right),
\label{eq:geom2}
\end{equation}
where $\delta x_i = \sqrt{(s_x^i - r_x^i)^2 + (s_y^i - r_y^i)^2}$ is the offset and $\delta z_i = z_r^i - z_s^i$ is the receiver elevation minus the source elevation. Both coordinates and geometric attributes are min-max normalized to $[0, 1]$ globally across all training files. For a shot gathers, $\mathbf{G} \in \mathbb{R}^{B \times L \times 2}$ denotes the tensor collecting all raw per-trace geometric attributes $\mathbf{g}_i$.

\subsection{Geometry-Aware Positional Encoding}

GeoFormer uses two complementary forms of positional encoding, namely 4D absolute coordinate encoding and rotary position embedding (RoPE). For trace $i$, the coordinate vector
\(\mathbf{c}_i=(s_x^i,s_y^i,r_x^i,r_y^i)\)
contains the source coordinates and receiver coordinates. These four coordinates define the absolute acquisition position of each trace. The model first encodes them with a 4D absolute positional encoding. The token dimension is divided into four coordinate groups with dimension $d_c=d_{\text{model}}/4$. For coordinate dimension $m \in \{1,2,3,4\}$ and frequency index $j$, the encoding is
\begin{equation}
\begin{aligned}
PE_{i,m,2j} &= \sin\left(\frac{\tilde{c}_{i,m}}{10000^{2j/d_c}}\right), \\
PE_{i,m,2j+1} &= \cos\left(\frac{\tilde{c}_{i,m}}{10000^{2j/d_c}}\right),
\end{aligned}
\label{eq:coord_abs_encoding}
\end{equation}
where $\tilde{c}_{i,m}$ denotes the normalized and scaled coordinate value. The four coordinate encodings are concatenated and added to the trace token,
\begin{equation}
\mathbf{x}_i \leftarrow \mathbf{x}_i + \gamma_{\mathrm{coord}} [\mathbf{PE}_{i,1};\mathbf{PE}_{i,2};\mathbf{PE}_{i,3};\mathbf{PE}_{i,4}],
\label{eq:coord_abs_add}
\end{equation}
where $\gamma_{\mathrm{coord}}$ is a learnable scale factor.

After the absolute coordinate information is injected into the token representation, RoPE is applied to the query and key vectors to encode sequential trace positions in self-attention. For a trace-order index $\ell_i$ and frequency band $\theta_j=10000^{-2j/d_k}$, RoPE rotates paired query and key channels as
\begin{equation}
\begin{aligned}
\mathbf{Q}'_{i} &= \mathbf{Q}_{i} \odot \cos(\ell_i\boldsymbol{\theta})
 + \mathrm{rotate\_half}(\mathbf{Q}_{i}) \odot \sin(\ell_i\boldsymbol{\theta}), \\
\mathbf{K}'_{i} &= \mathbf{K}_{i} \odot \cos(\ell_i\boldsymbol{\theta})
 + \mathrm{rotate\_half}(\mathbf{K}_{i}) \odot \sin(\ell_i\boldsymbol{\theta}),
\end{aligned}
\label{eq:rope}
\end{equation}
where $\mathrm{rotate\_half}$ swaps and negates paired channels. The rotated query and key vectors are then used in the self-attention computation (Equation~\ref{eq:geom_attn_bias}). By combining absolute acquisition coordinates with trace-order encoding, the attention module becomes spatially aware: it can distinguish traces not only by their waveforms, but also by their physical locations in the survey geometry. The geometry injection mechanisms described below further use offset and relative elevation as additional physical attributes.

\subsection{Geometry Injection Mechanisms}
\label{sec:geom_injection}

A seismic trace carries two complementary types of information: the waveform and the acquisition geometry. In the Vision Transformer, only the waveform enters the Transformer, while the geometry is ignored. GeoFormer restores this missing physical information by injecting geometric attributes into the Transformer at three complementary levels. These three levels correspond to three distinct questions that geometry can answer for a trace: \textit{what geometric representation does this trace carry?} (token level, GeomMLP), \textit{how should this trace's features be modulated given its spatial position?} (normalization level, GeomAdaLN), and \textit{which other traces should be attended to?} (attention level, GeomAttnBias). The resulting trace features are physically constrained because they combine learned waveform representations with geometric priors. When these geometry-aware features are processed by self-attention, the Transformer models inter-trace relationships using both waveform similarity and acquisition geometry. This results in a more complete global representation. The three geometry injection mechanisms are detailed below.

\begin{enumerate}
	\item \textbf{Token level (GeomMLP)}: Vision Transformer prepends a learned classification token that carries no physical meaning. GeomMLP replaces this with a per-trace geometric representation derived from offset and elevation. This geometric representation is additively injected at the encoder entry, giving every trace a physically meaningful representation before attention begins. When waveform is degraded by noise, this geometric representation provides a spatial prior that constrains the prediction.
	\item \textbf{Normalization level (GeomAdaLN)}: Standard layer normalization assumes all tokens share the same statistics. In seismic data, near-offset and far-offset traces typically exhibit different data distributions. GeomAdaLN replaces uniform normalization with geometry-conditioned modulation: it predicts channel-wise scale and shift parameters from per-trace geometric attributes $\mathbf{g}_i$, enabling the feature distribution of each trace to adapt to its spatial position.
	\item \textbf{Attention level (GeomAttnBias)}: Vision Transformer learns pairwise affinities purely from data. GeomAttnBias introduces an acquisition-aware relational bias that favors information exchange between traces with similar geometric attributes, while the learned attention scores retain the flexibility to emphasize waveform-dependent relationships. This prior is parameter-free: it requires no learning and is derived entirely from the survey geometry. 
\end{enumerate}

 GeomMLP supplies the geometric representation --- it answers what a trace is in physical space. GeomAdaLN supplies geometry-conditioned normalization --- it adapts how a trace's features are expressed based on where it sits. GeomAttnBias supplies a physical prior --- it determines which traces should communicate based on their geometric attributes. When all three are active, every trace feature passed into self-attention carries a fused representation of waveform and physical context, enabling the Transformer to model global relationships that respect both data-driven similarity and physics-driven proximity. The three mechanisms operate at complementary stages of the network, and their contributions are evaluated through controlled ablation experiments across the considered processing tasks.

\subsubsection{GeomMLP: Token-Level Geometric Representation}

GeomMLP operates at the token level. It maps the raw geometric attribute tensor $\mathbf{G}$ to a high-dimensional geometric representation:
\begin{equation}
\mathbf{E}_{\mathrm{geom}} = \mathrm{MLP}_{\mathrm{geom}}(\mathbf{G}) \in \mathbb{R}^{B \times L \times d_{\mathrm{model}}},
\label{eq:geom_mlp}
\end{equation}
where $\mathbf{G} \in \mathbb{R}^{B \times L \times 2}$ contains the raw per-trace geometric attributes, the MLP features hidden dimensions $[h, 2h, d_{\text{model}}]$ with $h = 256$, and GELU activations are used between linear layers. The resulting embedding is additively injected at the encoder:
\begin{equation}
\mathbf{X} \leftarrow \mathbf{X} + \gamma_{\text{geom}} \cdot \mathbf{E}_{\mathrm{geom}},
\label{eq:geom_injection}
\end{equation}
where $\gamma_{\text{geom}}$ is a learnable scalar initialized to $1.0$. This additive injection provides each trace token with a spatially conditioned bias. It provides each trace token with an explicit representation of its acquisition-dependent geometric context. When waveform evidence is incomplete or degraded, the geometric representation provides complementary spatial context for the task-specific prediction.

\subsubsection{GeomAdaLN: Normalization-Level Feature Modulation}

GeomAdaLN operates at the normalization level. It replaces standard RMSNorm with geometry-modulated variants throughout the encoder-decoder blocks. Unlike the additive injection of GeomMLP, GeomAdaLN enables multiplicative interaction between geometric attributes and features:
\begin{equation}
	\begin{aligned}
		\boldsymbol{\gamma}_c, \boldsymbol{\beta}_c
		&= \mathrm{chunk}\!\left(
		\mathrm{MLP}_{\text{adaln}}(\mathbf{g}_i)
		\right),\\
		\mathrm{GeomAdaLN}(\mathbf{x}_i, \mathbf{g}_i)
		&= (1+\boldsymbol{\gamma}_c)
		\odot \mathrm{RMSNorm}(\mathbf{x}_i)
		+\boldsymbol{\beta}_c.
	\end{aligned}
	\label{eq:adaln}
\end{equation}

where $\mathbf{x}_i$ is the feature vector of trace $i$, and $\boldsymbol{\gamma}_c, \boldsymbol{\beta}_c \in \mathbb{R}^{d_{\text{model}}}$ are channel-wise scale and shift parameters predicted from the raw geometric attributes $\mathbf{g}_i$. The modulation MLP output layer is zero-initialized, ensuring the modulation starts as an identity mapping. This mechanism dynamically rescales and shifts every feature channel based on trace spatial position, making the per-trace feature distribution geometry-dependent through normalization modulation.

\subsubsection{GeomAttnBias: Attention-Level Relational Bias}

GeomAttnBias operates at the attention level. It introduces an acquisition-aware relational bias into the self-attention computation, encouraging traces with similar geometric attributes to exchange information more directly. This is done by computing the distance between traces in the offset-elevation space and converting it to a negative bias:
\begin{equation}
\begin{aligned}
d_{ij} &= \|\mathbf{g}_i - \mathbf{g}_j\|_2, \\
B_{ij} &= -\frac{d_{ij}}{\tau}, \\
\mathbf{A} &= \mathrm{softmax}\left( \frac{\mathbf{Q}\mathbf{K}^{\top}}{\sqrt{d_k}} + \mathbf{B} \right),
\end{aligned}
\label{eq:geom_attn_bias}
\end{equation}
where $\tau = 1.0$ is a temperature parameter. Since $B_{ij} \leq 0$ for all pairs, traces with smaller pairwise geometric distances receive higher attention weights. GeomAttnBias operates on raw normalized geometric attributes and introduces no learnable parameters. It provides a geometry-dependent bias for inter-trace information aggregation without prescribing a task-specific output pattern. When the information in an individual trace is incomplete or corrupted, geometrically related traces can contribute complementary evidence through stronger attention interactions.

\subsection{Task-Specific Prediction Heads and Objectives}

Let $\mathbf{H}=F_{\theta}(\mathbf{W},\mathbf{C},\mathbf{G}) \in \mathbb{R}^{B \times L \times d}$ denote the geometry-aware trace representation produced by the shared GeoFormer backbone, where $B$ is the batch size, $L$ is the number of traces, and $d$ is the backbone feature dimension. Each processing task uses an independently trained prediction head $D_{\phi}^{(q)}$ and a task-specific objective, while the trace representation and geometry-conditioning mechanisms remain unchanged. This separation allows the same architectural design to be evaluated under different output spaces without assuming a common target or a multitask training objective.

\subsubsection{Seismic Noise Attenuation Head and Objective}

For seismic noise attenuation, the input waveform $\mathbf{W}^{\mathrm{noisy}}$ contains useful signal together with ground roll and random noise, and the target $\mathbf{W}^{\mathrm{clean}}$ is the corresponding clean or reference waveform. The noise-attenuation head maps the shared trace representation back to the temporal waveform domain:
\begin{equation}
\mathbf{W}^{\mathrm{pred}}
=D_{\phi}^{\mathrm{den}}(\mathbf{H})
\in \mathbb{R}^{B \times L \times T}.
\label{eq:denoise_head}
\end{equation}
The head is applied independently to the temporal samples of each trace after the backbone has aggregated information from geometrically related traces. We optimize the mean squared error (MSE) between the predicted and clean waveforms:
\begin{equation}
\mathcal{L}_{\mathrm{denoise}}
=\frac{1}{BLT}\sum_{b=1}^{B}\sum_{i=1}^{L}\sum_{t=1}^{T}
\left(w^{\mathrm{pred}}_{b,i,t}-
 w^{\mathrm{clean}}_{b,i,t}\right)^2.
\label{eq:denoise_loss}
\end{equation}
Here, $\mathbf{W}^{\mathrm{clean}}$ denotes the clean or reference waveform used by the corresponding noise-attenuation experiment.

\subsubsection{First-Arrival Picking Head and Objective}

For first-arrival picking, the task head produces one logit for each time sample of each trace:
\begin{equation}
\widehat{\mathbf{Y}}^{\mathrm{pick}}
=D_{\phi}^{\mathrm{pick}}(\mathbf{H})
\in \mathbb{R}^{B \times L \times T}.
\label{eq:pick_head}
\end{equation}
After applying the sigmoid function, $\widehat{m}_{b,i,t}=\sigma(\widehat{y}^{\mathrm{pick}}_{b,i,t})$ represents the probability of the post-arrival class. The step-label formulation assigns zero to samples before the manually interpreted first arrival and one to the arrival sample and subsequent samples. The first-arrival sample index $p_i$ is subsequently extracted from the probability sequence using Nearest-Point Picking (NPP) along the trace direction \citep{yuan2020robust}.

The picking instance is trained with a masked binary cross-entropy loss applied to the per-sample logits:
\begin{equation}
\mathcal{L}_{\mathrm{pick}}
=
\frac{
\sum_{b=1}^{B}\sum_{i=1}^{L}\sum_{t=1}^{T}
v_{b,i}\,\mathrm{BCE}\!\left(\widehat{y}^{\mathrm{pick}}_{b,i,t},m_{b,i,t}\right)
}{
T\sum_{b=1}^{B}\sum_{i=1}^{L}v_{b,i}+\epsilon
},
\label{eq:pick_loss}
\end{equation}
where $v_{b,i}\in\{0,1\}$ indicates whether trace $i$ in batch element $b$ has a valid first-arrival label, and $\epsilon$ prevents division by zero. This masking strategy excludes unlabeled traces from the supervised objective while retaining their waveform and geometry as contextual inputs to the backbone.

\section{Results and Discussion}

\subsection{Experimental Setup}

We evaluate GeoFormer on two representative prestack seismic processing tasks: seismic noise attenuation and first-arrival picking, representing regression and segmentation tasks, respectively. Seismic noise attenuation is formulated as a regression task, where the model reconstructs a clean seismic trace from each noisy input trace. First-arrival picking aims to identify the onset time of the first arrival on each trace and is commonly formulated as a segmentation task. Both tasks share the same geometry-aware GeoFormer backbone, with only the task-specific prediction head and training objective adjusted according to the output type.

\begin{table}[!htbp]
\centering
\small
\setlength{\tabcolsep}{3pt}
\caption{Summary of the four field datasets used for first-arrival picking.}
\label{tab:datasets}
\begin{adjustbox}{max width=\textwidth}
\begin{tabular}{lcccccc}
\toprule
Dataset & Shots & Total traces & Samples/trace & Sampling (ms) & Offset range (m) & Labeled traces (\%) \\
\midrule
Brunswick   & 1,541 & 4,490,714 & 751  & 2 & 2.0--7,188.2   & 83.1 \\
Halfmile    & 690   & 1,094,863 & 751  & 2 & 0.0--4,865.9   & 90.7 \\
Lalor       & 905   & 2,027,587 & 1,501 & 1 & 0.1--5,639.3   & 55.2 \\
Dongbei     & 10    & 266,270   & 2,751 & 2 & 26.9--14,029.5   & 97.9 \\
\bottomrule
\end{tabular}
\end{adjustbox}
\end{table}

\subsubsection{Seismic Noise Attenuation}

For the noise attenuation experiment, the synthetic dataset is constructed by generating the clean data, the ground roll, and the random noise separately and summing them to form the noisy input. The data are based on the open-source \href{https://wiki.seg.org/wiki/SEG_C3_45_shot}{SEG\_C3NA} velocity model, a three-dimensional volume of $676 \times 676 \times 201$ samples with a 10m spatial interval, covering $6.76~\mathrm{km} \times 6.76~\mathrm{km} \times 2.0~\mathrm{km}$ and containing velocities from 1500 to 4480~m/s. The clean data is generated by acoustic finite-difference modeling with a 15-Hz source wavelet, while the ground-roll is generated by elastic finite-difference modeling with free-surface boundary conditions and a 5-Hz source wavelet.
The acquisition geometry contains 72 shots arranged on a $9 \times 8$ grid with wide-azimuth coverage, and the recording interval is 8~ms. The shot gathers are divided by shot into training, validation, and test sets using a ratio of 58:7:7.

GeoFormer receives the complete set of traces in a shot gather together with their acquisition coordinates and derived geometric attributes. The regression head in Equation~\ref{eq:denoise_head} reconstructs the clean seismic trace from each noisy input trace, and the model is trained using the noise-attenuation objective defined in Equation~\ref{eq:denoise_loss}.

For noise attenuation, the GeoFormer backbone has a model width of 1024, 8 attention heads, 3 encoder--decoder stages, and an FFN width of 2048. The model is optimized with AdamW using an initial learning rate of $1\times10^{-4}$, a batch size of 12, and 500 training epochs on four NVIDIA A100 GPUs.

Noise attenuation is evaluated using three quantitative metrics. Let $\mathbf{W}^{\mathrm{pred}}$ and $\mathbf{W}^{\mathrm{clean}}$ denote the predicted and reference clean gathers, respectively. SNR and MAE are defined as
\begin{equation}
	\mathrm{SNR}=10\log_{10}\!\left(
	\frac{\left\|\mathbf{W}^{\mathrm{clean}}\right\|_2^2}
	{\left\|\mathbf{W}^{\mathrm{pred}}-\mathbf{W}^{\mathrm{clean}}\right\|_2^2}
	\right).
	\label{eq:ground_roll_snr}
\end{equation}

\begin{equation}
	\mathrm{MAE}=\frac{1}{BLT}\sum_{b=1}^{B}\sum_{i=1}^{L}\sum_{t=1}^{T}
	\left|w^{\mathrm{pred}}_{b,i,t}-w^{\mathrm{clean}}_{b,i,t}\right|.
	\label{eq:ground_roll_mae}
\end{equation}
Here, $B$, $L$, and $T$ denote the batch size, number of traces, and number of time samples, respectively. SSIM is computed over the 2D gathers as
\begin{equation}
\mathrm{SSIM}=\frac{(2\mu_{\mathrm{pred}}\mu_{\mathrm{clean}}+C_1)(2\sigma_{\mathrm{pred,clean}}+C_2)}
{(\mu_{\mathrm{pred}}^2+\mu_{\mathrm{clean}}^2+C_1)(\sigma_{\mathrm{pred}}^2+\sigma_{\mathrm{clean}}^2+C_2)},
\label{eq:ground_roll_ssim}
\end{equation}
where $\mu_{\mathrm{pred}}$ and $\mu_{\mathrm{clean}}$ are the local means, $\sigma_{\mathrm{pred}}^2$ and $\sigma_{\mathrm{clean}}^2$ are the local variances, $\sigma_{\mathrm{pred,clean}}$ is the local covariance, and $C_1$ and $C_2$ are numerical stabilization constants. 

GeoFormer is compared against two representative baselines: (1) SANet \citep{yang2024deep}, a convolutional network originally designed for ground-roll attenuation with a frequency-domain loss, and (2) Vision Transformer \citep{dosovitskiy2020image}, a standard Transformer applied to 2D seismic data to assess the performance of a general-purpose Transformer architecture. All deep learning models are trained using identical data splits. Training is conducted on four NVIDIA A100 GPUs using Distributed Data Parallel.

\subsubsection{First-Arrival Picking}

For the first-arrival picking experiments, we evaluate GeoFormer across four field datasets: Brunswick, Halfmile Lake, Lalor, and Dongbei. These datasets cover diverse survey scales. Offset ranges span from near-zero to over 14 km, and the proportion of labeled traces varies from 55.2\% to 97.9\%. Table~\ref{tab:datasets} summarizes the dataset statistics.

\begin{figure}[!htbp]
\centering
\begin{adjustbox}{max totalsize={\textwidth}{0.76\textheight},center}
\begin{minipage}{\textwidth}
\centering
\begin{minipage}[b]{0.92\columnwidth}
		\centering
		\includegraphics[width=\linewidth]{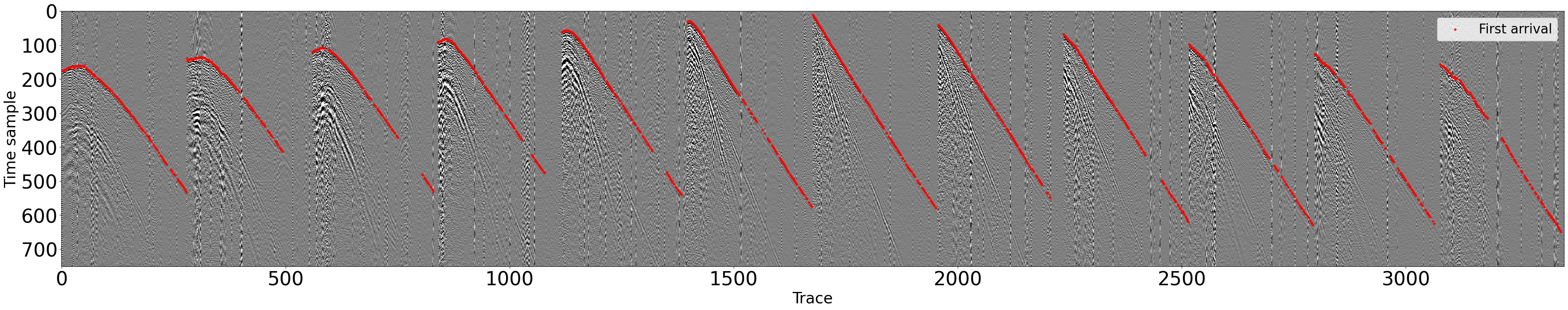}\\[-2pt]
		{\small (a) Seismic shot gathers with first-arrival}
	\end{minipage}
	
	\vspace{4pt}
	
	\begin{minipage}[b]{0.92\columnwidth}
		\centering
		\includegraphics[width=\linewidth]{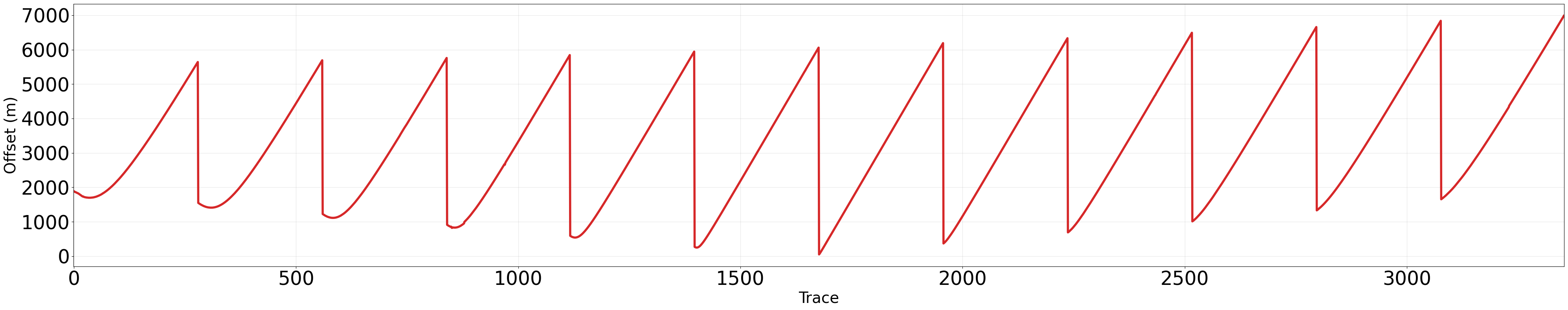}\\[-2pt]
		{\small (b) Offset curve}
	\end{minipage}
	
	\vspace{4pt}
	
	\begin{minipage}[b]{0.92\columnwidth}
		\centering
		\includegraphics[width=\linewidth]{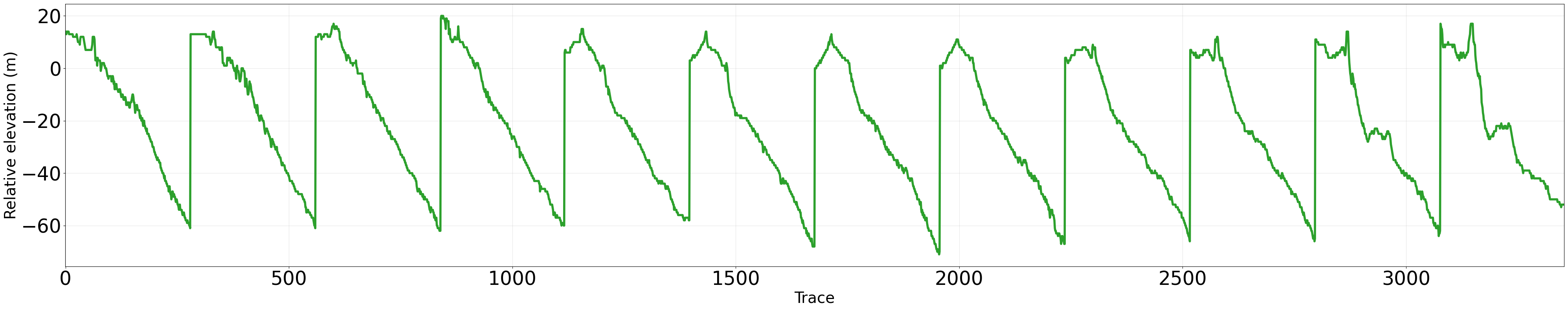}\\[-2pt]
		{\small (c) Relative-elevation curve}
	\end{minipage}
\end{minipage}
\end{adjustbox}
\caption{Relationship between seismic first-arrival structure and acquisition geometry. (a) Representative seismic shot gathers; (b) corresponding offset; (c) corresponding relative elevation. The consistent trace ordering across panels shows that offset and relative elevation are strongly structured with respect to the first-arrival moveout, motivating their use as geometry-aware inputs to GeoFormer.}
	\label{fig:geometry_prior_example}
\end{figure}

Figure~\ref{fig:geometry_prior_example} illustrates the relationship between geometric attributes and first-arrival traveltime. Offset and relative elevation provide complementary information about the spatial variation of first arrivals. These attributes can therefore serve as geometric context when waveform evidence is degraded by noise or missing traces.

Brunswick, Halfmile, and Lalor are from the seismic benchmark dataset released by \citet{stcharles2021deep}. Dongbei is acquired from northeastern China and contains densely annotated first-arrival labels. First-arrival ground truth is converted to binary 0/1 mask labels along the time axis. Across the four sites, the proportion of traces with valid manual picks varies considerably, ranging from 55.2\% to 97.9\%.

Because GeoFormer explicitly leverages shot-receiver coordinates and geometric attributes, adjacent traces within the same shot gather share highly correlated spatial information. Random trace-level splitting would cause information leakage between training and test sets and produce overly optimistic performance. To avoid this, each dataset is partitioned by shot, ensuring that all traces belonging to the same shot are assigned exclusively to one of the training, validation, or test splits. The training/validation/test ratios are approximately 80\%, 10\%, and 10\% for each site.

Picking accuracy is assessed using the hit rate (HR) at multiple sample tolerances (HR@1, HR@3, HR@5, HR@7, HR@9), defined as the percentage of traces for which the absolute picking error does not exceed a specified number of time samples:
\begin{equation}
\mathrm{HR@}k = \frac{1}{N} \sum_{i=1}^{N} \mathbf{1}\!\left[\,|p_i - \hat{p}_i| \leq k\,\right] \times 100\%,
\label{eq:hr}
\end{equation}
where $N$ is the number of traces with valid labels, $p_i$ and $\hat{p}_i$ are the ground-truth and predicted first-arrival sample indices for trace $i$, and $\mathbf{1}[\cdot]$ is the indicator function. In addition, root mean square error (RMSE), mean absolute error (MAE), and mean bias error (MBE) are reported, defined as
\begin{equation}
\begin{aligned}
\mathrm{RMSE} &= \sqrt{\frac{1}{N}\sum_{i=1}^{N}\left(p_i - \hat{p}_i\right)^{2}},\\
\mathrm{MAE} &= \frac{1}{N}\sum_{i=1}^{N}\left|p_i - \hat{p}_i\right|,\\
\mathrm{MBE} &= \frac{1}{N}\sum_{i=1}^{N}\left(\hat{p}_i - p_i\right).
\end{aligned}
\label{eq:metrics}
\end{equation}
All metrics are computed exclusively on traces with valid labels. HR is reported as a percentage, whereas RMSE, MAE, and MBE are reported in units of time samples.

GeoFormer is compared against three representative baselines: (1) Attention UNet, which augments the standard U-Net with an attention mechanism, (2) HU-Net \citep{pu2021improving}, a U-Net variant with hierarchical feature refinement designed for prestack seismic data processing, and (3) Vision Transformer \citep{dosovitskiy2020image}, a standard Transformer applied to 2D seismic data to assess the performance of a general-purpose Transformer architecture. All deep learning models are trained with identical data splits. Training is conducted on 4 NVIDIA A100 GPUs using Distributed Data Parallel.

\subsection{Seismic Noise Attenuation}

\begin{figure}[!htbp]
\centering
\begin{adjustbox}{max totalsize={\textwidth}{0.76\textheight},center}
\begin{minipage}{\textwidth}
\centering
\begin{minipage}[b]{0.35\columnwidth}
\centering
\includegraphics[width=\linewidth]{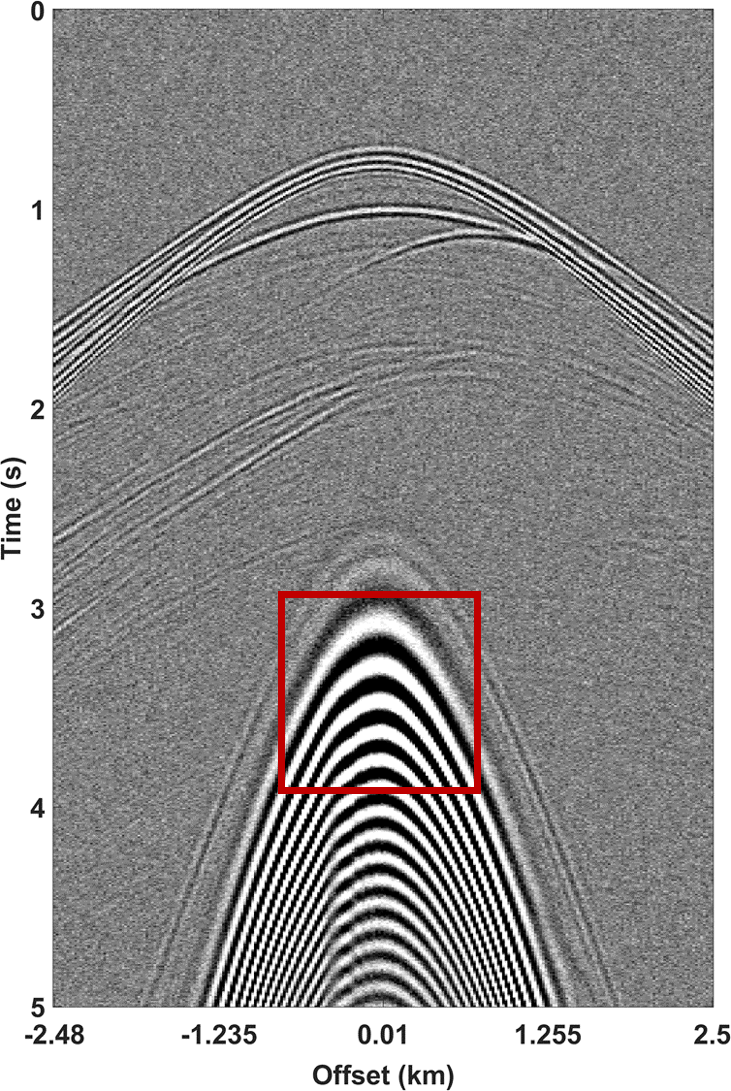}\\[-3pt]
{\small (a) Noisy data}
\end{minipage}

\vspace{3pt}

\begin{minipage}[b]{0.35\columnwidth}
\centering
\includegraphics[width=\linewidth]{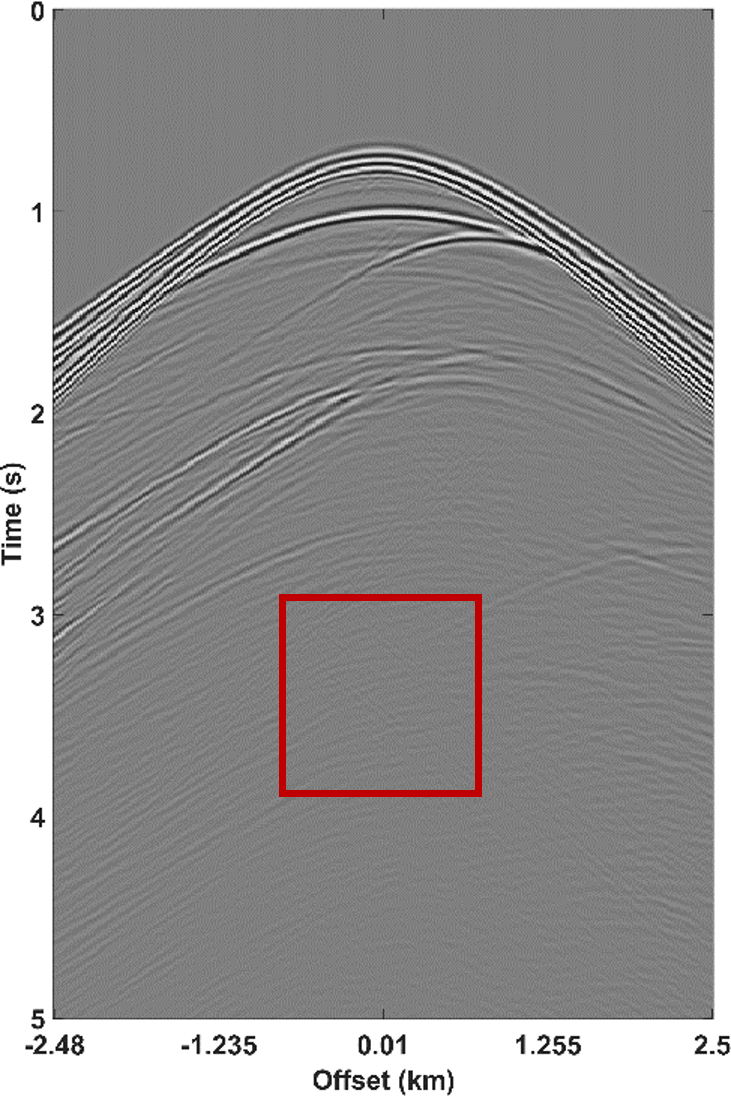}\\[-3pt]
{\small (b) Label}
\end{minipage}\hspace{0.02\columnwidth}\begin{minipage}[b]{0.35\columnwidth}
\centering
\includegraphics[width=\linewidth]{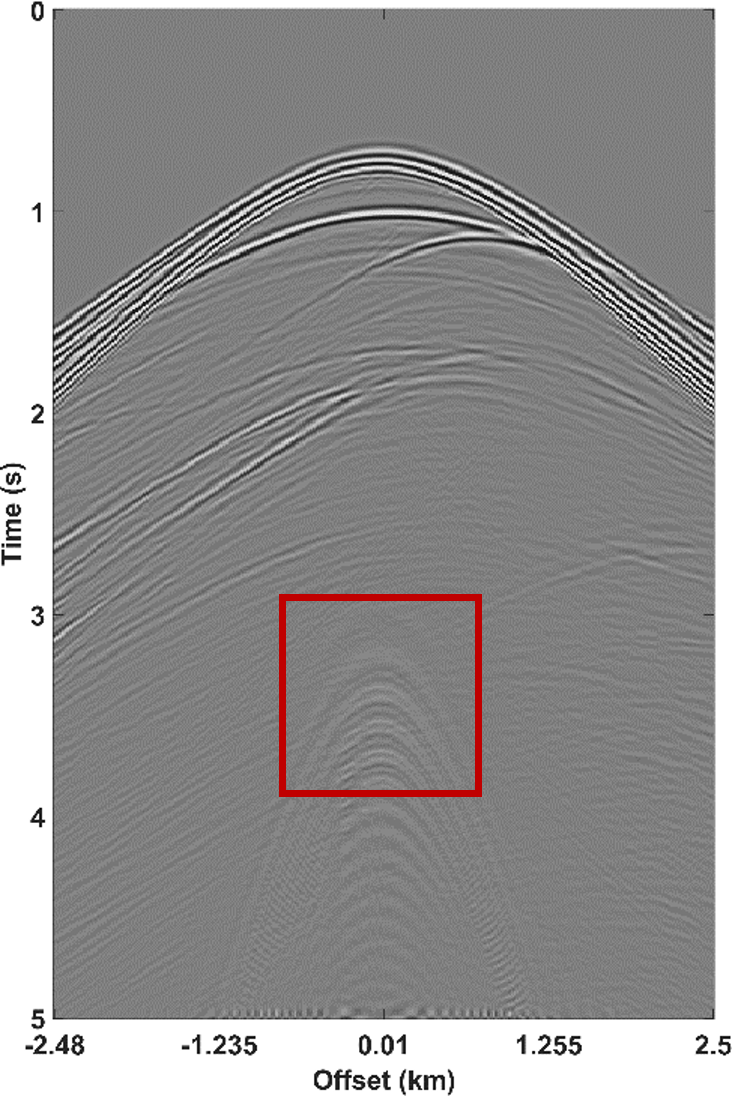}\\[-3pt]
{\small (c) SANet}
\end{minipage}

\vspace{3pt}

\begin{minipage}[b]{0.35\columnwidth}
\centering
\includegraphics[width=\linewidth]{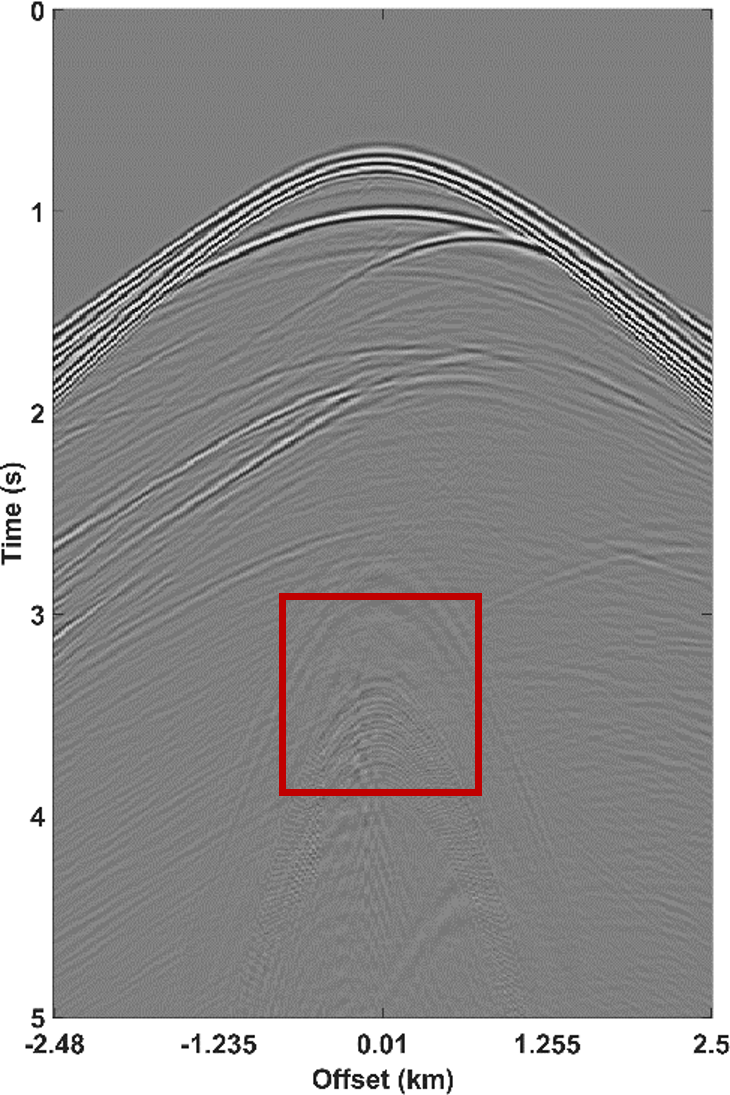}\\[-3pt]
{\small (d) Vision Transformer}
\end{minipage}\hspace{0.02\columnwidth}\begin{minipage}[b]{0.35\columnwidth}
\centering
\includegraphics[width=\linewidth]{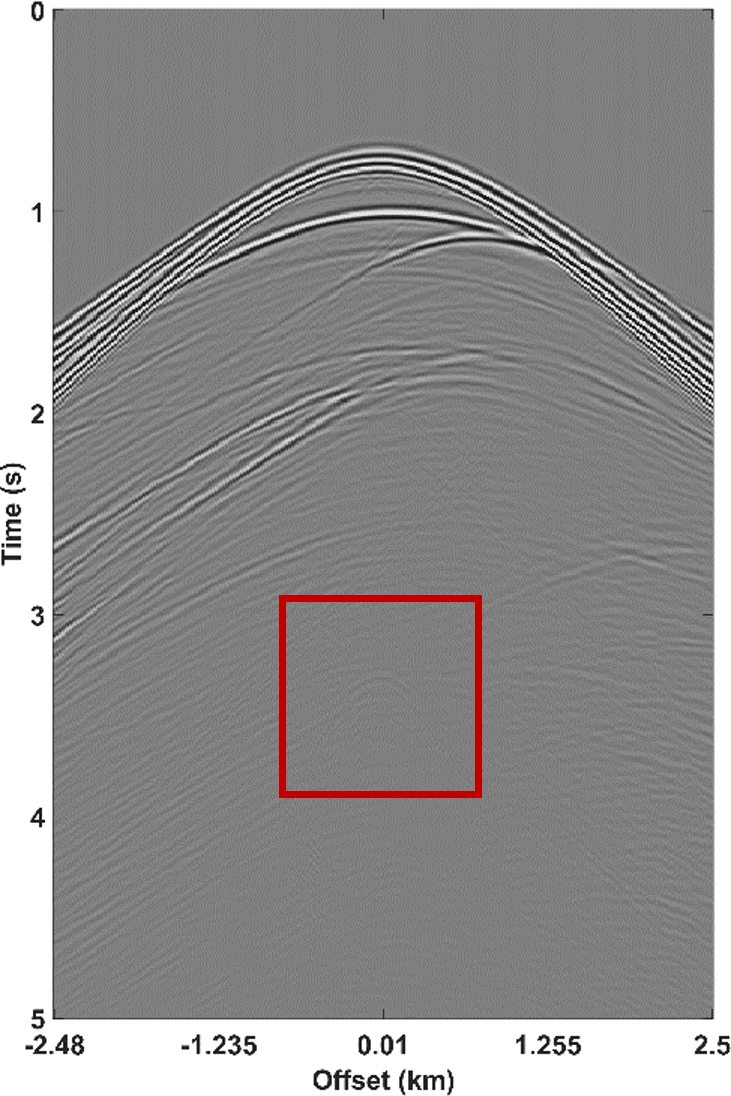}\\[-3pt]
{\small (e) GeoFormer}
\end{minipage}
\end{minipage}
\end{adjustbox}
\caption{Noise attenuation results on synthetic data for ground roll and random noise. (a) Noisy data; (b) Label; (c) SANet; (d) Vision Transformer; (e) GeoFormer.}
\label{fig:ground_roll_predictions}
\end{figure}

For a fair visual comparison, all results in Figure~\ref{fig:ground_roll_predictions} are displayed using the same enhancement settings. Figure~\ref{fig:ground_roll_predictions} presents the full 2D prediction comparison. The noisy input is strongly contaminated by ground roll and random noise, which obscure weaker reflection events. SANet and Vision Transformer suppress a substantial part of this interference, but visible residual ground roll and random noise remain in their predicted sections. In contrast, GeoFormer produces a prediction closer to the label, with the least residual ground roll and random noise among the compared learning-based methods while preserving the main reflection events.

\begin{figure}[!htbp]
\centering
\begin{adjustbox}{max totalsize={\textwidth}{0.76\textheight},center}
\begin{minipage}{\textwidth}
\centering
\begin{minipage}[b]{0.46\columnwidth}
\centering
\includegraphics[width=\linewidth]{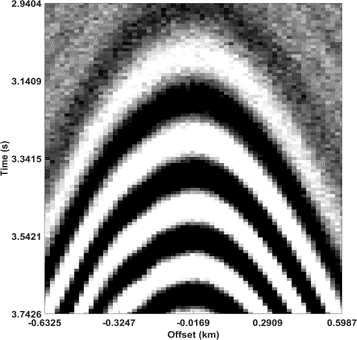}\\[-3pt]
{\small (a) Noisy data}
\end{minipage}

\vspace{3pt}

\begin{minipage}[b]{0.46\columnwidth}
\centering
\includegraphics[width=\linewidth]{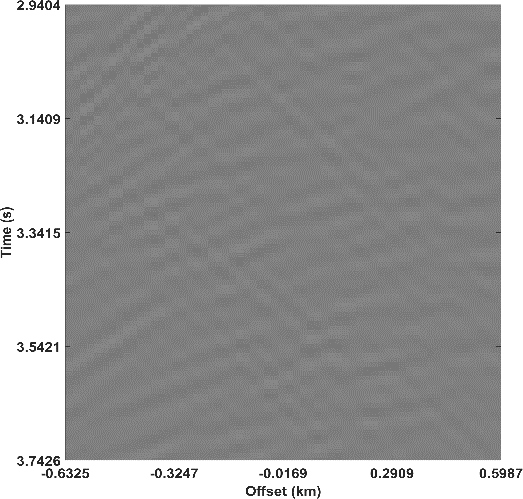}\\[-3pt]
{\small (b) Label}
\end{minipage}\hspace{0.02\columnwidth}\begin{minipage}[b]{0.46\columnwidth}
\centering
\includegraphics[width=\linewidth]{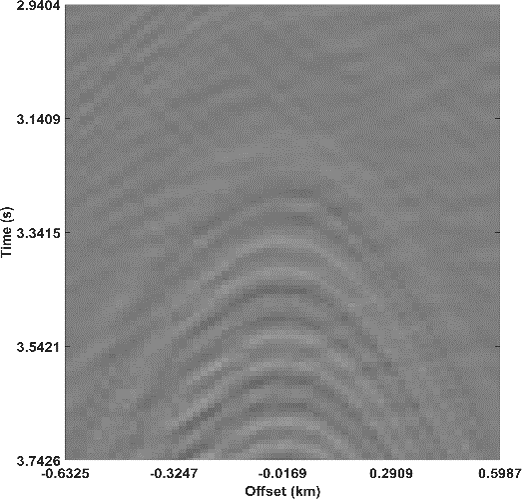}\\[-3pt]
{\small (c) SANet}
\end{minipage}

\vspace{3pt}

\begin{minipage}[b]{0.46\columnwidth}
\centering
\includegraphics[width=\linewidth]{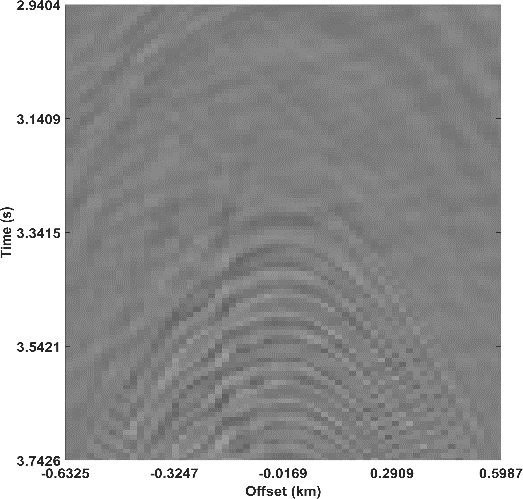}\\[-3pt]
{\small (d) Vision Transformer}
\end{minipage}\hspace{0.02\columnwidth}\begin{minipage}[b]{0.46\columnwidth}
\centering
\includegraphics[width=\linewidth]{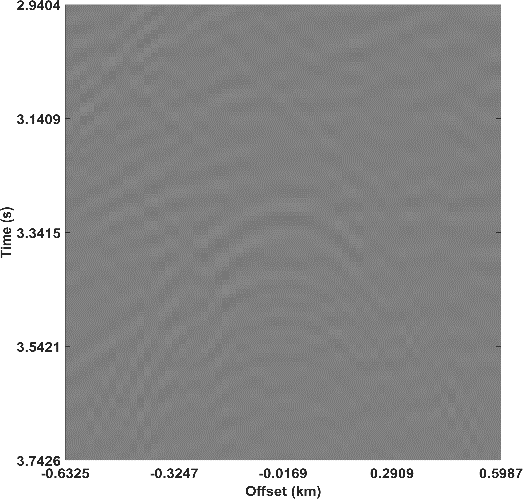}\\[-3pt]
{\small (e) GeoFormer}
\end{minipage}
\end{minipage}
\end{adjustbox}
\caption{Local enlargements of the noise-attenuation results on synthetic data. (a) Noisy data; (b) Label; (c) SANet; (d) Vision Transformer; (e) GeoFormer.}
\label{fig:ground_roll_local}
\end{figure}

The zoomed-in views in Figure~\ref{fig:ground_roll_local} make the differences in residual noise more apparent. SANet and Vision Transformer retain visible ground-roll and random-noise remnants within the selected region, whereas GeoFormer more effectively suppresses the residual ground roll and random noise, leaving almost no visible noise while preserving the local reflection. We attribute this improvement to GeoFormer's ability to model more complete relationships among seismic traces, which enables more effective separation of ground roll and random noise from the reflection events. The enlarged comparison therefore further supports the observations from the full gathers: GeoFormer achieves stronger attenuation without removing the useful seismic events.

\begin{figure}[!htbp]
\centering
\begin{adjustbox}{max totalsize={\textwidth}{0.76\textheight},center}
\begin{minipage}{\textwidth}
\centering
\begin{minipage}[b]{0.35\columnwidth}
\centering
\includegraphics[width=\linewidth]{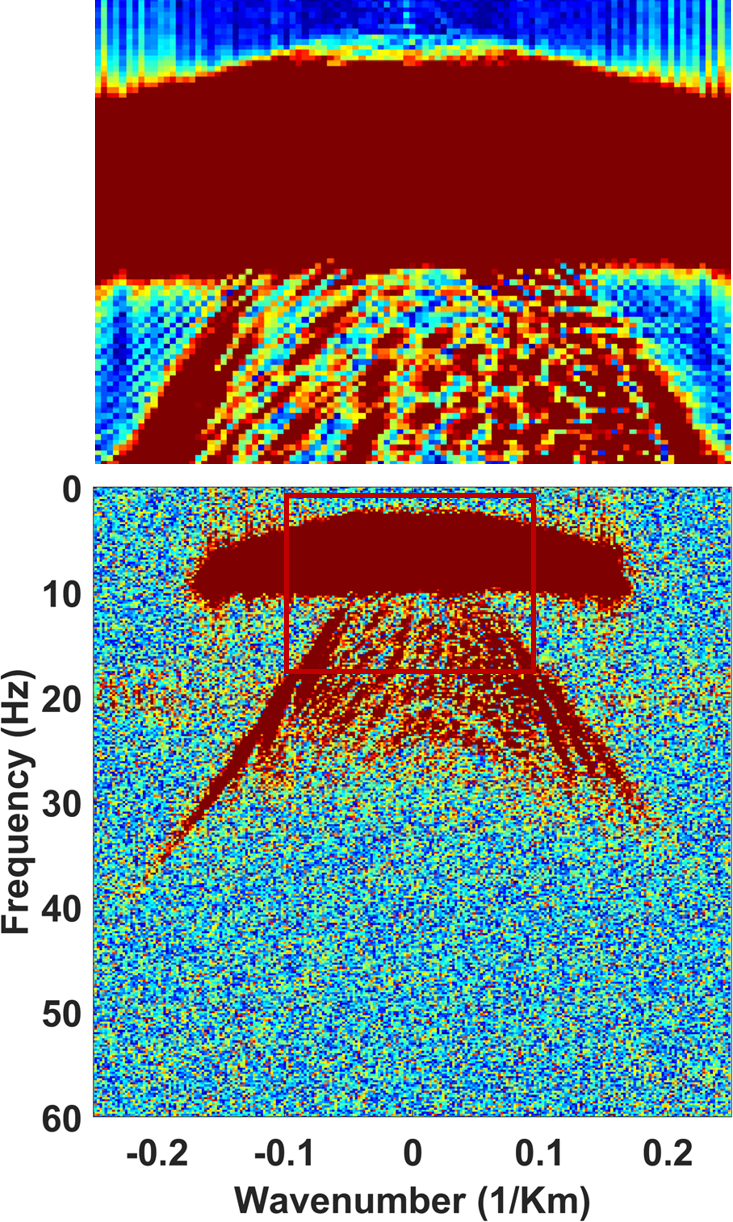}\\[-3pt]
{\small (a) Noisy data}
\end{minipage}

\vspace{3pt}

\begin{minipage}[b]{0.35\columnwidth}
\centering
\includegraphics[width=\linewidth]{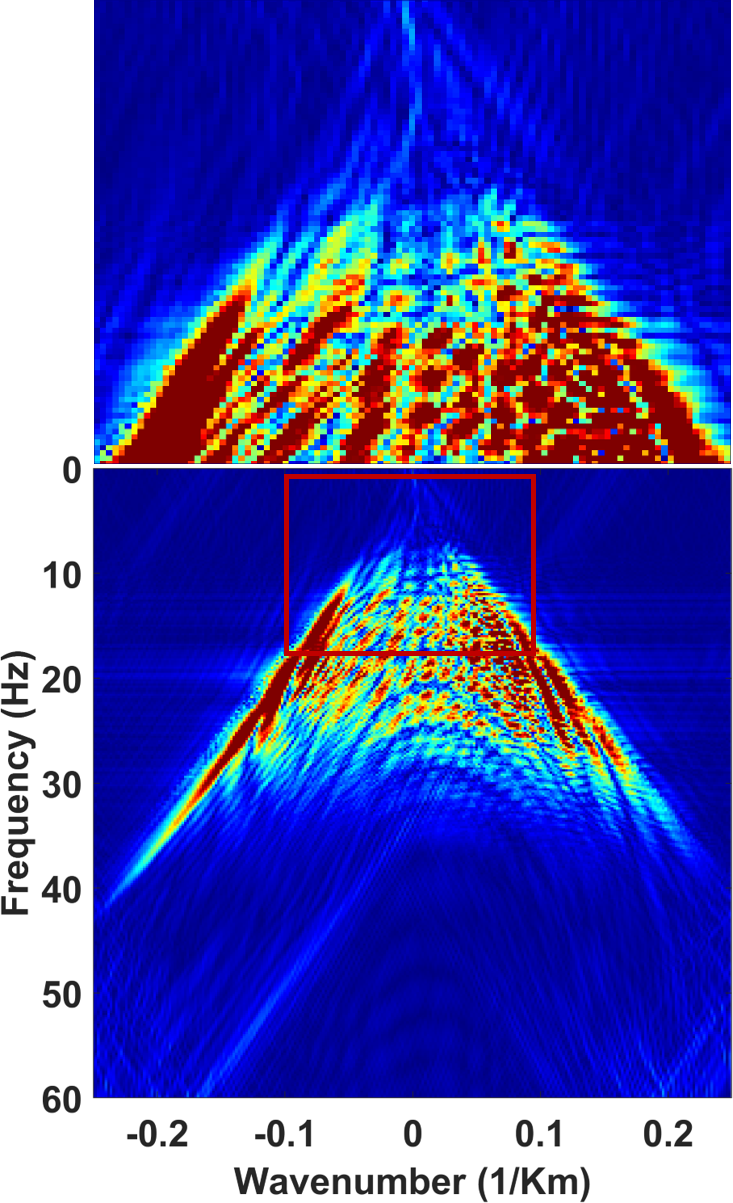}\\[-3pt]
{\small (b) Label}
\end{minipage}\hspace{0.02\columnwidth}\begin{minipage}[b]{0.35\columnwidth}
\centering
\includegraphics[width=\linewidth]{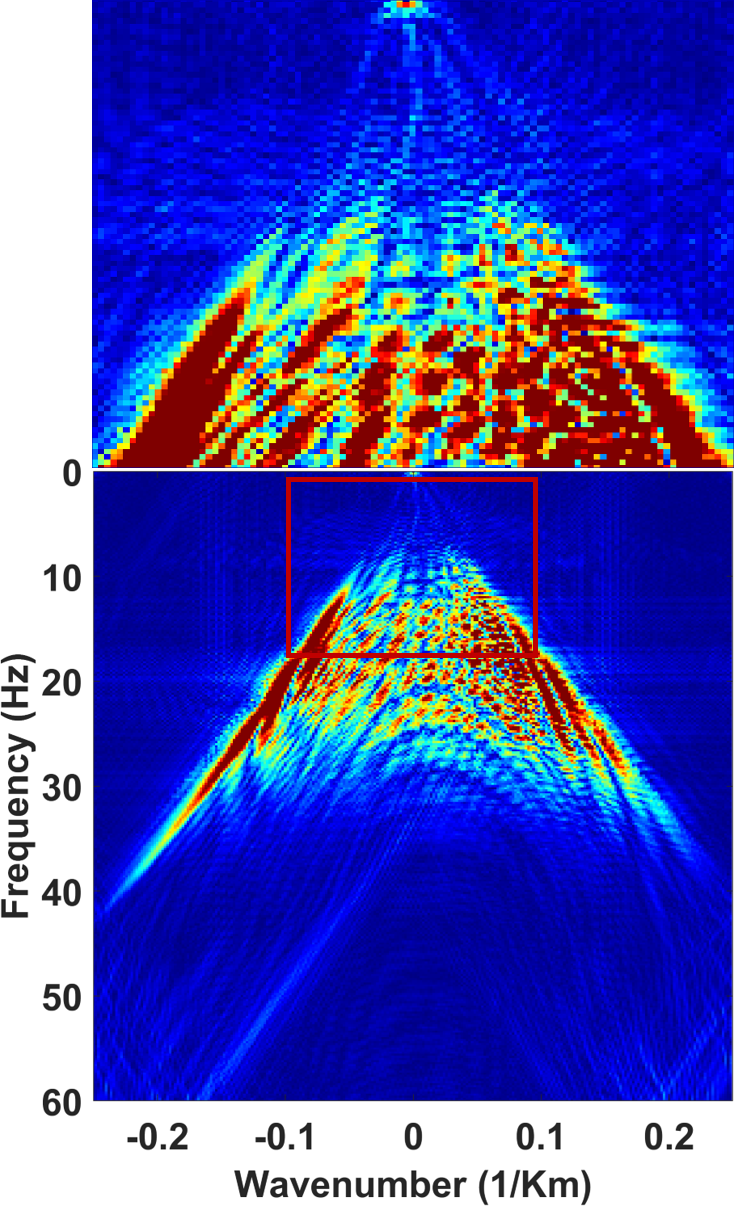}\\[-3pt]
{\small (c) SANet}
\end{minipage}

\vspace{3pt}

\begin{minipage}[b]{0.35\columnwidth}
\centering
\includegraphics[width=\linewidth]{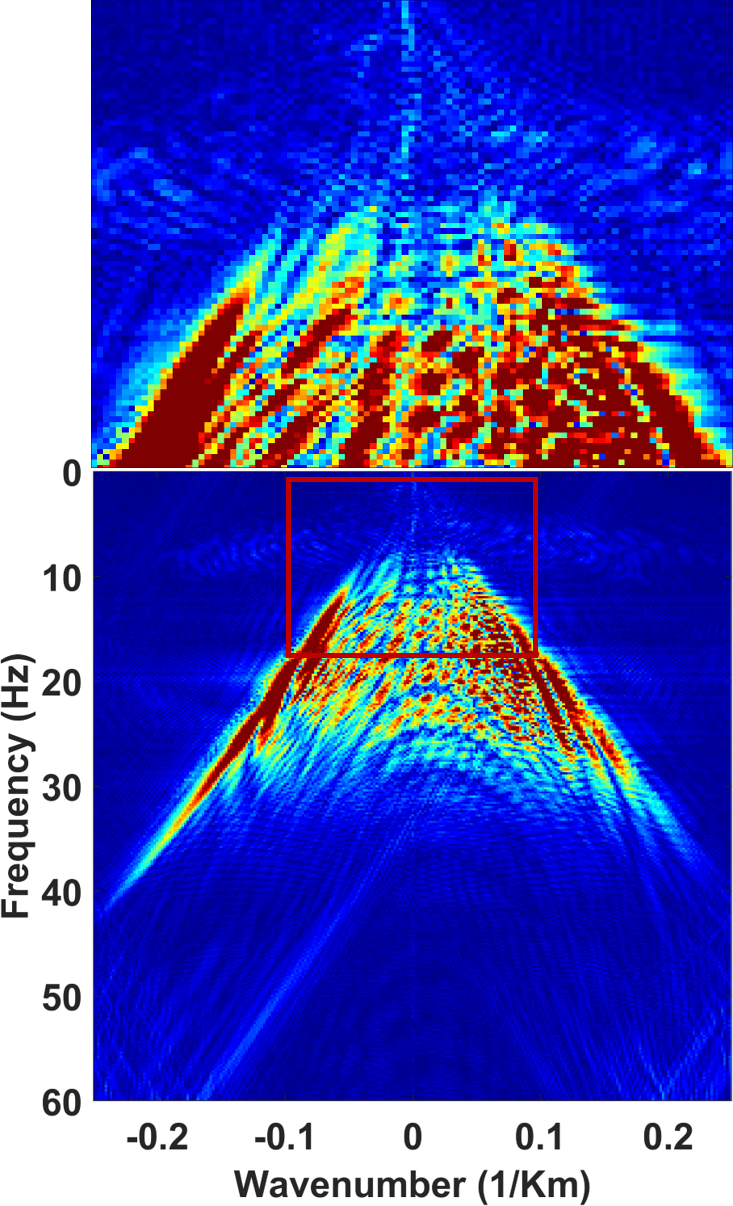}\\[-3pt]
{\small (d) Vision Transformer}
\end{minipage}\hspace{0.02\columnwidth}\begin{minipage}[b]{0.35\columnwidth}
\centering
\includegraphics[width=\linewidth]{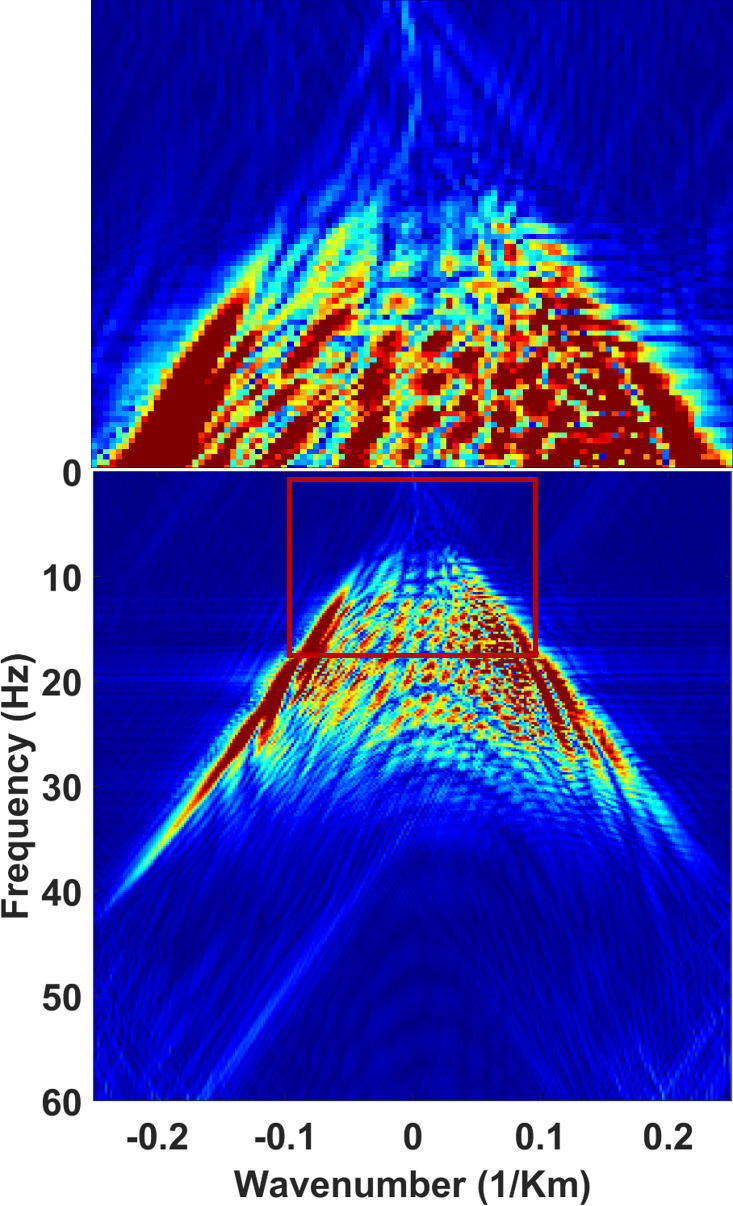}\\[-3pt]
{\small (e) GeoFormer}
\end{minipage}
\end{minipage}
\end{adjustbox}
\caption{The $f$-$k$ spectrum comparison for noise attenuation. (a) Noisy data; (b) Label; (c) SANet; (d) Vision Transformer; (e) GeoFormer.}
\label{fig:ground_roll_field_predictions}
\end{figure}

The $f$-$k$ spectrum comparison in Figure~\ref{fig:ground_roll_field_predictions} provides a complementary spectral evaluation of noise attenuation. Panels (a) and (b) show the $f$-$k$ spectra of the noisy input and clean label, respectively, while panels (c)--(e) present the corresponding spectra of SANet, Vision Transformer, and GeoFormer. A zoomed-in view of the ground-roll-dominated region is provided above each spectrum to facilitate a more detailed comparison. The ground roll is mainly distributed within the 0--12 Hz frequency range, whereas the random noise is broadband and contaminates the entire frequency band. In the zoomed-in views, SANet and Vision Transformer retain clear ground-roll frequency components and elevated broadband random-noise energy, whereas GeoFormer exhibits substantially less residual energy. Overall, the $f$-$k$ spectrum of GeoFormer most closely matches that of the label, indicating more effective suppression of ground-roll and random-noise energy with less spectral contamination. This spectral observation is consistent with the 2D prediction results.

\begin{figure}[!htbp]
\centering
\begin{adjustbox}{max totalsize={\textwidth}{0.76\textheight},center}
\begin{minipage}{\textwidth}
\centering
\begin{minipage}[b]{0.4\columnwidth}
		\centering
		\includegraphics[width=\linewidth]{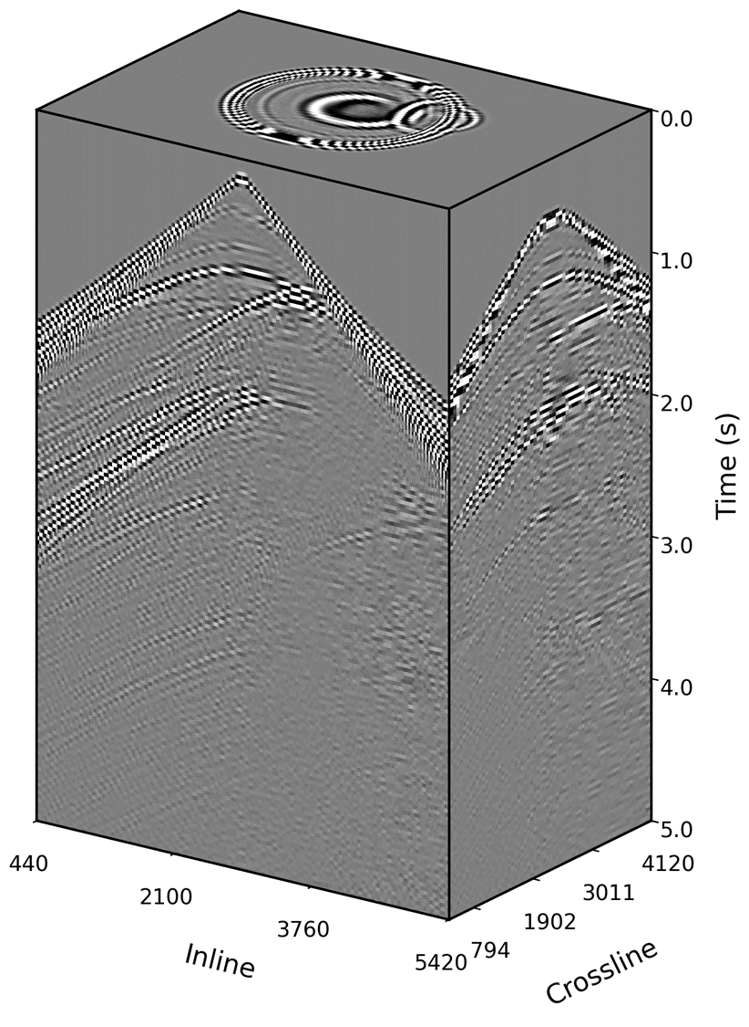}\\[-3pt]
		{\small (a) Label}
	\end{minipage}\hspace{0.02\columnwidth}\begin{minipage}[b]{0.4\columnwidth}
		\centering
		\includegraphics[width=\linewidth]{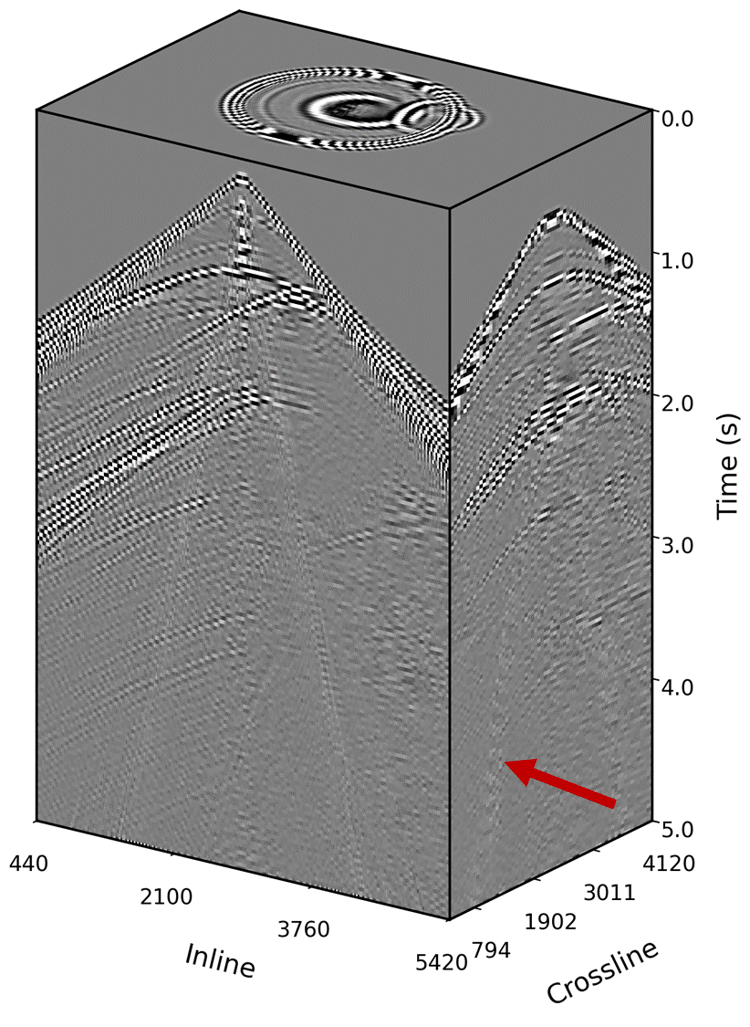}\\[-3pt]
		{\small (b) SANet}
	\end{minipage}

\vspace{3pt}

\begin{minipage}[b]{0.4\columnwidth}
		\centering
		\includegraphics[width=\linewidth]{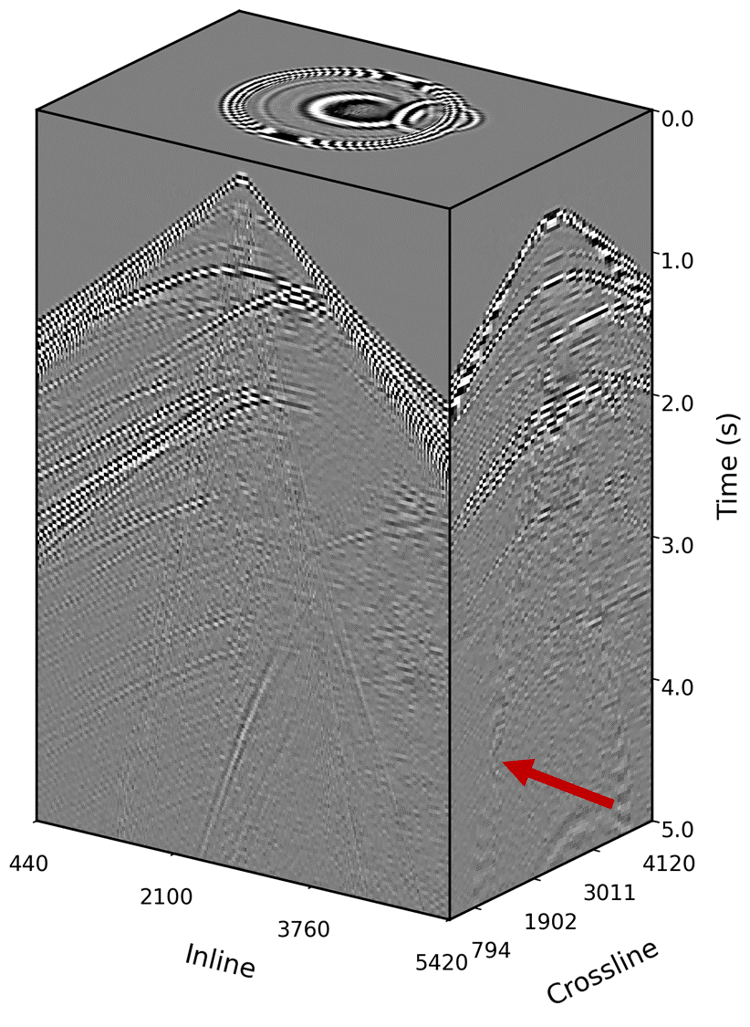}\\[-3pt]
		{\small (c) Vision Transformer}
	\end{minipage}\hspace{0.02\columnwidth}\begin{minipage}[b]{0.4\columnwidth}
		\centering
		\includegraphics[width=\linewidth]{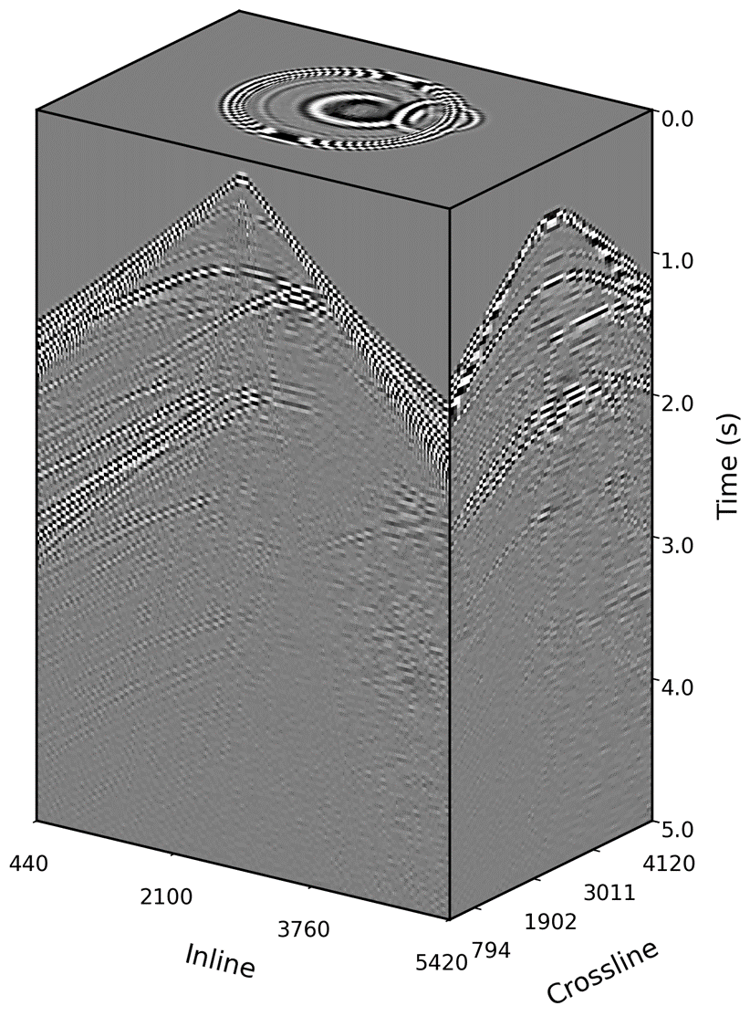}\\[-3pt]
		{\small (d) GeoFormer}
	\end{minipage}
\end{minipage}
\end{adjustbox}
\caption{3D noise-attenuation results on synthetic data. (a) Label; (b) SANet; (c) Vision Transformer; (d) GeoFormer.}
	\label{fig:ground_roll_predictions3D}
\end{figure}

\begin{figure}[!htbp]
\centering
\begin{adjustbox}{max totalsize={\textwidth}{0.76\textheight},center}
\begin{minipage}{\textwidth}
\centering
\begin{minipage}[b]{0.4\columnwidth}
	\vspace{0pt}
	\centering
	\includegraphics[width=\linewidth]{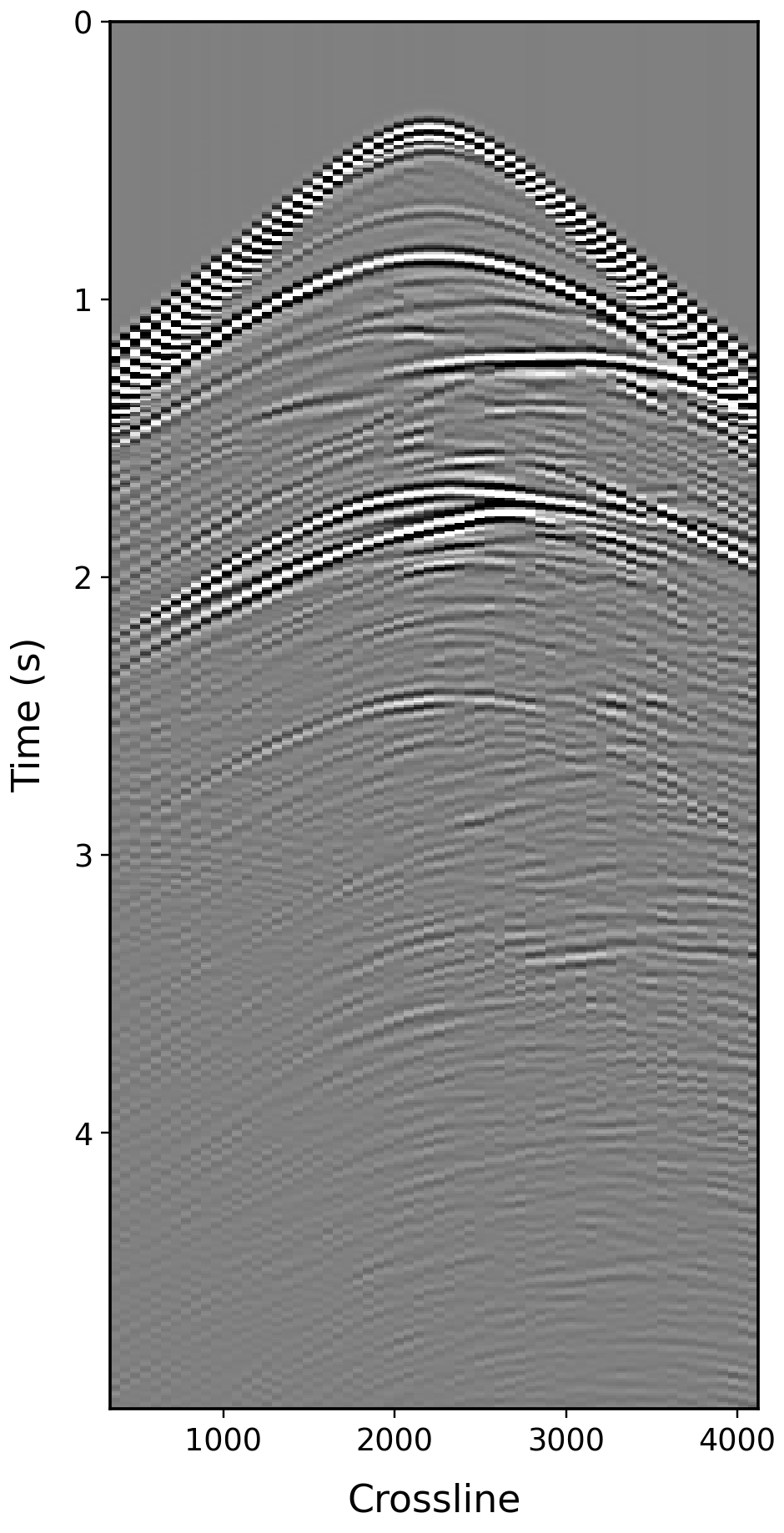}\\[-3pt]
	{\small (a) Label}
\end{minipage}\hspace{0.02\columnwidth}\begin{minipage}[b]{0.4\columnwidth}
	\vspace{0pt}
	\centering
	\includegraphics[width=\linewidth]{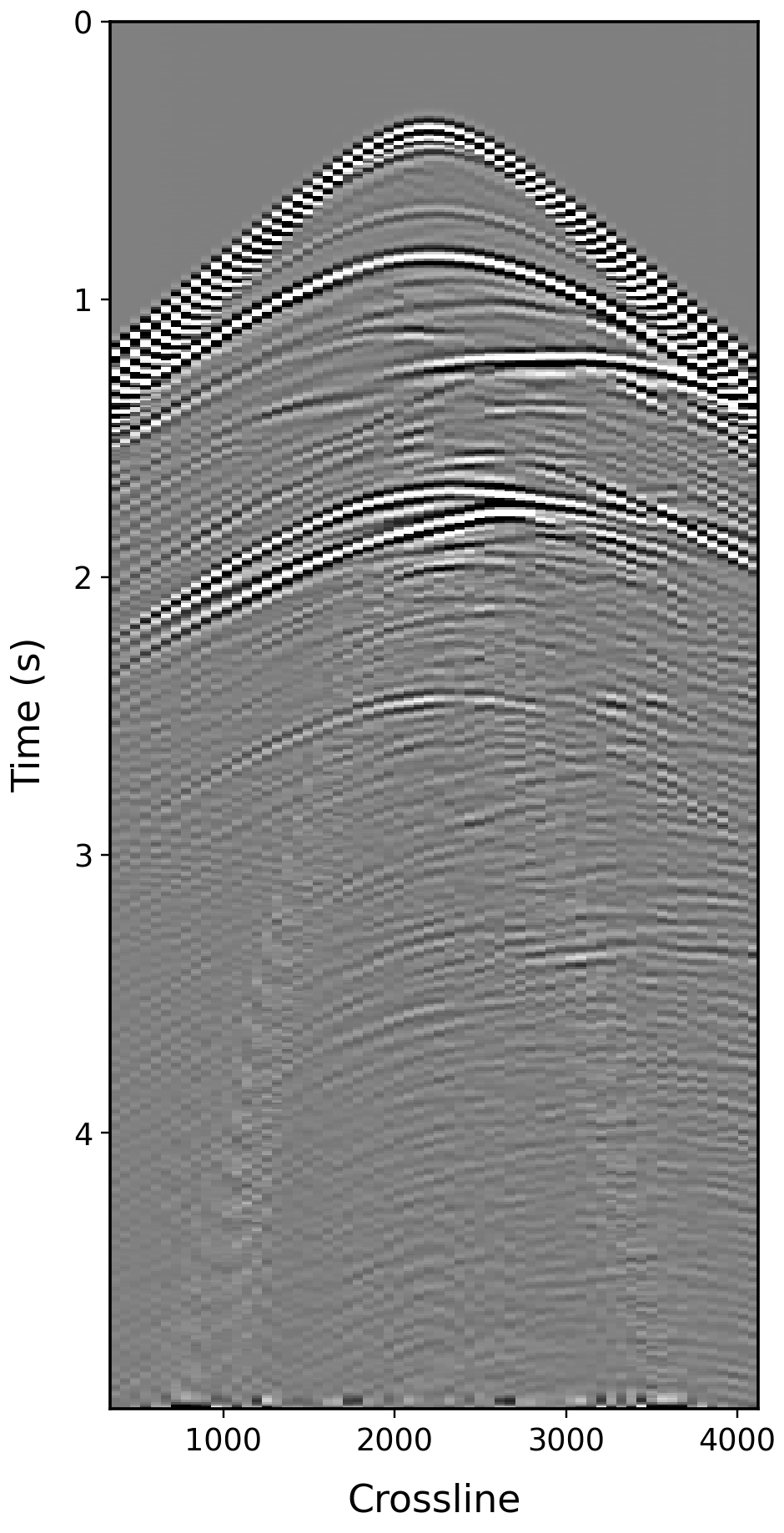}\\[-3pt]
	{\small (b) SANet}
\end{minipage}

\vspace{3pt}

\begin{minipage}[b]{0.4\columnwidth}
	\vspace{0pt}
	\centering
	\includegraphics[width=\linewidth]{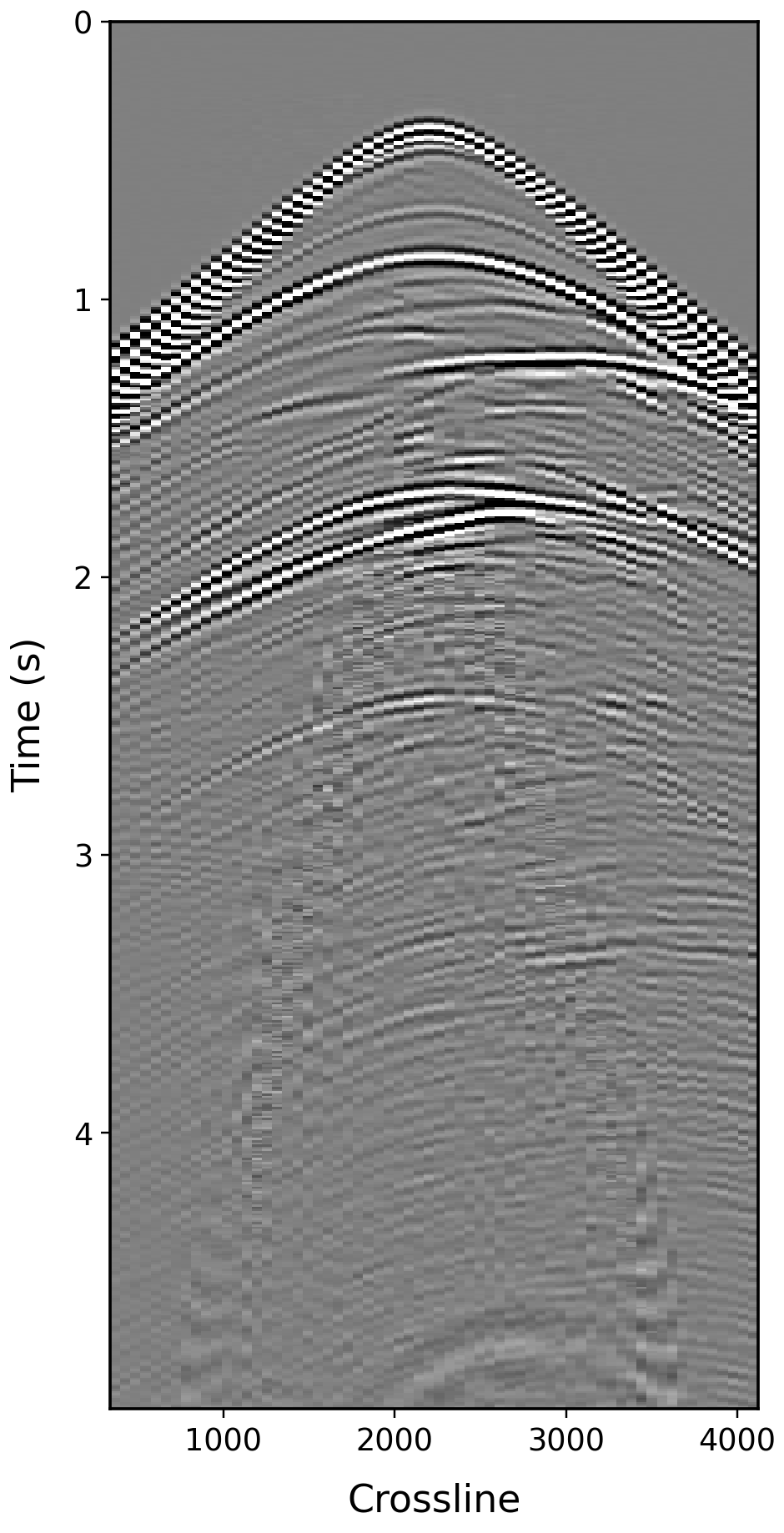}\\[-3pt]
	{\small (c) Vision Transformer}
\end{minipage}\hspace{0.02\columnwidth}\begin{minipage}[b]{0.4\columnwidth}
	\vspace{0pt}
	\centering
	\includegraphics[width=\linewidth]{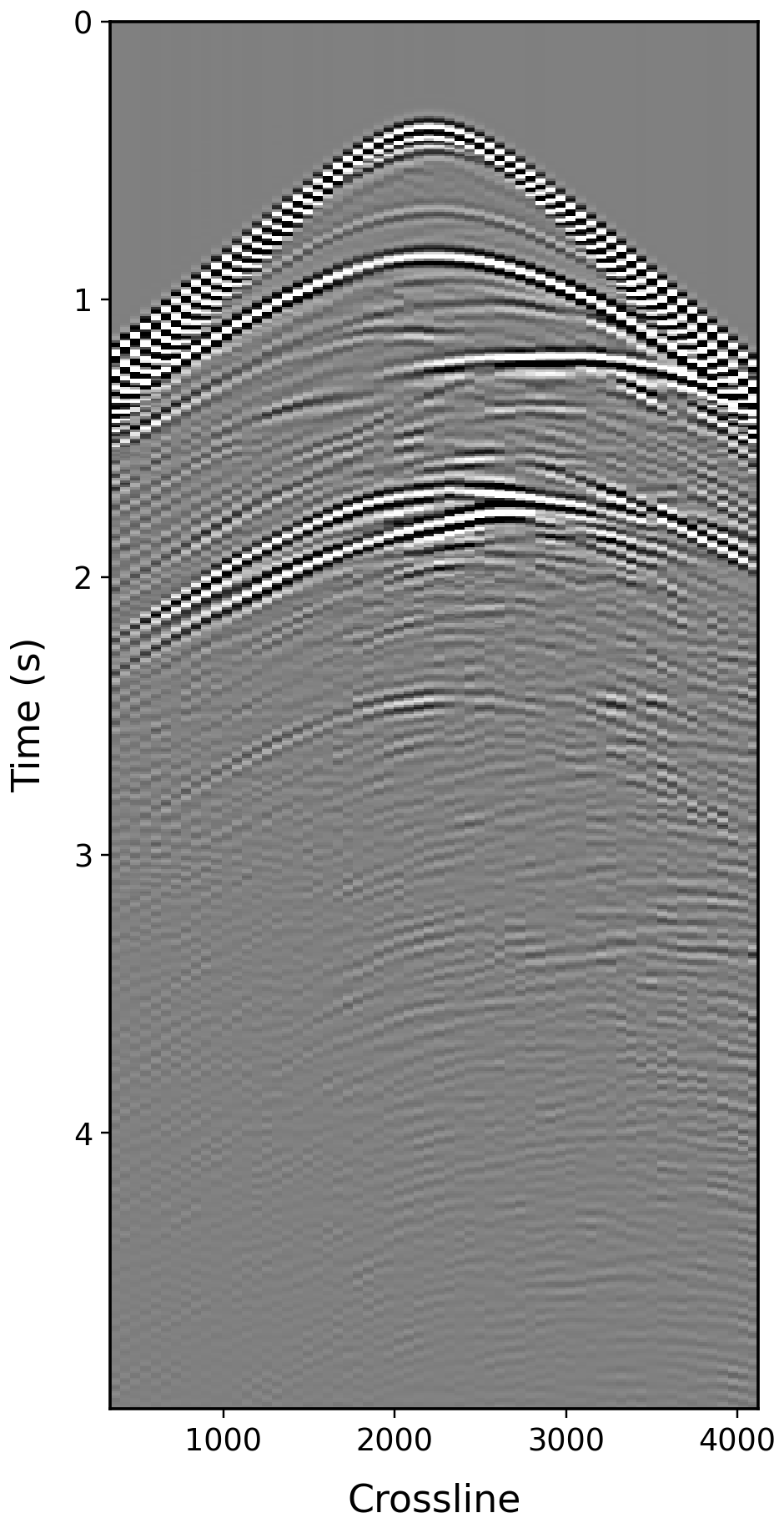}\\[-3pt]
	{\small (d) GeoFormer}
\end{minipage}
\end{minipage}
\end{adjustbox}
\caption{3D noise-attenuation results on synthetic data in the crossline direction. (a) Label; (b) SANet; (c) Vision Transformer; (d) GeoFormer.}
	\label{fig:ground_roll_predictions3D_crossline}
\end{figure}

Figure~\ref{fig:ground_roll_predictions3D} compares the clean reference volume with the denoised volumes predicted by SANet, Vision Transformer, and GeoFormer for a single shot, providing a larger-scale view of the noise-attenuation performance. SANet and Vision Transformer attenuate the ground roll but retain visible residual ground-roll and random-noise energy and local artifacts, whereas GeoFormer produces the cleanest 3D result with better preservation of the spatial continuity of the reflection events. Figure~\ref{fig:ground_roll_predictions3D_crossline} further shows crossline-direction slices extracted from these 3D results. Residual ground roll, random noise, and artifacts remain visible in the SANet and Vision Transformer results, particularly around the strong events in the upper part of the sections, whereas GeoFormer substantially reduces the residual ground roll and random noise while preserving the lateral continuity and amplitude patterns of the reflection events. These larger-scale comparisons further demonstrate the effectiveness of GeoFormer in jointly modeling seismic waveforms and acquisition geometry. By exploiting inter-trace spatial relationships, GeoFormer can distinguish seismic signals from noise while preserving the continuity and structural details of seismic events.

\begin{table}[!htbp]
	\centering
\small
\setlength{\tabcolsep}{3pt}
	\caption{Quantitative seismic noise attenuation results.}
	\label{tab:ground_roll_snr_ssim}
	\begin{adjustbox}{max width=\textwidth}
\begin{tabular}{lccc}
		\toprule
		Method & SNR (dB) & SSIM & MAE \\
		\midrule
		Noisy & $-8.63$ & 0.15 & 0.3400 \\
		SANet & 21.25 & 0.95 & 0.0064 \\
		Vision Transformer & 23.72 & 0.94 & 0.0075 \\
		GeoFormer & \textbf{26.91} & \textbf{0.97} & \textbf{0.0048} \\
		\bottomrule
	\end{tabular}
\end{adjustbox}
\end{table}

Table~\ref{tab:ground_roll_snr_ssim} summarizes the quantitative seismic noise attenuation results using SNR, SSIM, and MAE. Compared with the noisy input, GeoFormer achieves substantial improvements in all metrics, with a 35.54 dB increase in SNR, a 0.82 improvement in SSIM, and a 0.3352 reduction in MAE. Compared with both SANet and Vision Transformer, GeoFormer consistently achieves superior performance, obtaining the highest SNR and SSIM values and the lowest MAE.

\subsection{First-Arrival Picking}

Table~\ref{tab:main_results} reports the picking results of all methods on the four test sets.
GeoFormer achieves the lowest RMSE and MAE on all four datasets and the highest HR@3--HR@9 in most cases. HU-Net retains the highest HR@1 on Brunswick and Lalor, showing that the advantage of GeoFormer is more evident in reducing large errors and improving overall picking stability than in uniformly dominating every tolerance level. Compared with Vision Transformer, GeoFormer reduces RMSE from 14.62 to 1.73 on Brunswick, from 14.23 to 1.84 on Halfmile, and from 14.39 to 3.52 on Dongbei. The difference between the two Transformer models indicates that self-attention alone does not account for the observed improvement; explicitly providing acquisition geometry is beneficial for this task.

\begin{table}[!htbp]
\centering
\small
\setlength{\tabcolsep}{3pt}
\caption{Picking accuracy of GeoFormer and baseline methods across four field datasets. Best results are in \textbf{bold}. HR@$k$ is reported as a percentage with a tolerance of $k$ time samples; RMSE, MAE, and MBE are in units of time samples.}
\label{tab:main_results}
\begin{adjustbox}{max width=\textwidth}
\begin{tabular}{llcccccccc}
\toprule
Site & Method & RMSE & MAE & MBE & HR@1 & HR@3 & HR@5 & HR@7 & HR@9 \\
\midrule
\midrule
\multirow{4}{*}{Brunswick}
 & Vision Transformer & 14.62 & 1.88 & $-$0.19 & 80.9 & 92.8 & 96.0 & 97.7 & 98.5 \\
 & Attention UNet & 114.96 & 42.07 & $-$41.62 & 82.0 & 83.2 & 83.6 & 83.8 & 83.9 \\
 & HU-Net & 12.06 & 1.19 & $+$0.69 & \textbf{94.8} & \textbf{97.5} & 98.2 & 98.5 & 98.7 \\
 & \textbf{GeoFormer} & \textbf{1.73} & \textbf{0.47} & \textbf{$+$0.09} & 93.8 & \textbf{97.5} & \textbf{98.5} & \textbf{99.0} & \textbf{99.3} \\
\midrule
\multirow{4}{*}{Halfmile}
 & Vision Transformer & 14.23 & 3.06 & $-$0.25 & 55.4 & 84.2 & 92.2 & 95.5 & 97.0 \\
 & Attention UNet & 83.74 & 32.58 & $-$31.97 & 73.1 & 77.0 & 79.2 & 80.5 & 81.2 \\
 & HU-Net & 15.29 & 2.57 & $+$2.11 & 86.7 & 92.0 & 94.5 & 95.9 & 96.6 \\
 & \textbf{GeoFormer} & \textbf{1.84} & \textbf{0.80} & \textbf{$+$0.12} & \textbf{88.3} & \textbf{94.0} & \textbf{97.2} & \textbf{98.7} & \textbf{99.4} \\
\midrule
\multirow{4}{*}{Lalor}
 & Vision Transformer & 37.31 & 11.43 & $+$1.74 & 23.5 & 48.2 & 64.2 & 73.9 & 80.0 \\
 & Attention UNet & 78.07 & 23.63 & $-$22.70 & 76.4 & 80.0 & 83.1 & 85.8 & 87.5 \\
 & HU-Net & 13.56 & 2.02 & $+$0.72 & \textbf{78.4} & 83.0 & 84.2 & 89.3 & 93.6 \\
 & \textbf{GeoFormer} & \textbf{5.32} & \textbf{1.78} & \textbf{$+$0.18} & 75.6 & \textbf{85.2} & \textbf{90.4} & \textbf{94.2} & \textbf{96.7} \\
\midrule
\multirow{4}{*}{Dongbei}
 & Vision Transformer & 14.39 & 5.20 & $-$0.22 & 39.9 & 66.8 & 77.2 & 82.6 & 86.3 \\
 & Attention UNet & 158.57 & 26.66 & $-$23.94 & 63.3 & 85.0 & 88.3 & 89.2 & 89.9 \\
 & HU-Net & 9.94 & 2.40 & $-$1.50 & 65.7 & 88.0 & 92.1 & 93.7 & 94.6 \\
 & \textbf{GeoFormer} & \textbf{3.52} & \textbf{1.21} & \textbf{$-$0.09} & \textbf{80.9} & \textbf{93.3} & \textbf{94.3} & \textbf{94.8} & \textbf{95.2} \\
\bottomrule
\end{tabular}
\end{adjustbox}
\end{table}

\begin{table}[!htbp]
\centering
\small
\setlength{\tabcolsep}{3pt}
\caption{Ablation study of geometry injection mechanisms on the Halfmile dataset.  Best results are in \textbf{bold}.}
\label{tab:ablation}
\begin{adjustbox}{max width=\textwidth}
\begin{tabular}{lcccccccc}
\toprule
Configuration & RMSE & MAE & MBE & HR@1 & HR@3 & HR@5 & HR@7 & HR@9 \\
\midrule
Full GeoFormer & \textbf{1.84} & \textbf{0.80} & $+$0.12 & \textbf{88.3} & \textbf{94.0} & \textbf{97.2} & \textbf{98.7} & \textbf{99.4} \\
$-$ GeomAttnBias & 2.87 & 0.91 & $+$0.12 & 85.1 & 92.9 & 96.1 & 98.5 & 99.0 \\
$-$ GeomAdaLN & 2.90 & 0.94 & $+$0.11 & 85.2 & 92.8 & 96.0 & 98.6 & 99.3 \\
$-$ GeomMLP & 5.55 & 1.16 & $+$0.11 & 83.6 & 92.5 & 96.0 & 97.9 & 98.7 \\
$-$ All geometry modules & 10.30 & 2.10 & \textbf{$+$0.01} & 84.3 & 90.9 & 93.9 & 95.6 & 96.4  \\
\bottomrule
\end{tabular}
\end{adjustbox}
\end{table}

\begin{figure}[!htbp]
\centering
\begin{adjustbox}{max totalsize={\textwidth}{0.76\textheight},center}
\begin{minipage}{\textwidth}
\centering
\begin{minipage}[b]{0.4\columnwidth}
		\centering
		\includegraphics[width=\linewidth]{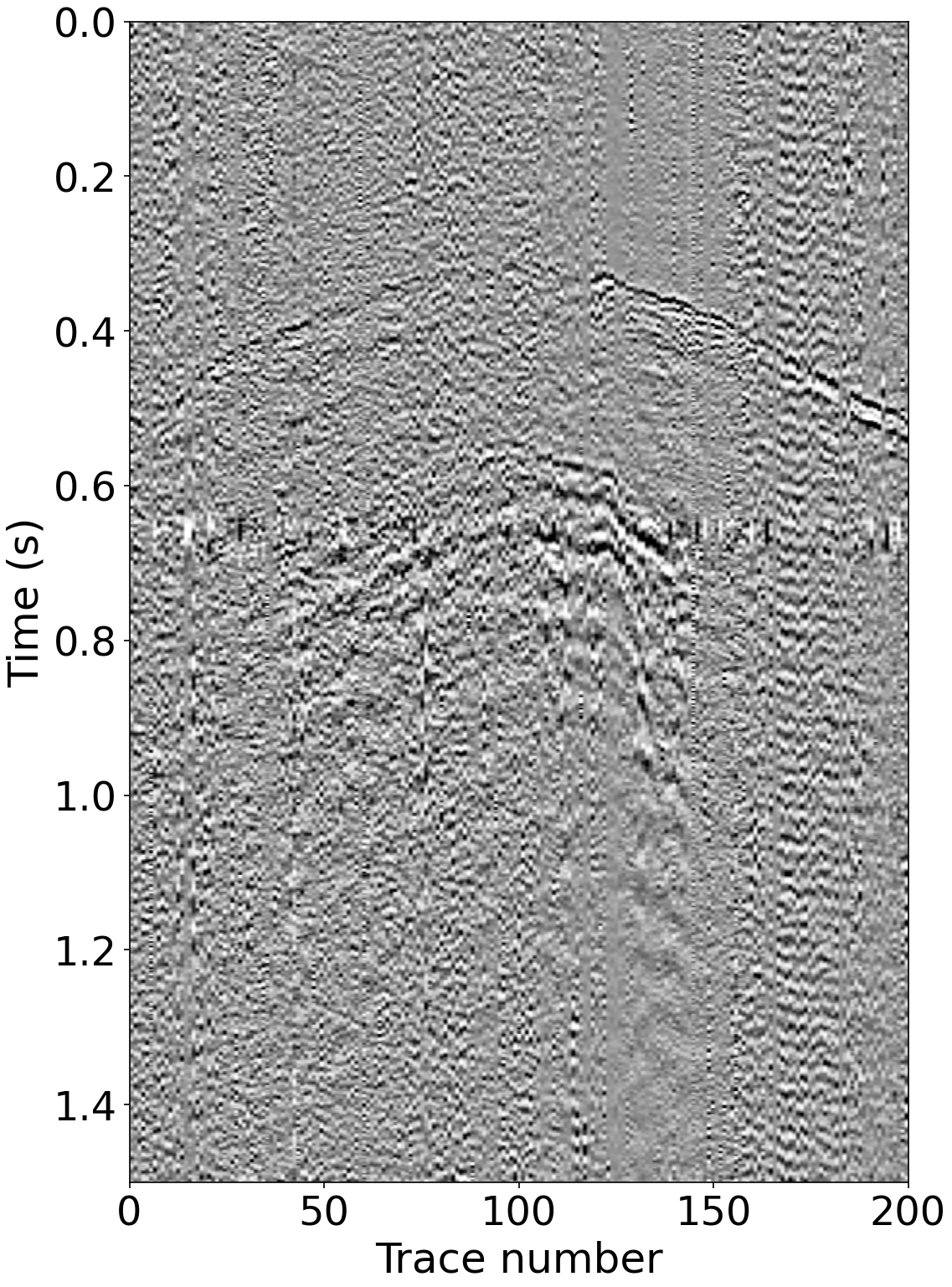}\\[-1pt]
		{\small (a) Seismic}
		\label{fig:pick_hm_a}
	\end{minipage}\hspace{0.02\columnwidth}\begin{minipage}[b]{0.4\columnwidth}
		\centering
		\includegraphics[width=\linewidth]{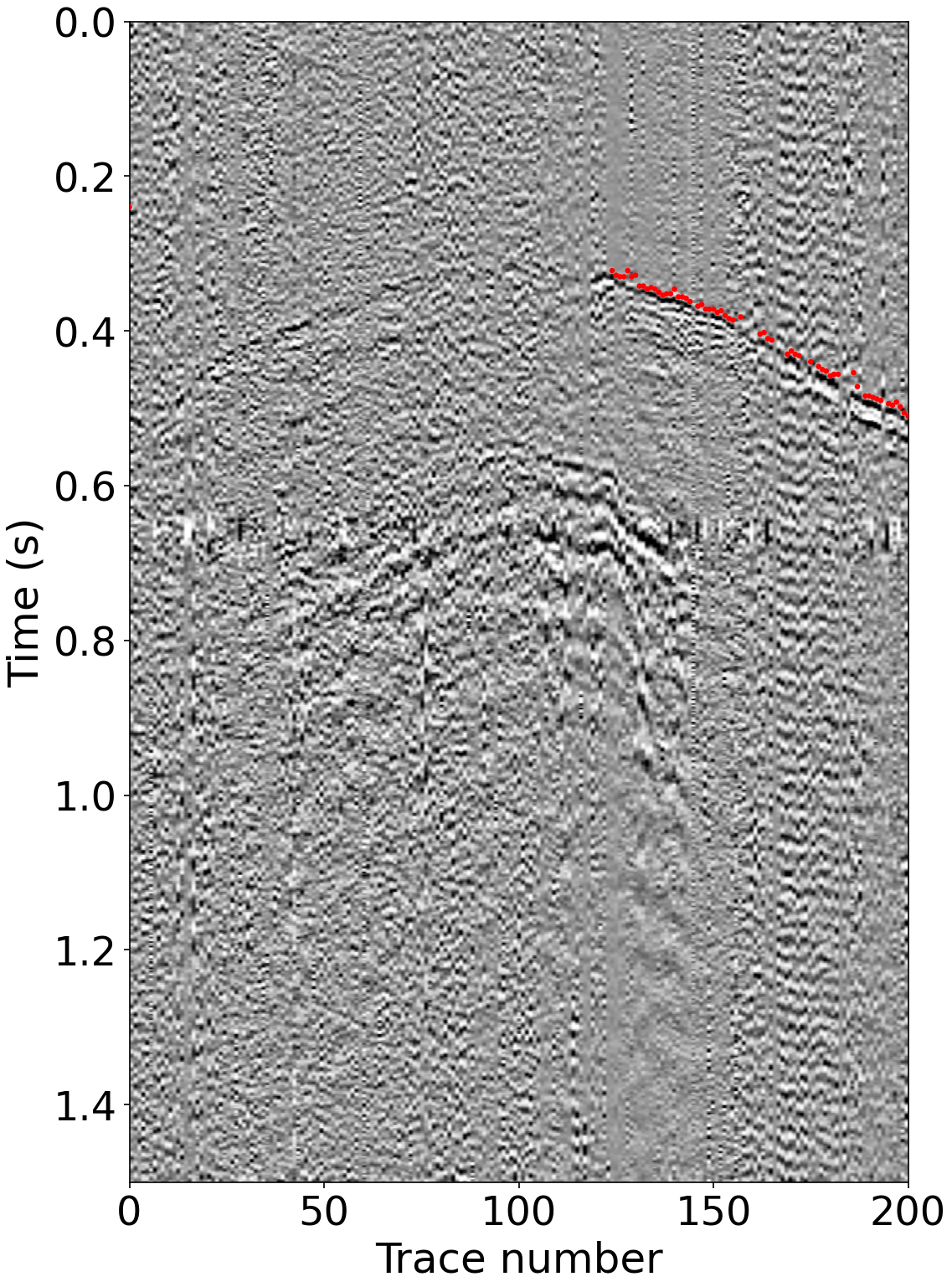}\\[-1pt]
		{\small (b) Label}
		\label{fig:pick_hm_b}
	\end{minipage}

\vspace{3pt}

\begin{minipage}[b]{0.4\columnwidth}
		\centering
		\includegraphics[width=\linewidth]{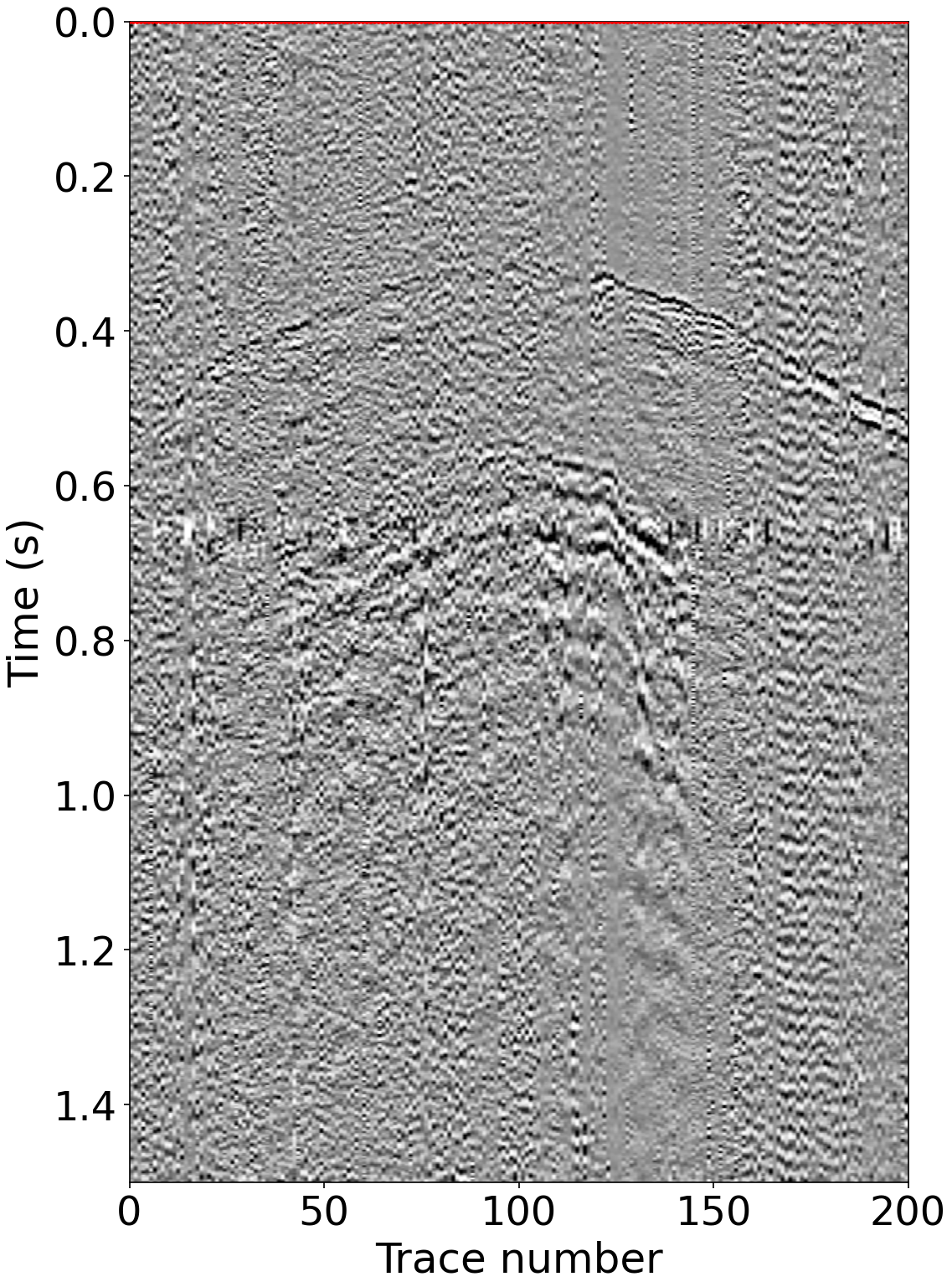}\\[-1pt]
		{\small (c) AttnUNet}
		\label{fig:pick_hm_c}
	\end{minipage}\hspace{0.02\columnwidth}\begin{minipage}[b]{0.4\columnwidth}
		\centering
		\includegraphics[width=\linewidth]{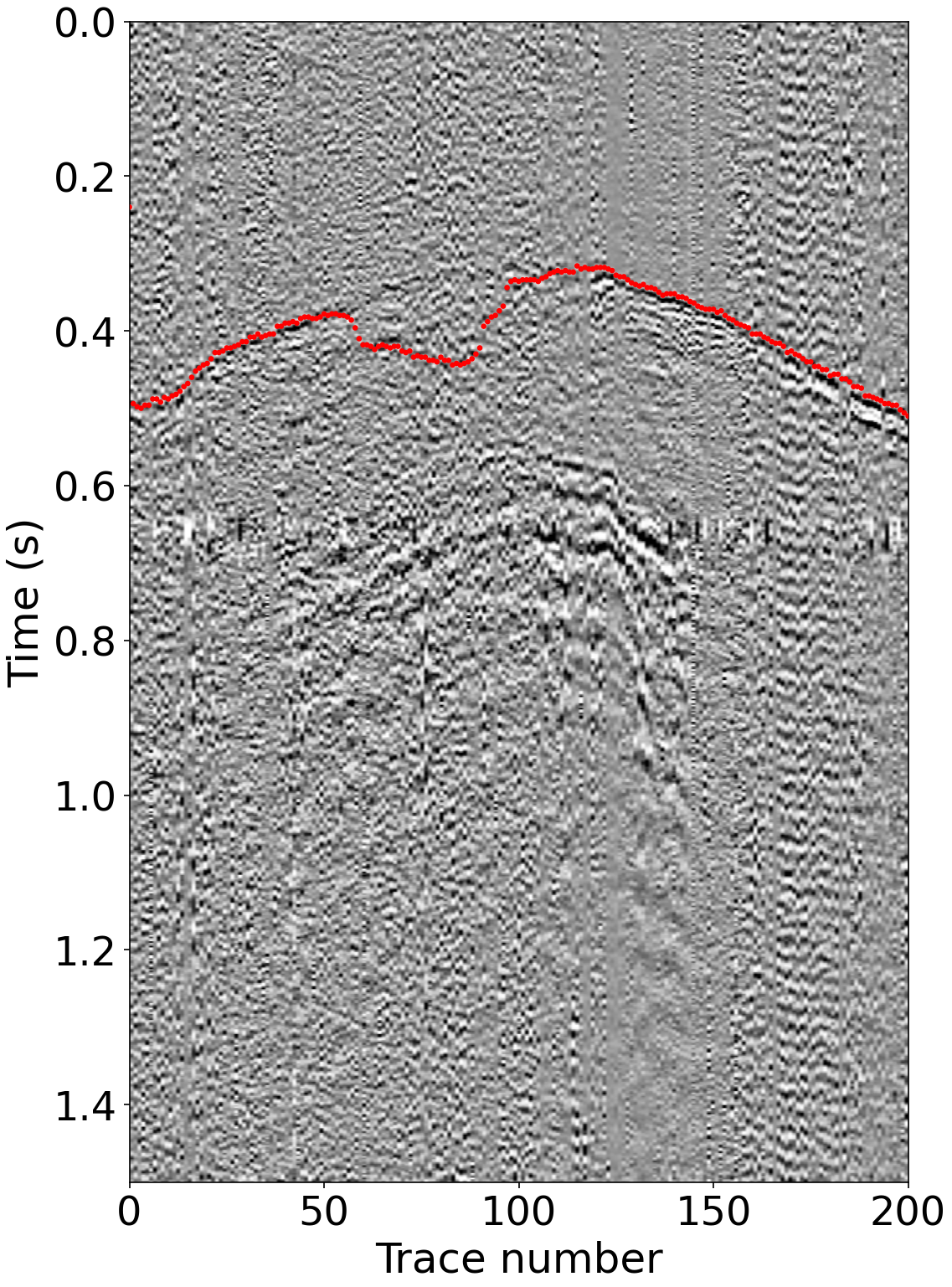}\\[-1pt]
		{\small (d) HU-Net}
		\label{fig:pick_hm_d}
	\end{minipage}

\vspace{3pt}

\begin{minipage}[b]{0.4\columnwidth}
		\centering
		\includegraphics[width=\linewidth]{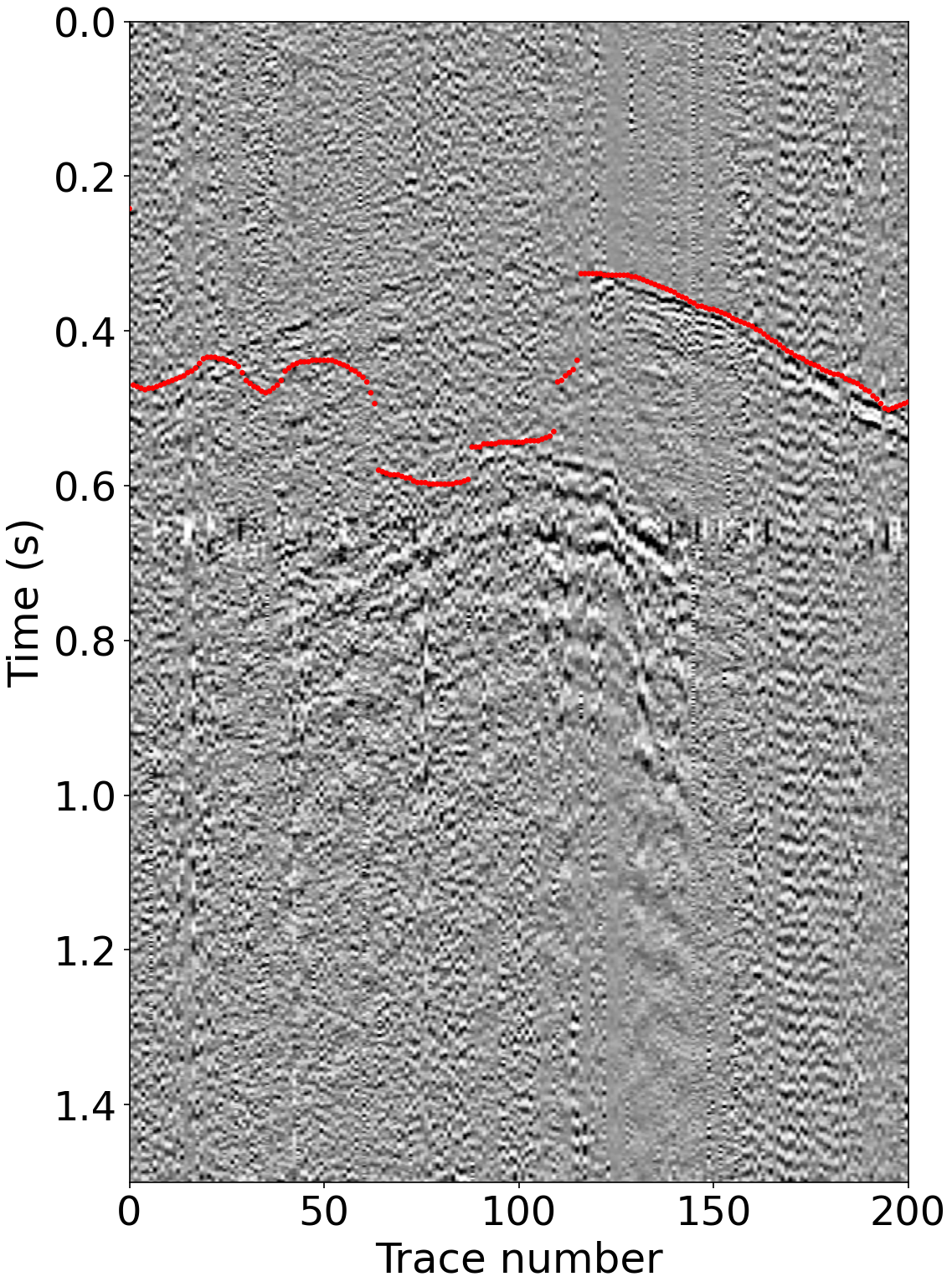}\\[-1pt]
		{\small (e) Vision Transformer}
		\label{fig:pick_hm_e}
	\end{minipage}\hspace{0.02\columnwidth}\begin{minipage}[b]{0.4\columnwidth}
		\centering
		\includegraphics[width=\linewidth]{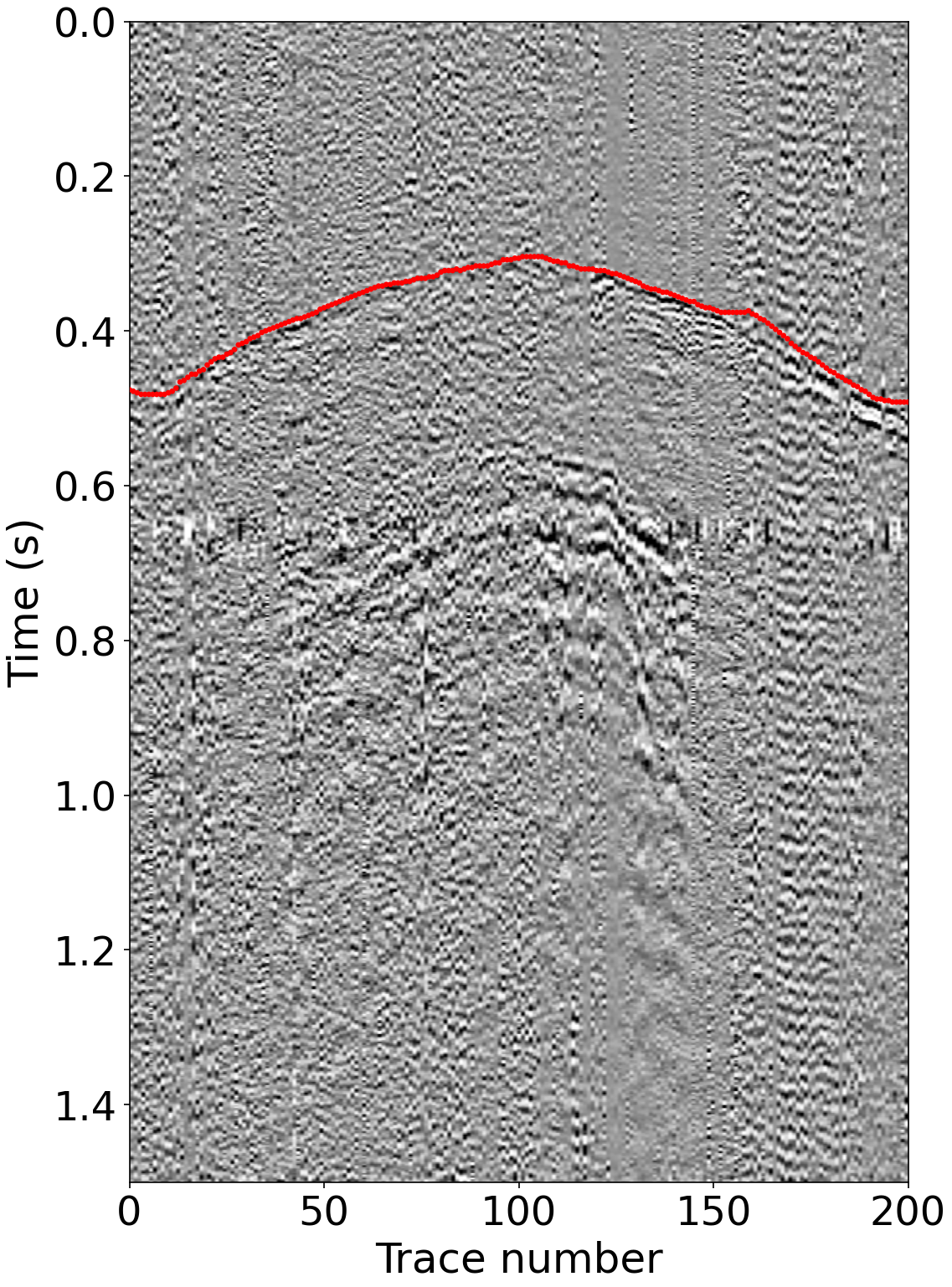}\\[-1pt]
		{\small (f) GeoFormer}
		\label{fig:pick_hm_f}
	\end{minipage}
\end{minipage}
\end{adjustbox}
\caption{Representative first-arrival picking results on the Halfmile test set. (a) Seismic shot gather; (b) manual picks; (c) AttnUNet; (d) HU-Net; (e) Vision Transformer; (f) GeoFormer.}
	\label{fig:picking_halfmile}
\end{figure}

\begin{figure}[!htbp]
\centering
\begin{adjustbox}{max totalsize={\textwidth}{0.76\textheight},center}
\begin{minipage}{\textwidth}
\centering
\begin{minipage}[b]{0.4\columnwidth}
		\centering
		\includegraphics[width=\linewidth]{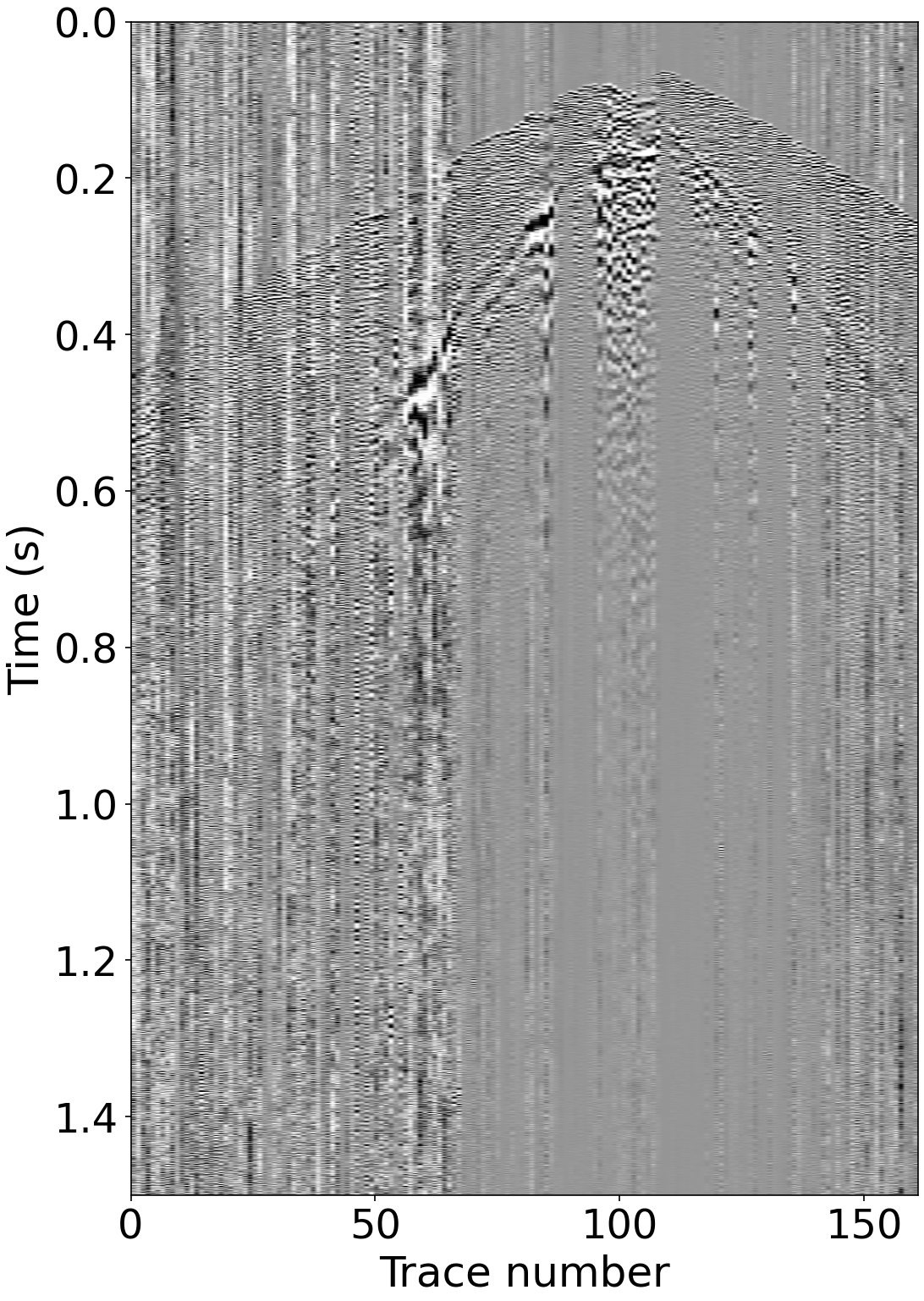}\\[-1pt]
		{\small (a) Seismic}
		\label{fig:pick_lr_a}
	\end{minipage}\hspace{0.02\columnwidth}\begin{minipage}[b]{0.4\columnwidth}
		\centering
		\includegraphics[width=\linewidth]{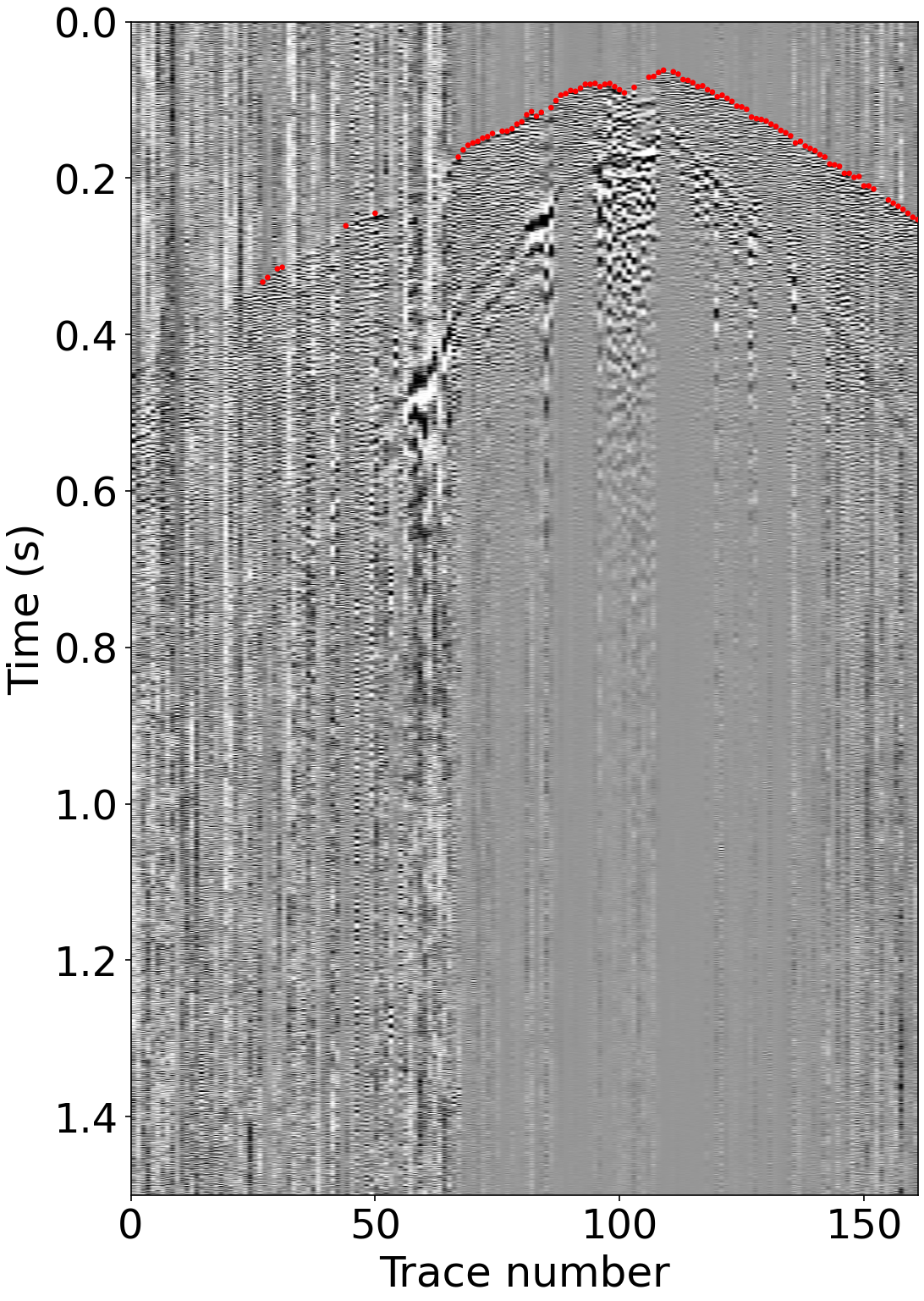}\\[-1pt]
		{\small (b) Label}
		\label{fig:pick_lr_b}
	\end{minipage}

\vspace{3pt}

\begin{minipage}[b]{0.4\columnwidth}
		\centering
		\includegraphics[width=\linewidth]{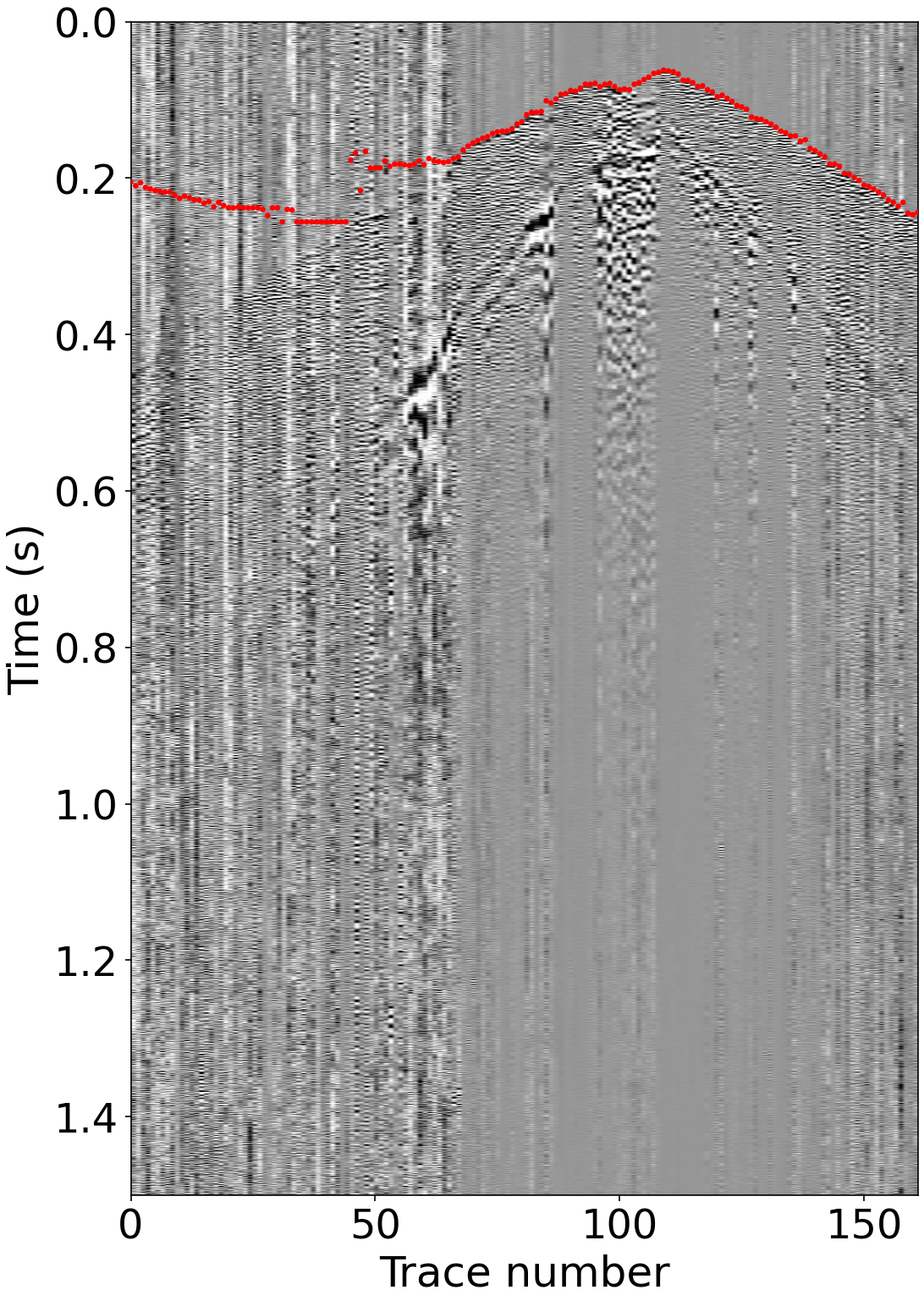}\\[-1pt]
		{\small (c) AttnUNet}
		\label{fig:pick_lr_c}
	\end{minipage}\hspace{0.02\columnwidth}\begin{minipage}[b]{0.4\columnwidth}
		\centering
		\includegraphics[width=\linewidth]{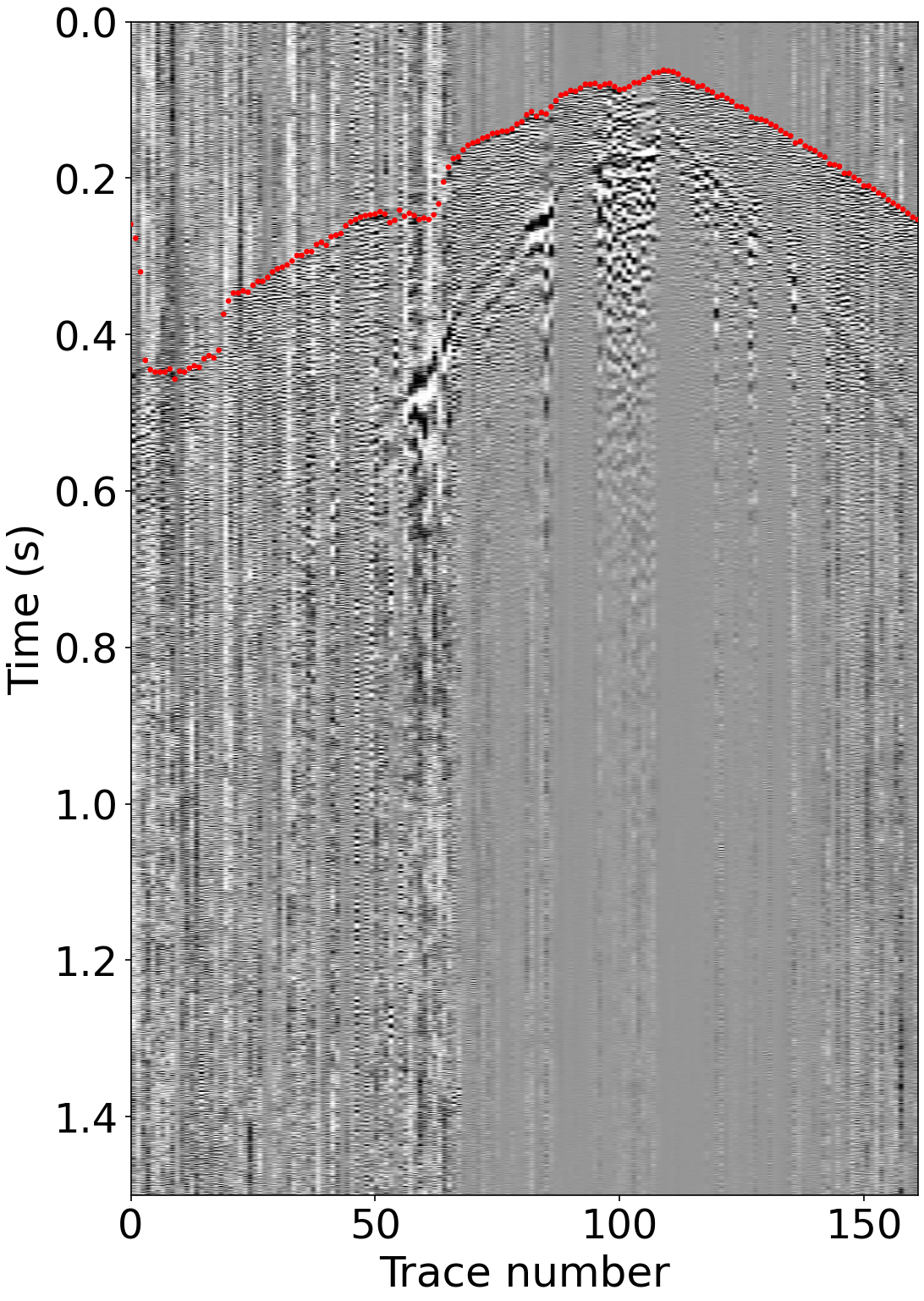}\\[-1pt]
		{\small (d) HU-Net}
		\label{fig:pick_lr_d}
	\end{minipage}

\vspace{3pt}

\begin{minipage}[b]{0.4\columnwidth}
		\centering
		\includegraphics[width=\linewidth]{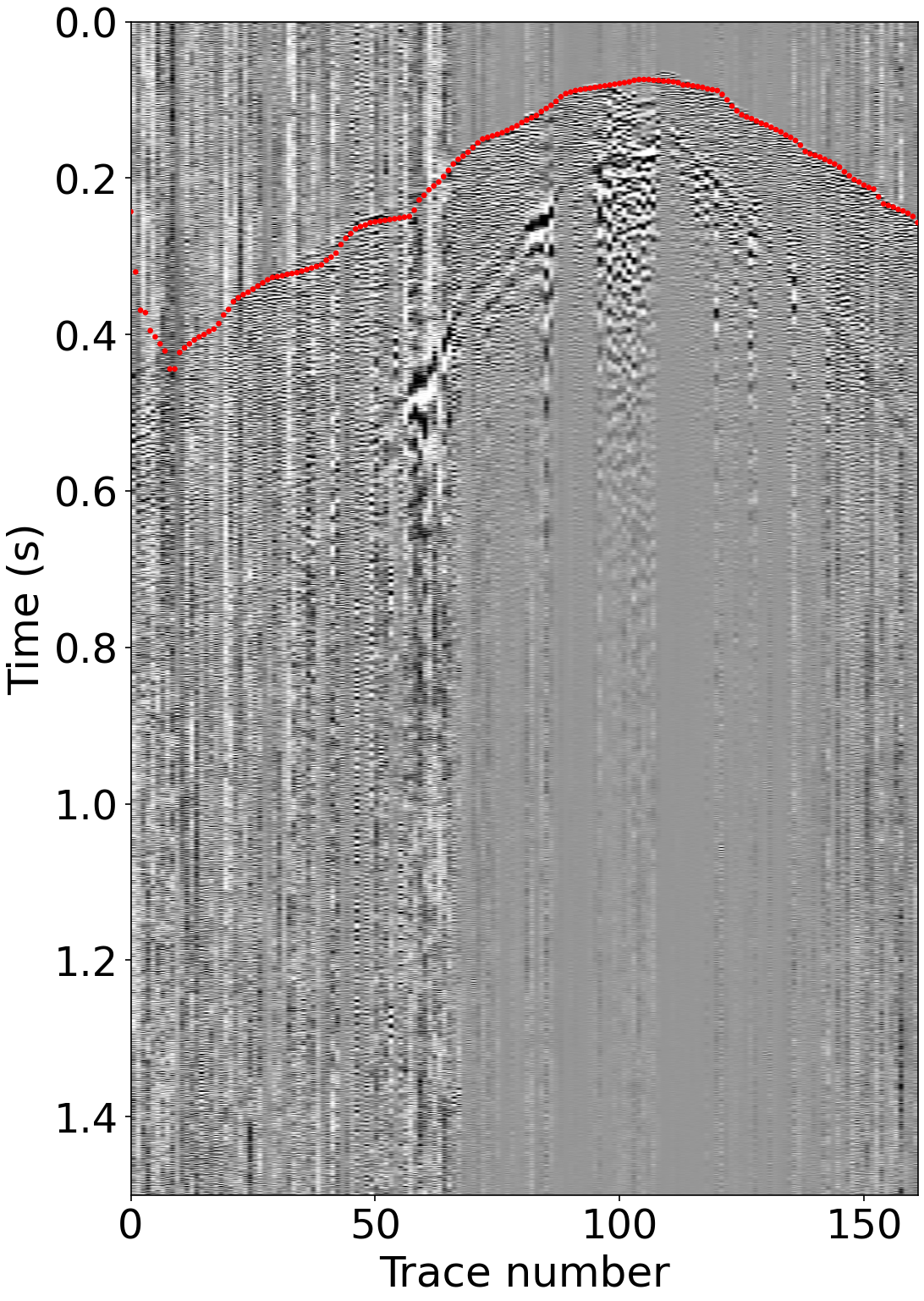}\\[-1pt]
		{\small (e) Vision Transformer}
		\label{fig:pick_lr_e}
	\end{minipage}\hspace{0.02\columnwidth}\begin{minipage}[b]{0.4\columnwidth}
		\centering
		\includegraphics[width=\linewidth]{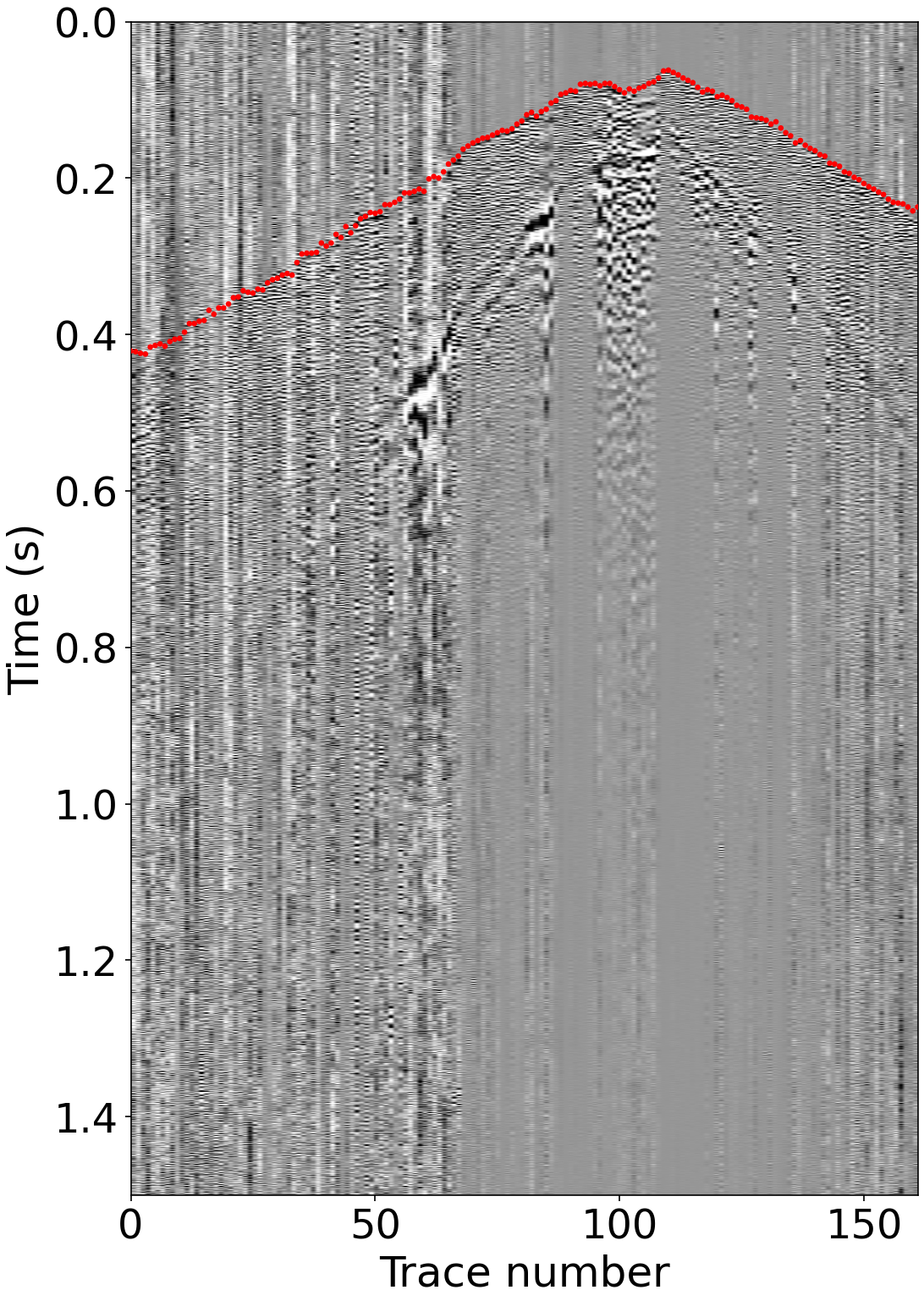}\\[-1pt]
		{\small (f) GeoFormer}
		\label{fig:pick_lr_f}
	\end{minipage}
\end{minipage}
\end{adjustbox}
\caption{Representative first-arrival picking results on the Lalor test set. (a) Seismic shot gather; (b) manual picks; (c) AttnUNet; (d) HU-Net; (e) Vision Transformer; (f) GeoFormer.}
	\label{fig:picking_lalor}
\end{figure}

Figures~\ref{fig:picking_halfmile} and~\ref{fig:picking_lalor} show representative 2D picking results from Halfmile and Lalor. Both shot gathers contain strong noise and missing traces, especially at far offsets. Under these challenging conditions, GeoFormer produces a more continuous picking curve with fewer visible deviations from the manual picks. The geometry injection mechanisms use offset and elevation to reinforce neighboring traces with similar acquisition geometry, which helps stabilize the picks when waveform is degraded. In contrast, Attention UNet, HU-Net, and Vision Transformer exhibit more trace-level discontinuities and large-deviation outliers. Vision Transformer, lacking built-in geometry-aware design, performs competitively on clean traces but degrades sharply under noise, confirming that a Vision Transformer architecture without geometry-aware design is insufficient for robust first-arrival picking.

\begin{figure}[!htbp]
\centering
\begin{adjustbox}{max totalsize={\textwidth}{0.76\textheight},center}
\begin{minipage}{\textwidth}
\centering
\begin{minipage}[b]{0.31\columnwidth}
		\centering
		\includegraphics[width=\linewidth]{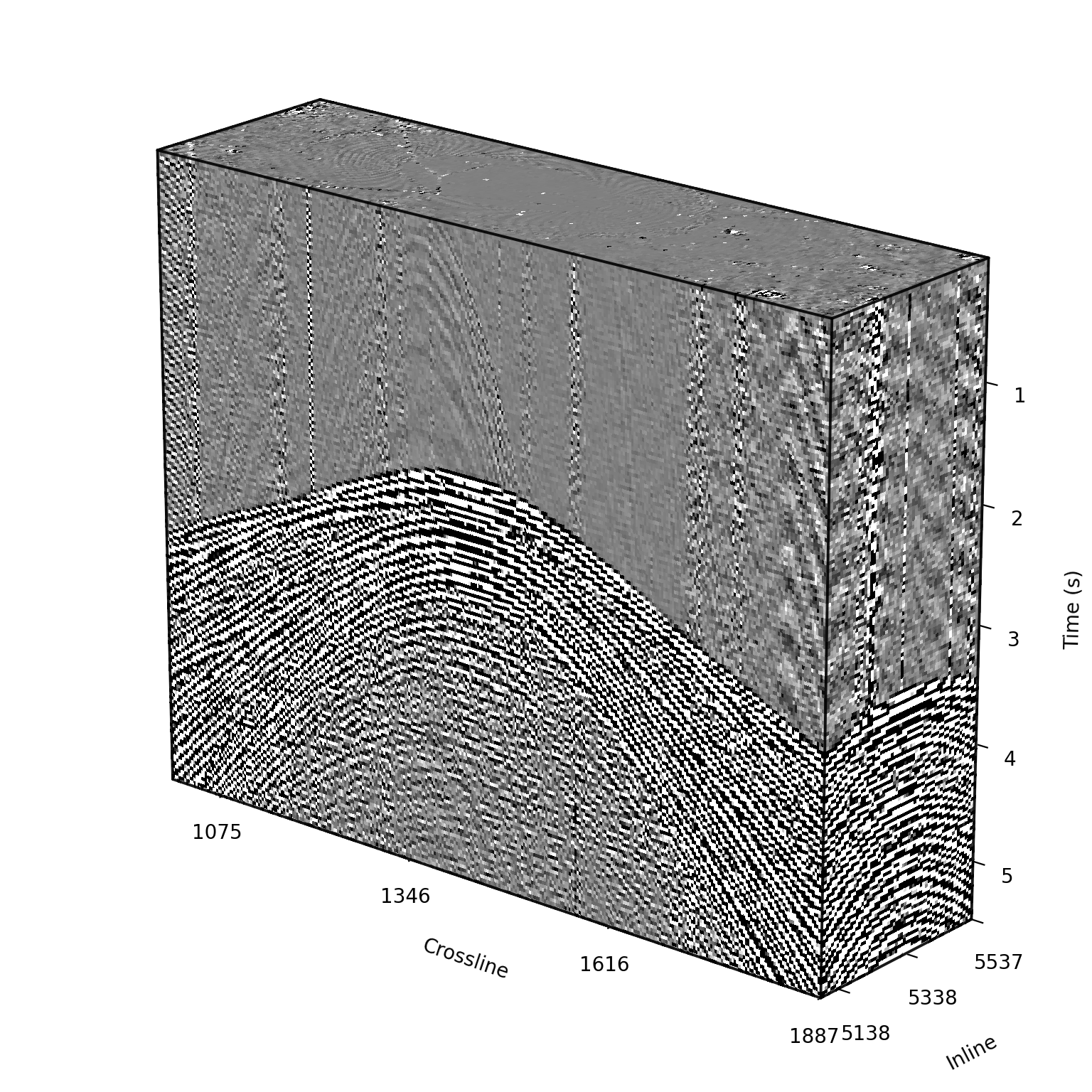}\\[-2pt]
		{\scriptsize (a) Seismic volume}
		\label{fig:sparse_a}
	\end{minipage}\hspace{0.015\columnwidth}\begin{minipage}[b]{0.31\columnwidth}
		\centering
		\includegraphics[width=\linewidth]{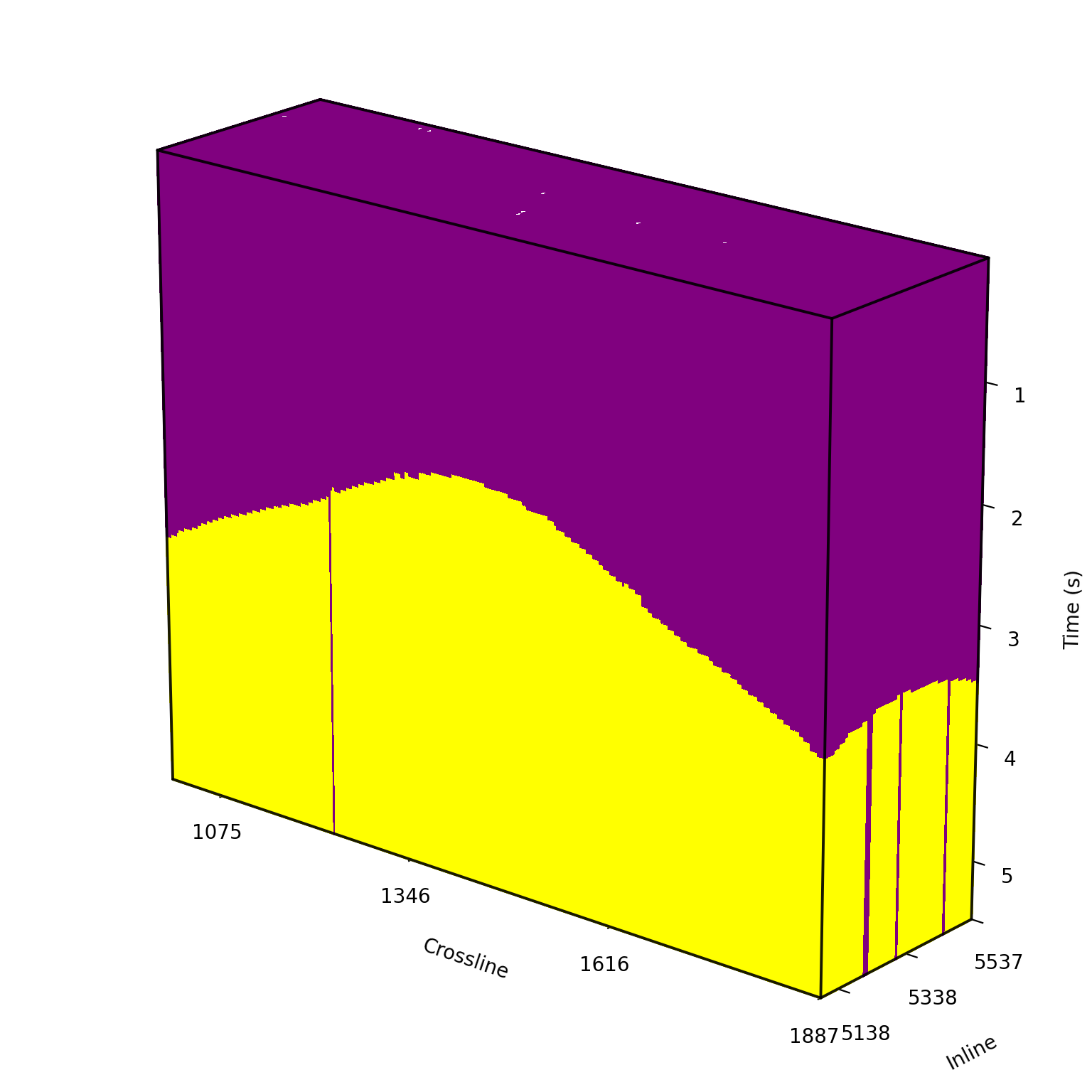}\\[-2pt]
		{\scriptsize (b) Mask, 100\%}
		\label{fig:sparse_b}
	\end{minipage}\hspace{0.015\columnwidth}\begin{minipage}[b]{0.31\columnwidth}
		\centering
		\includegraphics[width=\linewidth]{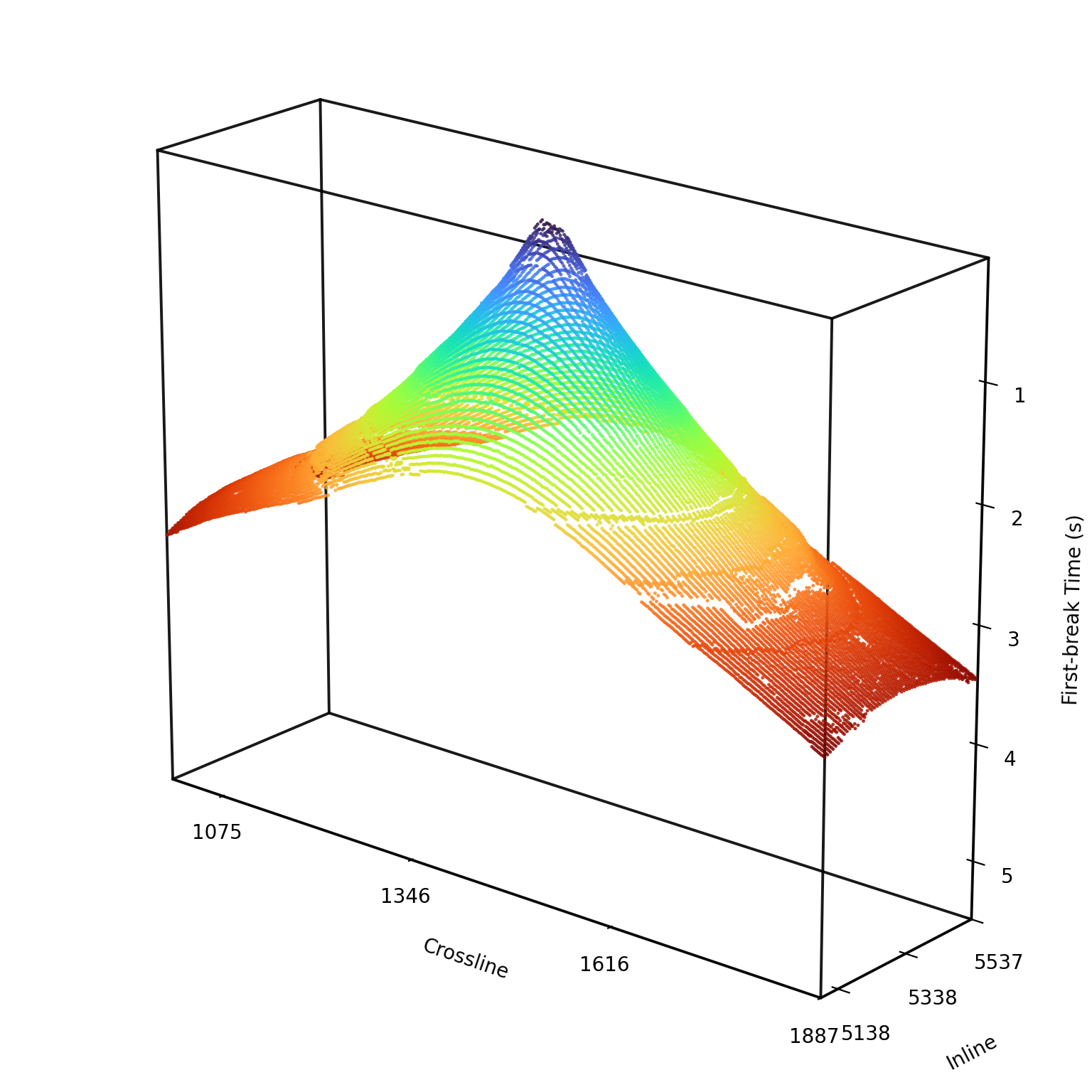}\\[-2pt]
		{\scriptsize (c) Pick, 100\%}
		\label{fig:sparse_c}
	\end{minipage}
\end{minipage}
\end{adjustbox}
\caption{Full-label 3D visualization on the Dongbei dataset. (a) Seismic volume; (b) 100\%-label mask view; (c) 100\%-label first-arrival pick view.}
	\label{fig:dongbei_full_label_3d}
\end{figure}

\begin{figure}[!htbp]
\centering
\begin{adjustbox}{max totalsize={\textwidth}{0.76\textheight},center}
\begin{minipage}{\textwidth}
\centering
\begin{minipage}[b]{0.48\columnwidth}
    \centering
    \includegraphics[width=\linewidth]{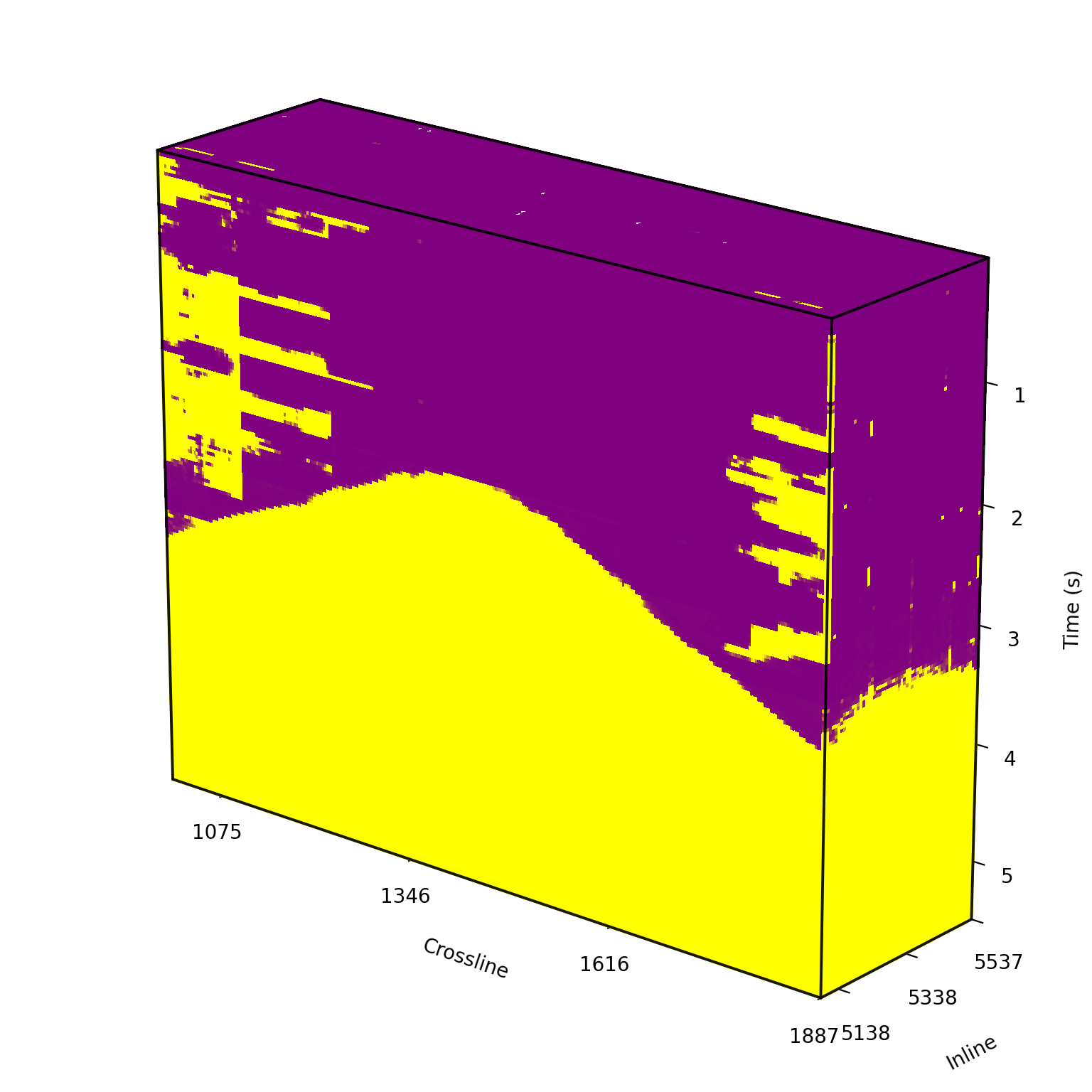}\\[-2pt]
    {\scriptsize (a) Attention UNet}
\end{minipage}\hfill\begin{minipage}[b]{0.48\columnwidth}
    \centering
    \includegraphics[width=\linewidth]{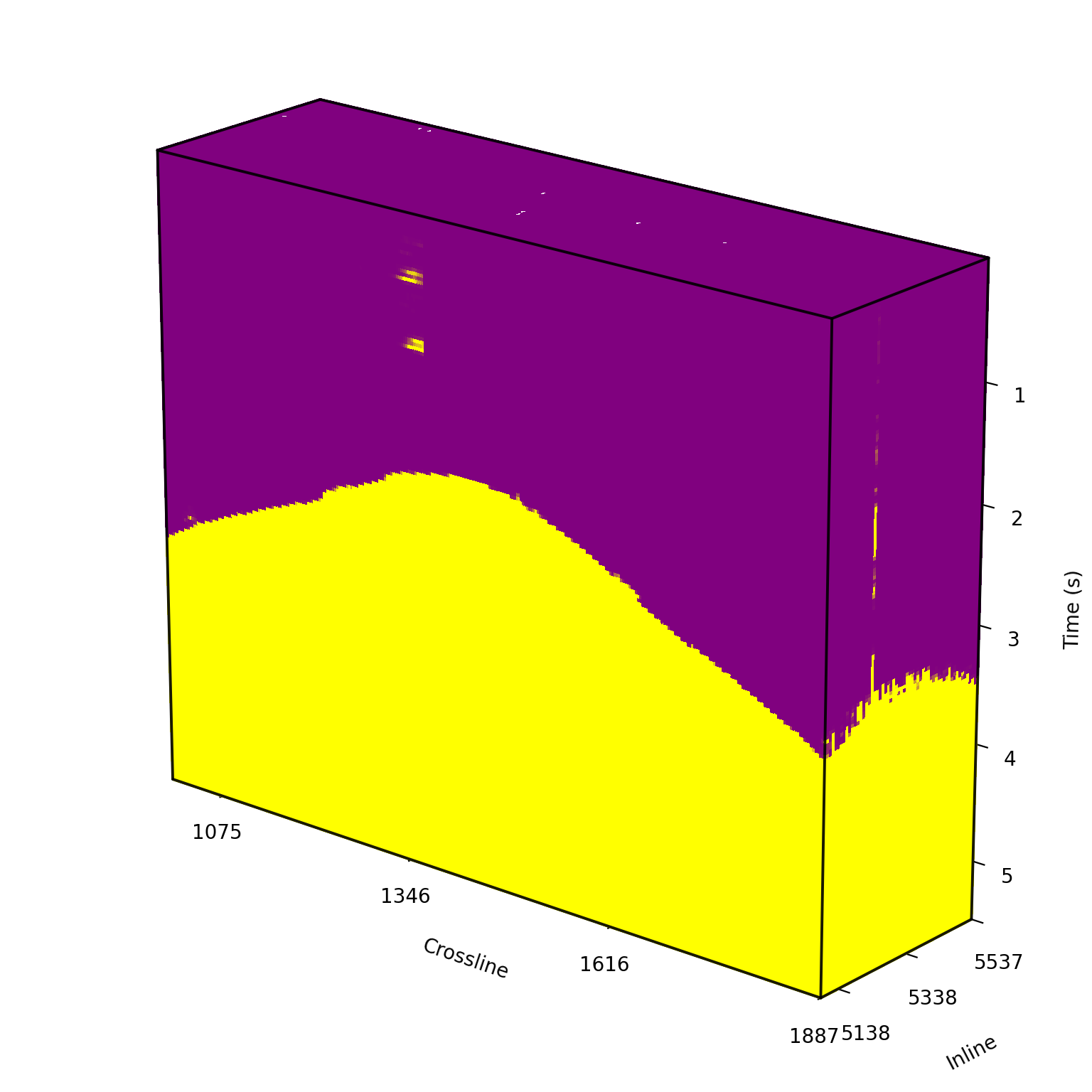}\\[-2pt]
    {\scriptsize (b) HU-Net}
\end{minipage}

\vspace{4pt}

\begin{minipage}[b]{0.48\columnwidth}
    \centering
    \includegraphics[width=\linewidth]{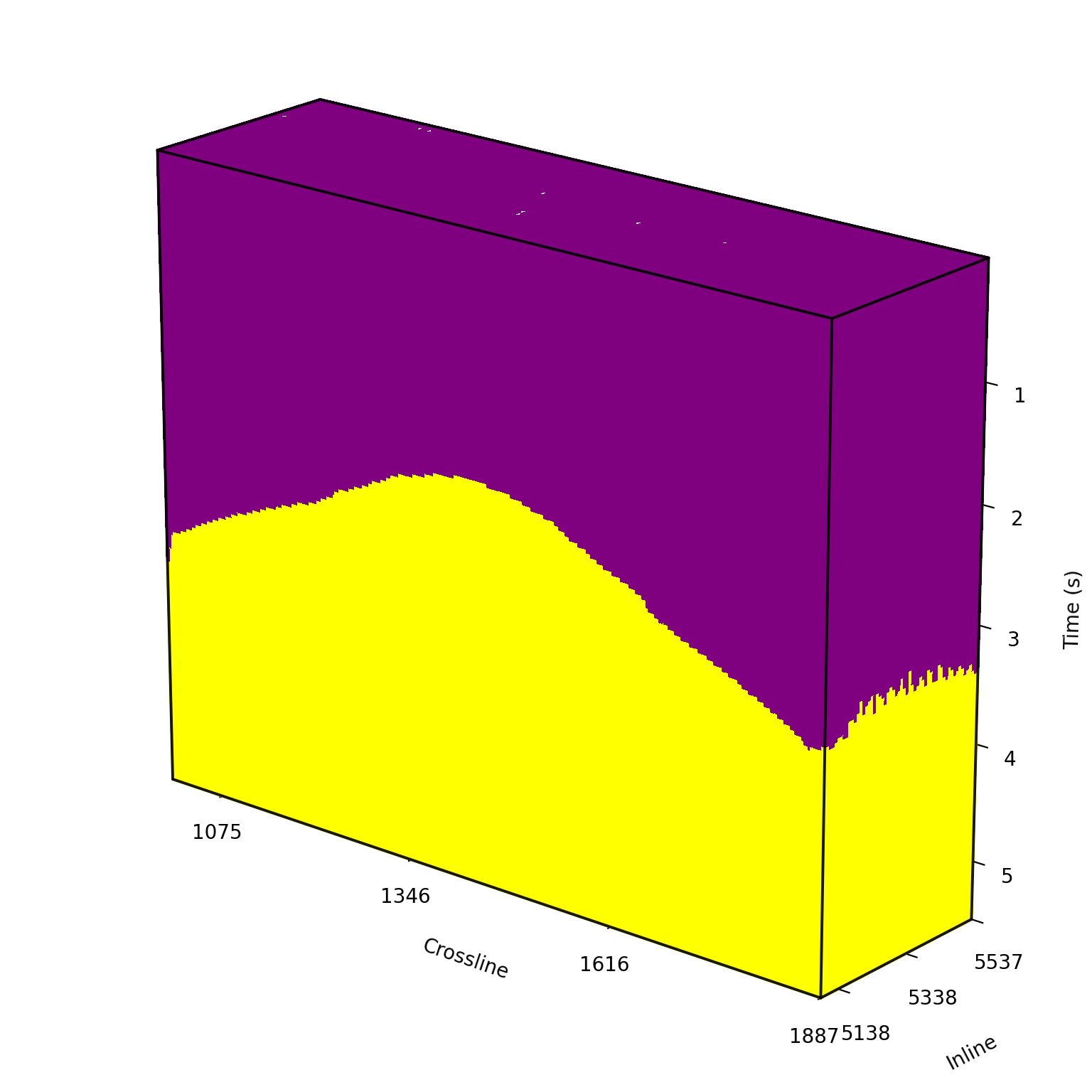}\\[-2pt]
    {\scriptsize (c) Vision Transformer}
\end{minipage}\hfill\begin{minipage}[b]{0.48\columnwidth}
    \centering
    \includegraphics[width=\linewidth]{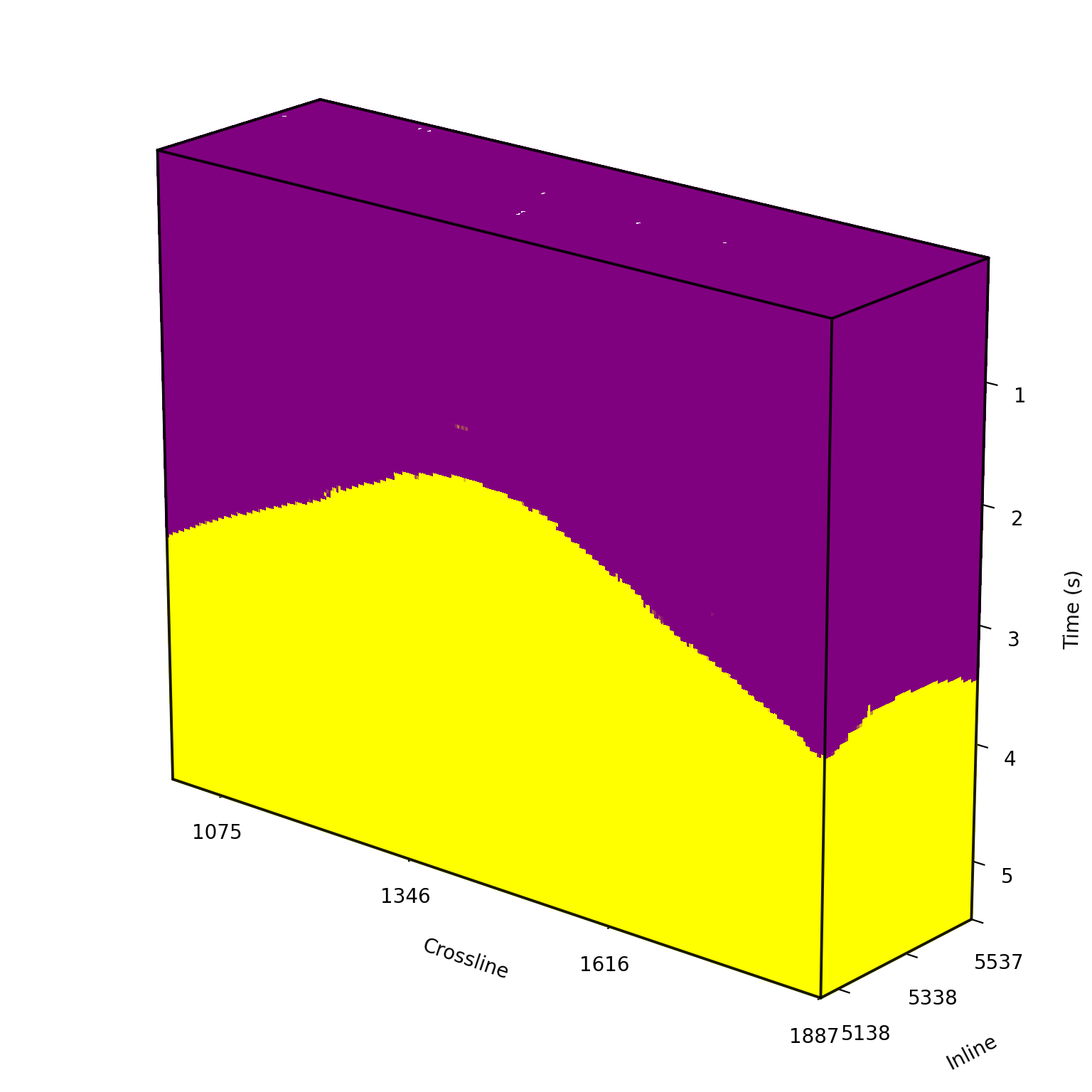}\\[-2pt]
    {\scriptsize (d) GeoFormer}
\end{minipage}
\end{minipage}
\end{adjustbox}
\caption{3D mask prediction comparison on the Dongbei dataset. (a) Attention UNet; (b) HU-Net; (c) Vision Transformer; (d) GeoFormer. GeoFormer produces a cleaner first-arrival mask that remains continuous across the 3D acquisition space.}
\label{fig:dongbei_mask3d_baseline}
\end{figure}

\begin{figure}[!htbp]
\centering
\begin{adjustbox}{max totalsize={\textwidth}{0.76\textheight},center}
\begin{minipage}{\textwidth}
\centering
\begin{minipage}[b]{0.48\columnwidth}
    \centering
    \includegraphics[width=\linewidth]{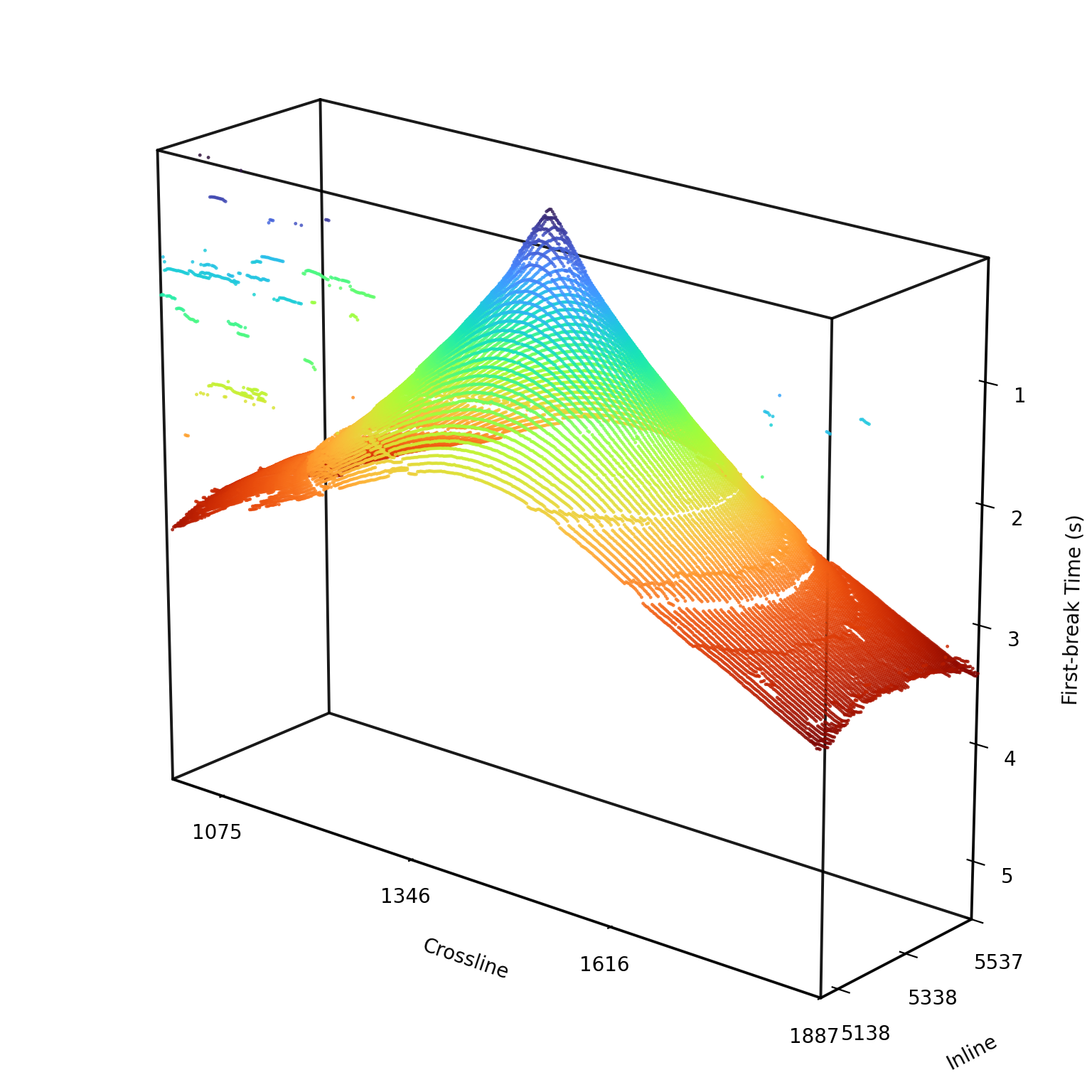}\\[-2pt]
    {\scriptsize (a) Attention UNet}
\end{minipage}\hfill\begin{minipage}[b]{0.48\columnwidth}
    \centering
    \includegraphics[width=\linewidth]{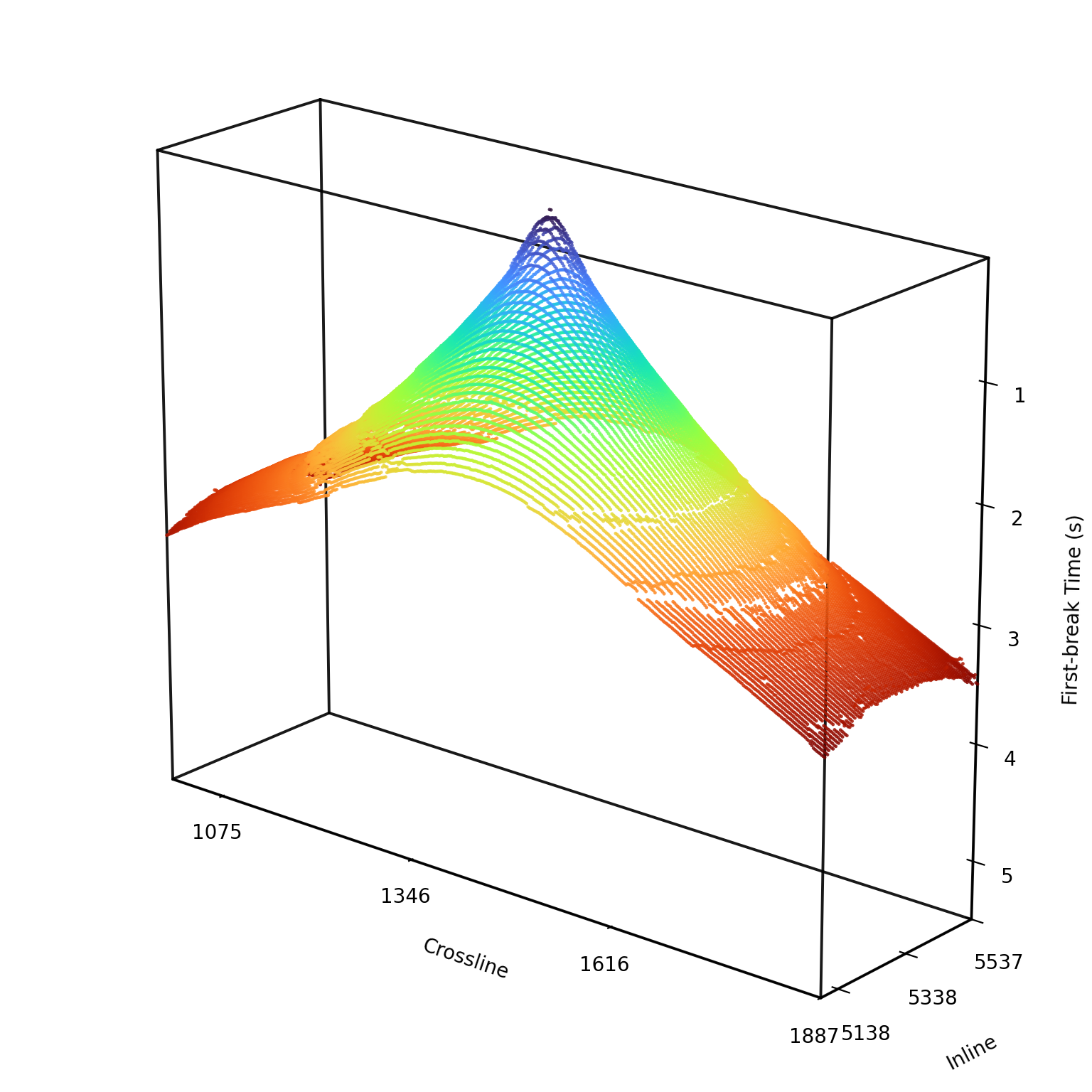}\\[-2pt]
    {\scriptsize (b) HU-Net}
\end{minipage}

\vspace{4pt}

\begin{minipage}[b]{0.48\columnwidth}
    \centering
    \includegraphics[width=\linewidth]{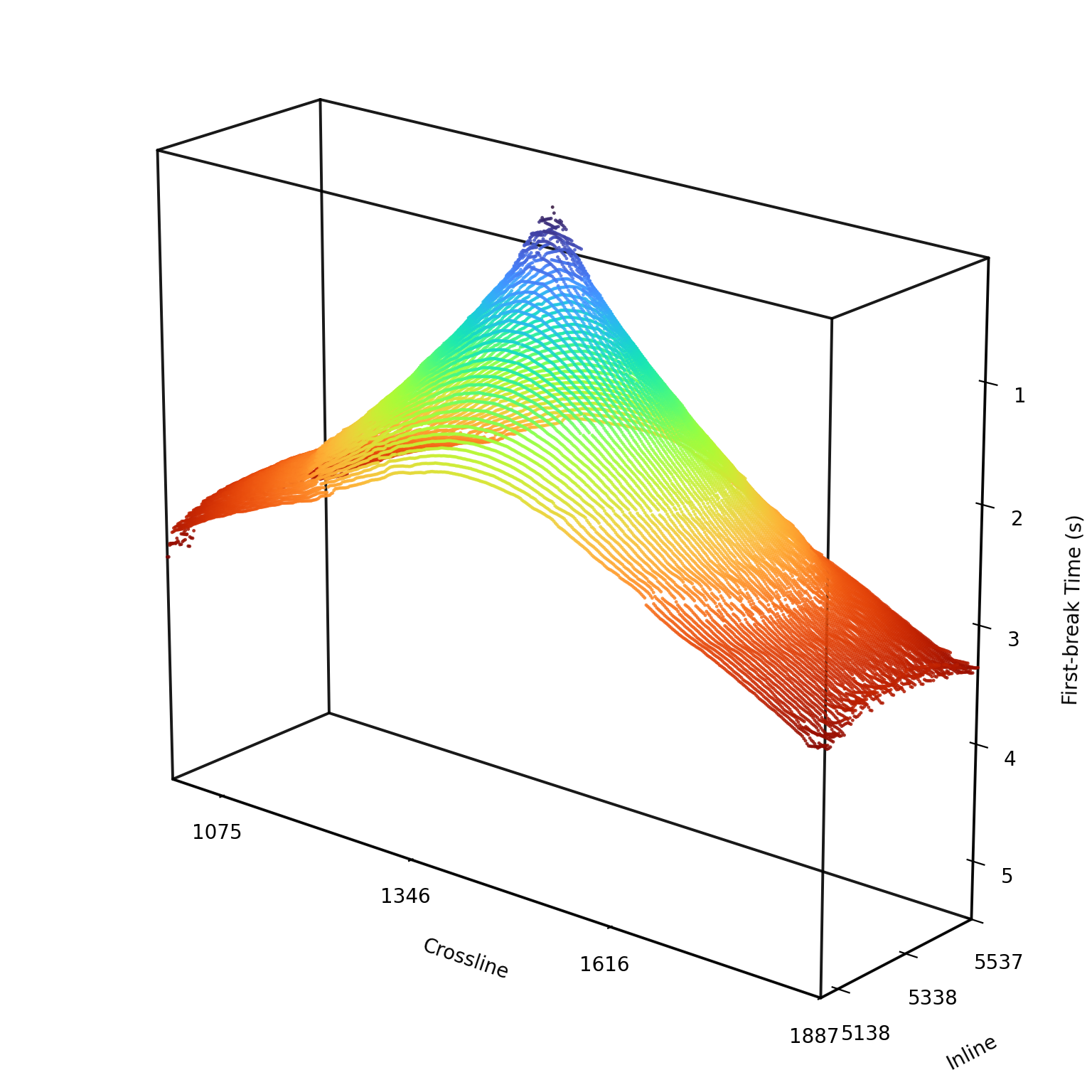}\\[-2pt]
    {\scriptsize (c) Vision Transformer}
\end{minipage}\hfill\begin{minipage}[b]{0.48\columnwidth}
    \centering
    \includegraphics[width=\linewidth]{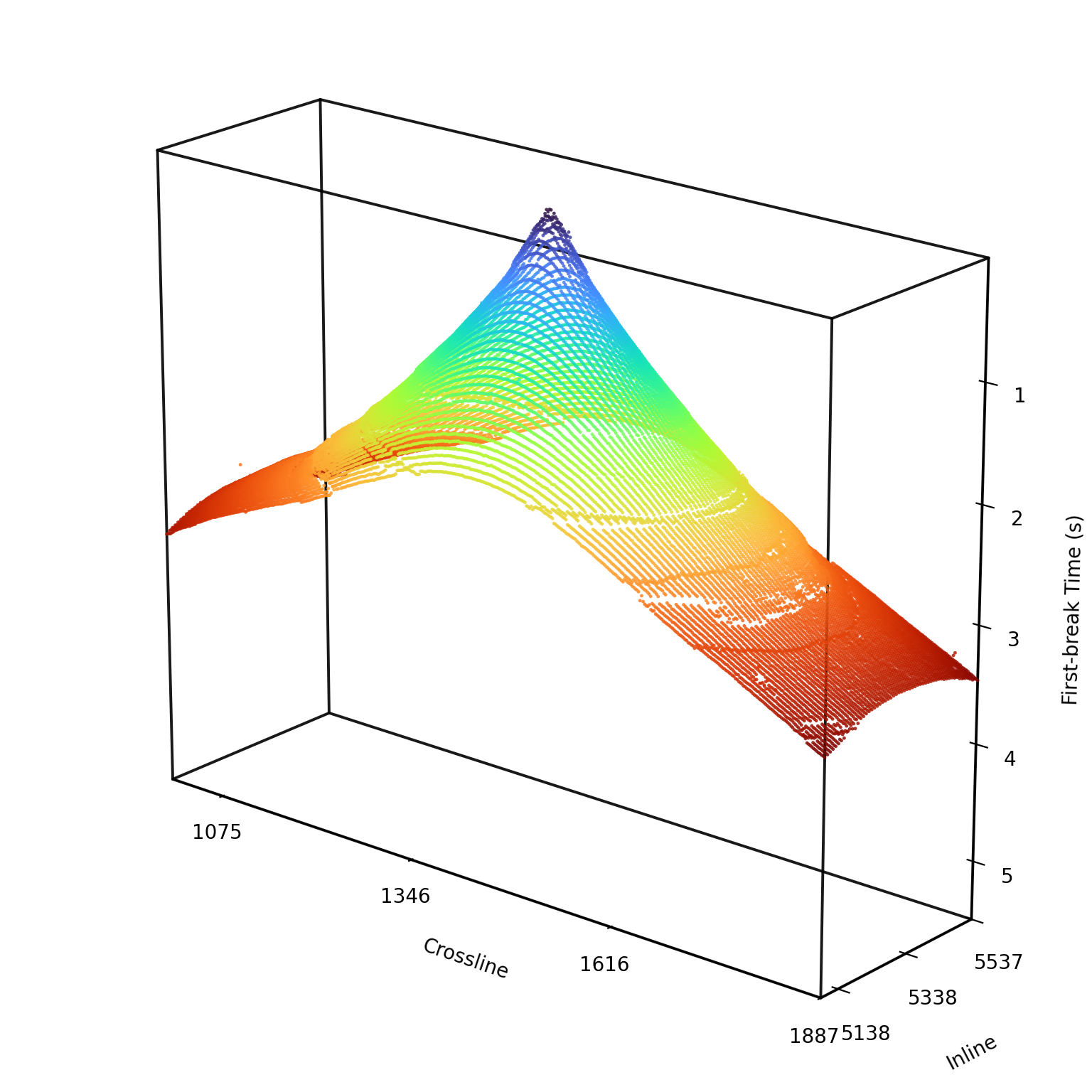}\\[-2pt]
    {\scriptsize (d) GeoFormer}
\end{minipage}
\end{minipage}
\end{adjustbox}
\caption{3D first-arrival picking comparison on the Dongbei dataset. (a) Attention UNet; (b) HU-Net; (c) Vision Transformer; (d) GeoFormer. GeoFormer better preserves the spatial continuity of the first-arrival surface and suppresses scattered outlier picks.}
\label{fig:dongbei_pick3d_baseline}
\end{figure}

\begin{figure}[!htbp]
\centering
\begin{adjustbox}{max totalsize={\textwidth}{0.76\textheight},center}
\begin{minipage}{\textwidth}
\centering
\begin{minipage}[b]{0.48\columnwidth}
    \centering
    \includegraphics[width=\linewidth]{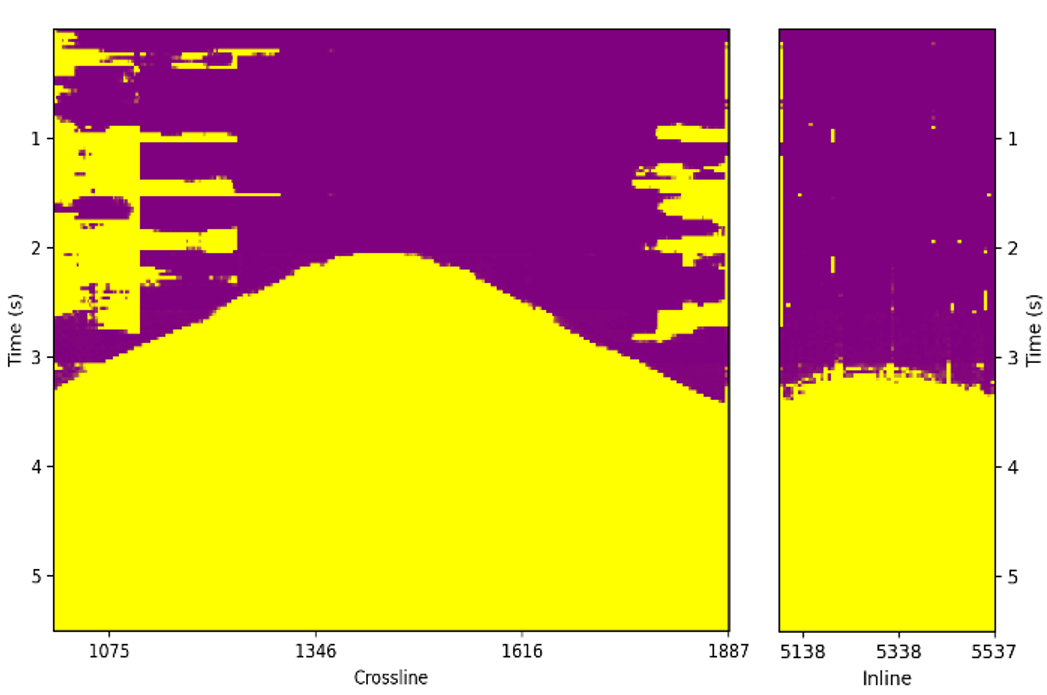}\\[-2pt]
    {\scriptsize (a) Attention UNet}
\end{minipage}\hfill\begin{minipage}[b]{0.48\columnwidth}
    \centering
    \includegraphics[width=\linewidth]{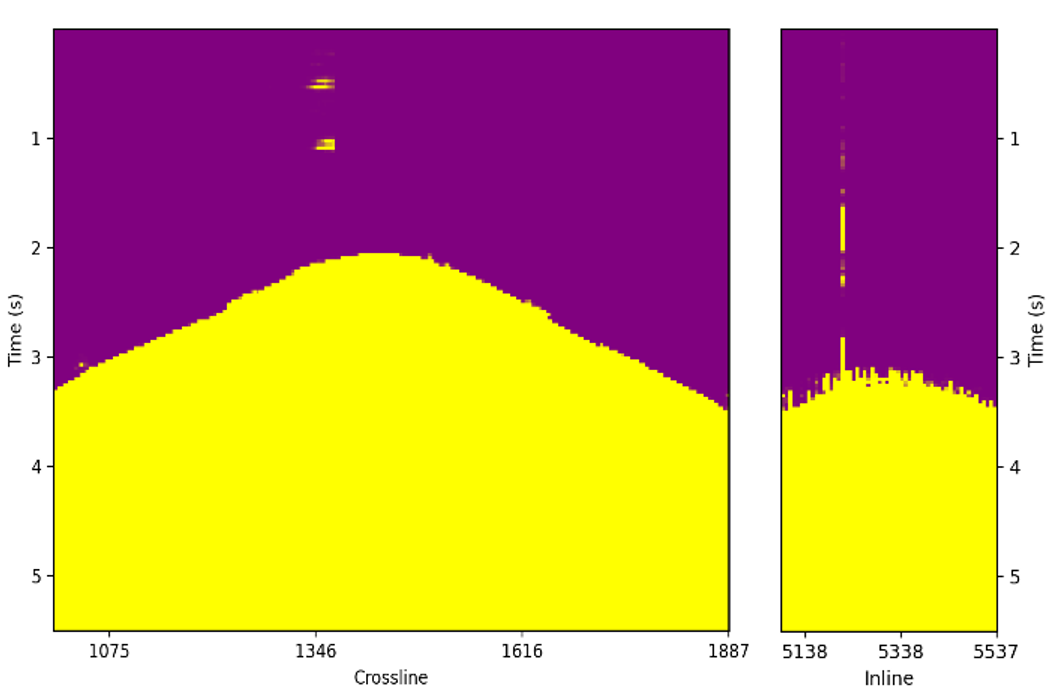}\\[-2pt]
    {\scriptsize (b) HU-Net}
\end{minipage}

\vspace{4pt}

\begin{minipage}[b]{0.48\columnwidth}
    \centering
    \includegraphics[width=\linewidth]{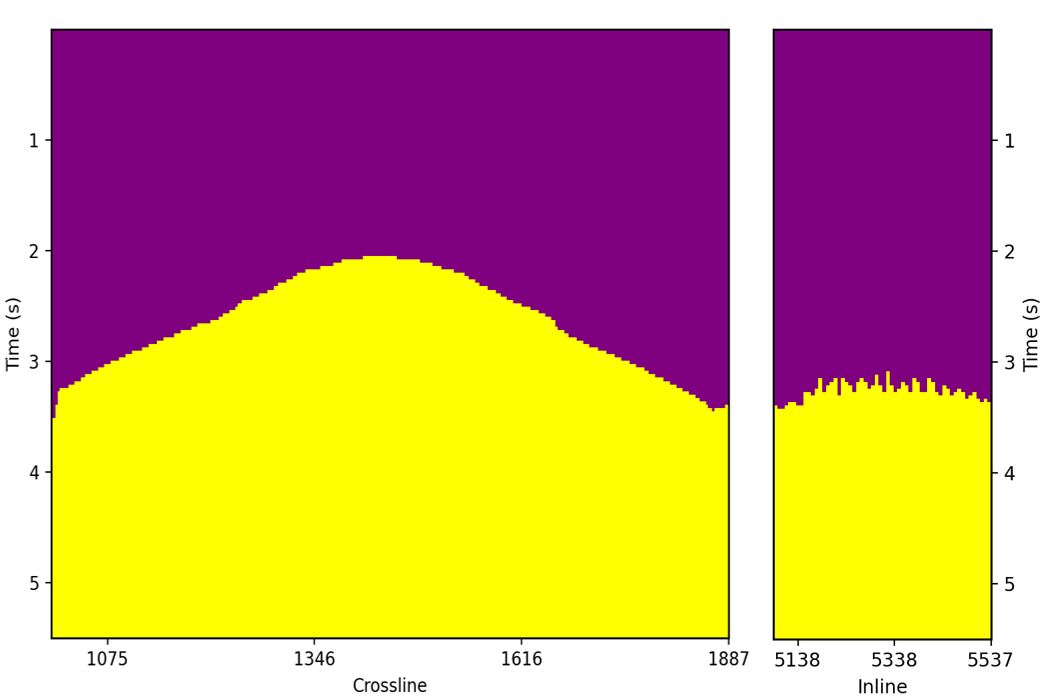}\\[-2pt]
    {\scriptsize (c) Vision Transformer}
\end{minipage}\hfill\begin{minipage}[b]{0.48\columnwidth}
    \centering
    \includegraphics[width=\linewidth]{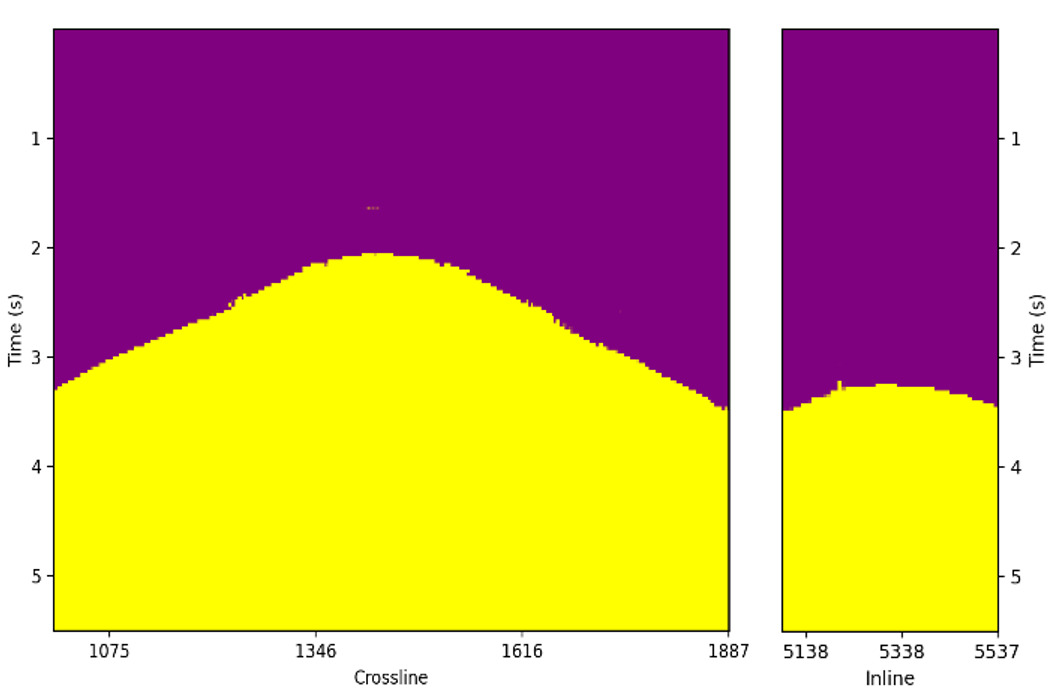}\\[-2pt]
    {\scriptsize (d) GeoFormer}
\end{minipage}
\end{minipage}
\end{adjustbox}
\caption{2D mask prediction comparison on representative Dongbei crossline sections. (a) Attention UNet; (b) HU-Net; (c) Vision Transformer; (d) GeoFormer. GeoFormer yields a smoother and more continuous mask boundary with fewer spurious high-confidence regions.}
\label{fig:dongbei_mask2d_baseline}
\end{figure}

\begin{figure}[!htbp]
\centering
\begin{adjustbox}{max totalsize={\textwidth}{0.76\textheight},center}
\begin{minipage}{\textwidth}
\centering
\begin{minipage}[b]{0.48\columnwidth}
    \centering
    \includegraphics[width=\linewidth]{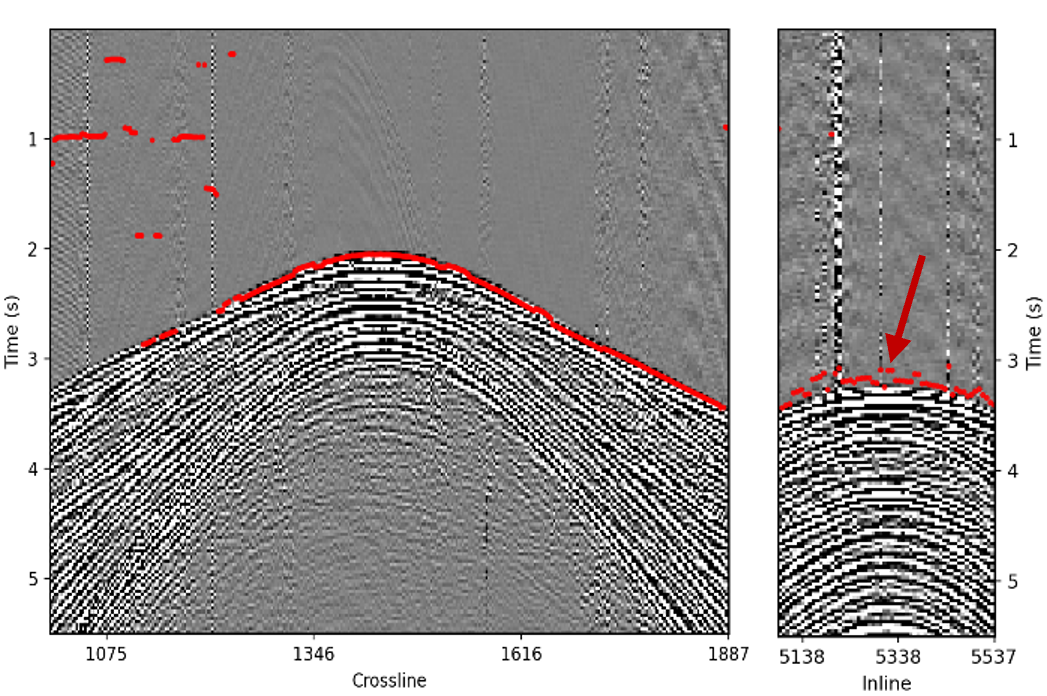}\\[-2pt]
    {\scriptsize (a) Attention UNet}
\end{minipage}\hfill\begin{minipage}[b]{0.48\columnwidth}
    \centering
    \includegraphics[width=\linewidth]{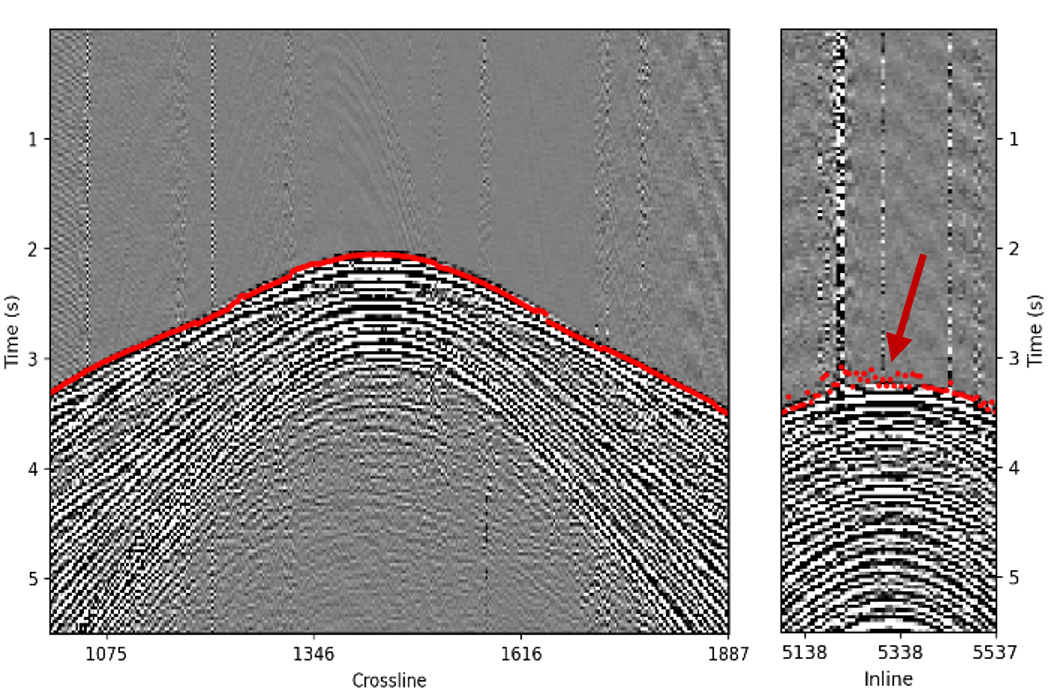}\\[-2pt]
    {\scriptsize (b) HU-Net}
\end{minipage}

\vspace{4pt}

\begin{minipage}[b]{0.48\columnwidth}
    \centering
    \includegraphics[width=\linewidth]{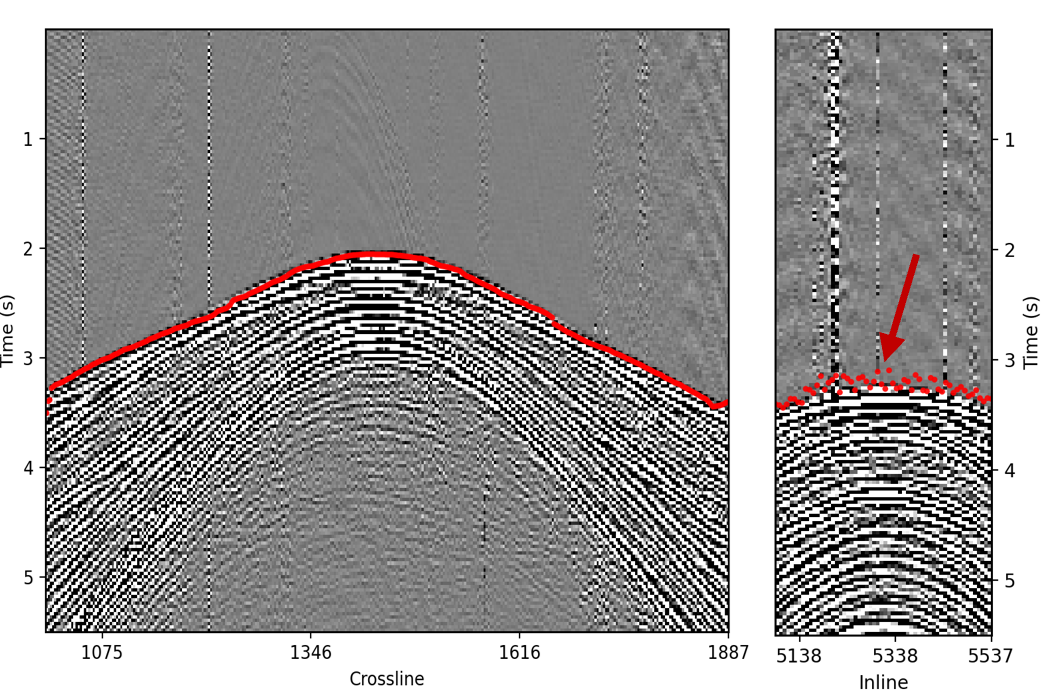}\\[-2pt]
    {\scriptsize (c) Vision Transformer}
\end{minipage}\hfill\begin{minipage}[b]{0.48\columnwidth}
    \centering
    \includegraphics[width=\linewidth]{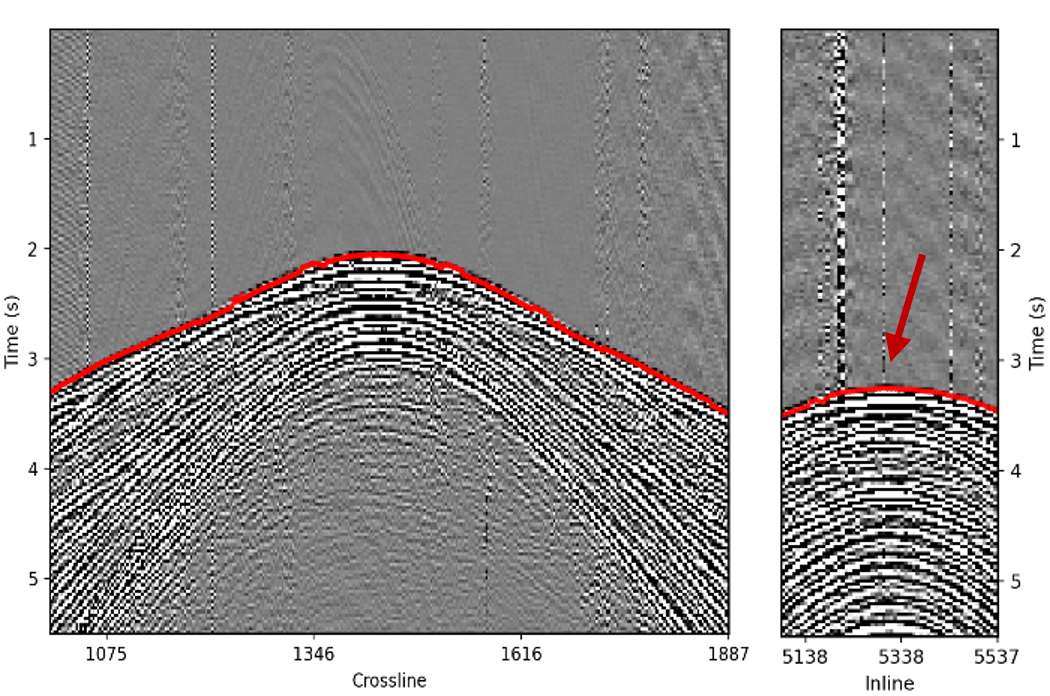}\\[-2pt]
    {\scriptsize (d) GeoFormer}
\end{minipage}
\end{minipage}
\end{adjustbox}
\caption{2D first-arrival picking comparison on representative Dongbei crossline sections. (a) Attention UNet; (b) HU-Net; (c) Vision Transformer; (d) GeoFormer. GeoFormer produces the most continuous crossline picking curve and reduces local outliers in noisy regions.}
\label{fig:dongbei_pick2d_baseline}
\end{figure}

Figure~\ref{fig:dongbei_full_label_3d} shows the full-label (100\%) 3D visualization of the Dongbei dataset as a reference. Figures~\ref{fig:dongbei_mask3d_baseline}--\ref{fig:dongbei_pick2d_baseline} compare the four methods on Dongbei in both 3D volumetric and 2D section. In the 3D mask visualizations, Attention UNet produces artifacts above the first-arrival boundary. HU-Net reduces some artifacts but still leaves isolated false positives. GeoFormer yields a cleaner mask that follows the expected first-arrival surface more closely and remains continuous across the 3D acquisition space. Vision Transformer, lacking geometry-aware design, produces artifacts comparable to Attention UNet in the 3D mask and yields scattered picks in the 3D visualization. The 3D picking visualizations show a similar pattern: the baseline methods produce scattered or locally unstable picks, whereas GeoFormer preserves a smoother and more spatially continuous first-arrival surface. This advantage is also visible in the 2D slices extracted from the 3D result. Along the crossline direction, the baseline methods show jittered pick trajectories, and Vision Transformer exhibits similar instability, while GeoFormer maintains a more continuous curve, indicating that geometry-aware design, rather than the self-attention mechanism per se, improves global consistency over 2D picking models.

\begin{figure}[!htbp]
\centering
\begin{adjustbox}{max totalsize={\textwidth}{0.76\textheight},center}
\begin{minipage}{\textwidth}
\centering
\begin{minipage}[b]{0.48\columnwidth}
		\centering
		\includegraphics[width=\linewidth]{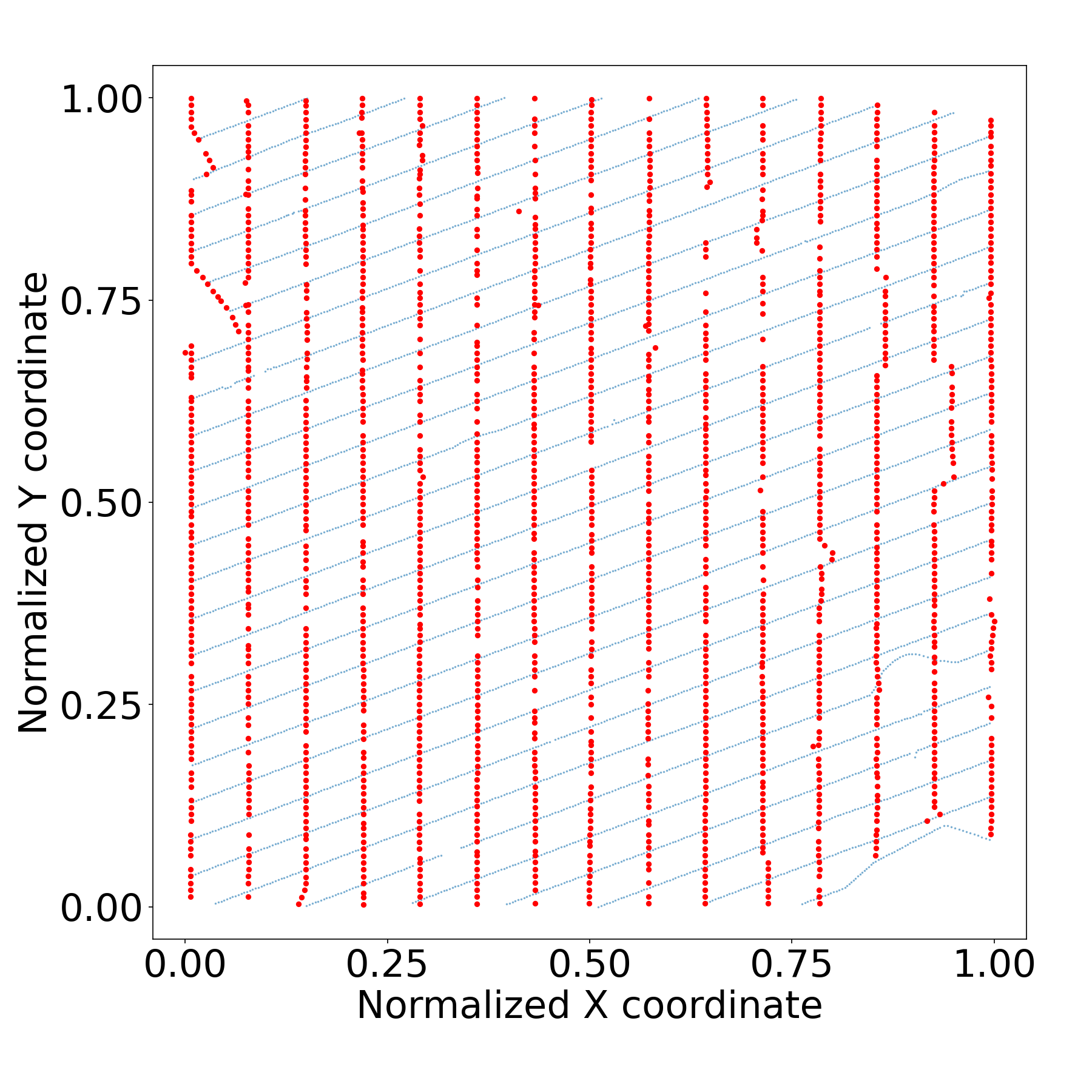}\\[-2pt]
		{\small (a) Brunswick}
		\label{fig:brunswick_geometry}
	\end{minipage}\hspace{0.02\columnwidth}\begin{minipage}[b]{0.48\columnwidth}
		\centering
		\includegraphics[width=\linewidth]{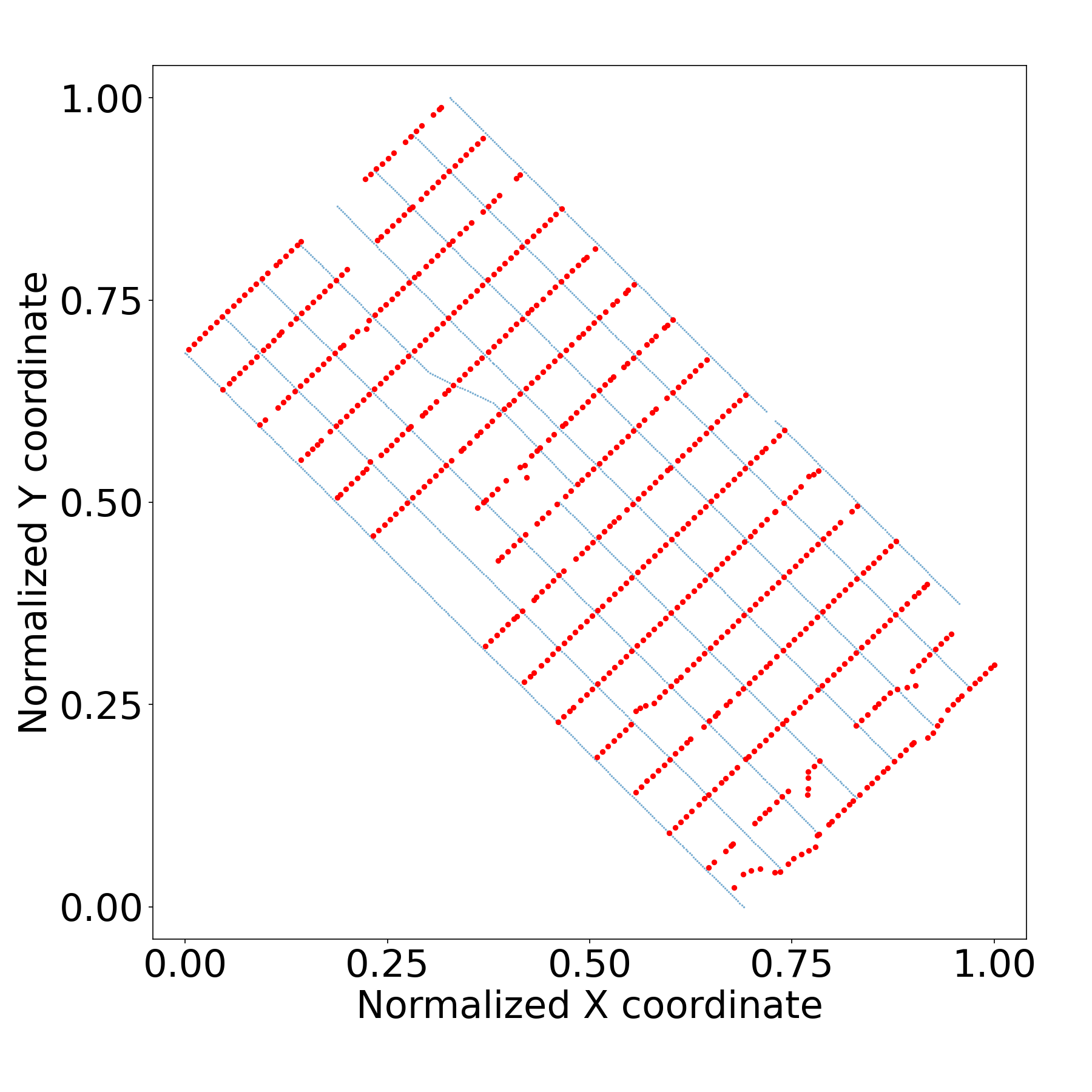}\\[-2pt]
		{\small (b) Halfmile}
		\label{fig:halfmile_geometry}
	\end{minipage}
\end{minipage}
\end{adjustbox}
\caption{Acquisition geometry of Brunswick and Halfmile used in cross-site transfer experiments.}
	\label{fig:survey_geometry}
\end{figure}

\subsection{Ablation Study}

The geometry injection mechanisms are analyzed on the first-arrival picking task as a controlled setting with dense trace-level supervision and stable evaluation metrics. Noise attenuation and first-arrival picking remain equally weighted task-level evaluations representing regression and segmentation, respectively. Restricting the component-level ablation to one controlled task avoids repeating the same architectural analysis and does not assign greater importance to either application.

To quantify the individual contribution of each geometry injection mechanism, systematic ablation experiments are conducted. Starting from the full GeoFormer configuration, we individually disable GeomMLP, GeomAdaLN, and GeomAttnBias. Table~\ref{tab:ablation} summarizes the results.

The ablation results quantify how each geometric mechanism contributes to picking accuracy. Removing all three mechanisms effectively reverts GeoFormer to a standard trace-level Transformer without geometry awareness; RMSE increases from 1.84 to 10.30, a factor of 5.6. 
Among individual mechanisms, GeomMLP removal produces the largest degradation, with HR@1 dropping by 4.7 points and RMSE increases from 1.84 to 5.55. This outsized impact reflects its role as the entry point of geometric attributes: without GeomMLP, the downstream GeomAdaLN and GeomAttnBias operate on tokens that carry no geometric representation, effectively disabling the entire geometry pipeline. Removing only GeomAttnBias or GeomAdaLN produces more moderate degradation (RMSE 2.87 and 2.90), with comparable impacts, indicating that normalization-level modulation and attention-level bias are roughly equally important once the geometric representation is established. The progressive degradation from Full GeoFormer (1.84) through single-mechanism removal (2.87--5.55) to all-mechanism removal (10.30) directly validates the contribution of each design choice.

We further evaluate cross-site transfer and parameter-efficient adaptation using limited target-site annotations.

\begin{table}[!htbp]
\centering
\small
\setlength{\tabcolsep}{3pt}
\caption{Cross-site transfer results from Brunswick to Halfmile. Best results among transfer settings are in \textbf{bold}.}
\label{tab:cross_b2h}
\begin{adjustbox}{max width=\textwidth}
\begin{tabular}{lcccccccc}
\toprule
Setting & RMSE & MAE & MBE & HR@1 & HR@3 & HR@5 & HR@7 & HR@9 \\
\midrule
Zero-shot (Brunswick $\rightarrow$ Halfmile) & 53.05 & 26.38 & $+$26.19 & 59.0 & 63.6 & 65.8 & 67.1 & 67.9 \\
$+$ FT 5\% target data & \textbf{3.18} & \textbf{1.39} & $+$0.24 & \textbf{79.1} & \textbf{88.6} & \textbf{92.9} & \textbf{95.4} & \textbf{97.0} \\
$+$ FT 10\% target data & 3.27 & 1.49 & \textbf{$+$0.05} & 77.3 & 87.5 & 92.2 & 95.0 & 96.8 \\
$+$ FT 15\% target data & 3.25 & 1.52 & \textbf{$-$0.05} & 76.5 & 87.1 & 92.0 & 95.0 & 96.8 \\
\midrule
Target-site supervised (Halfmile) & 1.84 & 0.80 & $+$0.12 & 88.3 & 94.0 & 97.2 & 98.7 & 99.4 \\
\bottomrule
\end{tabular}
\end{adjustbox}
\end{table}

\begin{table}[!htbp]
\centering
\small
\setlength{\tabcolsep}{3pt}
\caption{Cross-site transfer results from Halfmile to Brunswick. Best results among transfer settings are in \textbf{bold}.}
\label{tab:cross_h2b}
\begin{adjustbox}{max width=\textwidth}
\begin{tabular}{lcccccccc}
\toprule
Setting & RMSE & MAE & MBE & HR@1 & HR@3 & HR@5 & HR@7 & HR@9 \\
\midrule
Zero-shot (Halfmile $\rightarrow$ Brunswick) & 57.74 & 21.18 & $-$21.09 & 53.3 & 66.4 & 68.8 & 68.9 & 68.9 \\
$+$ FT 5\% target data & 3.13 & 1.23 & $+$0.46 & 83.5 & 90.3 & 93.1 & 95.3 & 97.0 \\
$+$ FT 10\% target data & \textbf{2.70} & \textbf{0.99} & \textbf{$+$0.30} & 86.6 & \textbf{92.4} & \textbf{94.7} & \textbf{96.4} & \textbf{97.8} \\
$+$ FT 15\% target data & 2.84 & \textbf{0.99} & $+$0.48 & \textbf{87.3} & 92.3 & 94.3 & 95.9 & 97.3 \\
\midrule
Target-site supervised (Brunswick) & 1.73 & 0.47 & $+$0.09 & 93.8 & 97.5 & 98.5 & 99.0 & 99.3 \\
\bottomrule
\end{tabular}
\end{adjustbox}
\end{table}

\subsection{Cross-Site Transfer}
\label{sec:cross_site}

GeoFormer relies on acquisition geometry. A natural concern is whether this dependence limits generalization to sites with different acquisition geometry. Because first arrivals are governed by acquisition geometry, explicitly encoding geometric attributes couples part of the model to the training-site acquisition geometry. The question is not whether this coupling exists. It is whether the coupling can be managed in practice.

We test this issue with transfer experiments between Brunswick and Halfmile. These two sites have substantially different acquisition geometries as shown in Figure~\ref{fig:survey_geometry}, which makes direct model transfer difficult. For each transfer direction, a model is first trained on the source site with full supervision and then evaluated directly on the target site (zero-shot). The same pre-trained model is subsequently fine-tuned on the target site using 5\%, 10\%, and 15\% of the target-site training data. All fine-tuning follows a geometry-only strategy. Only the geometry-related modules are updated. These include GeomMLP, the multi-stage geometry projections and scaling factors, and the global trace context MLP. The remaining modules, including the waveform encoder and decoder, are frozen. The trainable modules account for only 4,363,784 out of 46,061,335 total parameters (9.5\%). Updating only these lightweight geometry modules is sufficient for cross-site adaptation. Full supervised training on the target site serves as the upper bound.

Zero-shot transfer degrades performance substantially in both directions (Tables~\ref{tab:cross_b2h} and~\ref{tab:cross_h2b}). From Brunswick to Halfmile, HR@1 drops from 88.3\% under target-site supervised training to 59.0\%, and RMSE increases from 1.84 to 53.05 samples with an MBE of $+$26.19. From Halfmile to Brunswick, HR@1 drops to 53.3\% and RMSE reaches 57.74 samples with an MBE of $-$21.09. The opposite MBE signs are physically informative: they reflect the different offset distributions between the two sites. A Brunswick-trained model systematically picks later on Halfmile (positive MBE), whereas a Halfmile-trained model picks earlier on Brunswick (negative MBE). This directional bias is exactly what one would expect from a model that has learned a site-specific geometry-traveltime mapping. These results confirm that the geometry modules learn a geometric calibration specific to the training site, and that this calibration does not transfer directly to a site with a different acquisition geometry.

It is important to clarify that GeoFormer is not designed as a geometry-agnostic zero-shot picker. Because first-arrival traveltimes are directly governed by survey geometry, explicitly encoding offset and elevation inevitably couples part of the model to the acquisition configuration of the training site. The degraded zero-shot performance is therefore not an unexpected failure, but a manifestation of this physically interpretable coupling. The same geometric sensitivity that degrades cross-site transfer is what enables the within-site accuracy gains demonstrated in Tables~\ref{tab:main_results} and~\ref{tab:ablation}. The results thus quantify an accuracy-transferability trade-off.

The same experiments show that this coupling can be recalibrated efficiently with limited target-site annotation. Geometry-only fine-tuning with 5\% target-site data recovers most of the lost performance in both directions. From Halfmile to Brunswick, HR@1 improves from 53.3\% to 83.5\%, and RMSE decreases from 57.74 to 3.13. The MBE magnitude drops from 21.09 to 0.46 samples, indicating that the systematic bias is largely corrected. With 10\% data, HR@1 reaches 86.6\% and RMSE decreases further to 2.70. From Brunswick to Halfmile, 5\% fine-tuning restores HR@1 to 79.1\% and reduces RMSE from 53.05 to 3.18. In both directions, the 5\% setting already approaches the performance of the 10\% and 15\% settings. 

These results support an architectural interpretation with practical implications. GeoFormer's geometry modules function as a site-specific geometric calibrator layered on a transferable waveform feature extractor. Only 9.5\% of the model parameters require updating during cross-site adaptation, and the pre-trained encoder-decoder preserves its waveform representation capability across sites. This design fits routine deployment workflows: acquiring a small number of labeled shot gathers at a new survey site is standard quality-control practice. Rather than relying on zero-shot generalization, GeoFormer provides a parameter-efficient pathway to adapt a geometry-aware picker to each new survey.

\section{Conclusion}

We propose GeoFormer, a geometry-aware Transformer for five-dimensional (5D) prestack seismic data. It performs trace-level tokenization, representing each seismic trace through its waveform and acquisition geometry. To incorporate this information, GeoFormer introduces coordinate-based positional encoding and three geometry injection modules: GeomMLP, GeomAdaLN, and GeomAttnBias, operating at the token, normalization, and attention levels, respectively.

We validate GeoFormer on two representative seismic processing tasks, seismic noise attenuation and first-arrival picking, covering both regression and segmentation. Extensive experiments demonstrate that GeoFormer consistently outperforms the baseline methods on both tasks. For noise attenuation, comprehensive comparisons are conducted on 3D seismic volumes, 2D slices, local magnified views, and $f$-$k$ spectra. GeoFormer achieves more effective suppression of ground roll and random noise and better preserves reflection signals than both the latest task-specific noise-attenuation method and Vision Transformer. For first-arrival picking, GeoFormer reduces RMSE by factors of 4.1--8.5 compared with the standard Vision Transformer across the four field datasets and also outperforms task-specific methods including Attention UNet and HU-Net. Ablation studies further demonstrate that GeomMLP, GeomAdaLN, and GeomAttnBias each contribute to the overall performance. GeoFormer also supports parameter-efficient adaptation across survey sites, where fine-tuning only 9.5\% of the parameters with 5\% of the target-site labeled traces recovers most of the fully supervised performance.

These results highlight the importance of explicitly incorporating acquisition geometry into Transformer-based prestack seismic processing. GeoFormer provides a unified, geometry-aware framework that can potentially be extended to other prestack tasks, such as deblending and interpolation, and adapted to diverse acquisition geometries.
\section*{Data and Code Availability}

The SEG C3NA velocity model used in the noise-attenuation
experiments is publicly available from the SEG Wiki
\citep{segc3na_velocity}.

The Brunswick, Halfmile Lake, and Lalor seismic datasets,
including first-arrival labels, used in the first-arrival
picking experiments are publicly available through the
Hardpicks project and hosted by Natural Resources Canada
\citep{hardpicks_dataset}.

Version 1.0.1 of the GeoFormer source code developed
for this study is archived in Zenodo
\citep{gao2026geoformer} and is also publicly available
through the GitHub development repository
\url{https://github.com/ZDDWLIG/GeoFormer}.

\clearpage
\bibliographystyle{seg}
\bibliography{references}
\end{document}